\documentclass[11pt]{article}
\usepackage{amsmath,amssymb}
\usepackage[all]{xy}

\newtheorem{definition}{\noindent \noindent {\bf Definition}}[section]
\newtheorem{lemma}{{\bf Lemma}}[section]
\newtheorem{theorem}{Theorem}[section]
\newtheorem{proposition}{{\bf Proposition}}[section]
\newtheorem{corollary}{{\bf Corollary}}[section]
\newtheorem{remark}{{\bf Remark}}[section]

\newtheorem{state}{{\bf Statement}}[section]

 1   
 1  

\def\qed{\ifvmode\Realemovelastskip\fi
{\unskip\nobreak\hfil\penalty50\hbox{}\nobreak\hfil \hbox{\vrule
height1.2ex width1.2ex}\parfillskip=0pt \finalhyphendemerits=0
\par\smallskip}}
\newenvironment{proof}{\noindent{(Proof)~}}{\qed\medskip}
\def\QED{\hskip0.1em\hfill\null\ \null\nobreak\hfill
\kern3pt\lower1.8pt\vbox{\hrule\hbox {\vrule\kern1pt\vbox{\kern1.7pt
\hbox{$\scriptstyle QED$}\kern0.2pt}\kern1pt\vrule}\hrule}}
\def\hook{\, \hbox to 15pt{\vbox{\vskip 6pt\hrule width 8pt height 1pt}
        \kern -5pt\vrule height 8pt width 1pt\hfil}}

\def\r{\ensuremath{\mathbb{R}}}
\def\rk{{\mathbb R}^{k}}
\def\rkq{\rk \times Q}

\def\rktkq{\rk\times T^1_kQ}

\def\bea{\begin{eqnarray}}
\def\eea{\end{eqnarray}}
\def\beann{\begin{eqnarray*}}
\def\eeann{\end{eqnarray*}}

\newcommand{\ds}{\displaystyle}

\def\Cinfty{{\rm C}^\infty}
\def\derpar#1#2{\displaystyle\frac{\partial{#1}}{\partial{#2}}}
\def\derpars#1#2#3{\displaystyle\frac{\partial^2{#1}}{\partial{#2}\partial{#3}}}

\font\frd=eufm10 scaled\magstep2

\def\vf{\mbox{\frd X}}

\begin{document}

\parskip=3pt

\begin{center}
{\Large \sc $k$-Cosymplectic Classical Field Theories: Tulckzyjew,
Skinner-Rusk and Lie algebroid formulations}
\end{center}

\begin{center}
\noindent
  {\sc Angel M. Rey} \\
   (angelmrey@edu.xunta.es) \\
   {\it Departamento de Xeometr\'{\i}a e Topolox\'{\i}a,
   Facultade de Matem\'{a}ticas,\\
   Universidade de Santia\-go de Compostela,
   15782-Santia\-go de Compostela, Spain}

\noindent
  {\sc Narciso Rom\'an-Roy} \\
  (nrr@ma4.upc.edu)\\
   {\it Departamento de Matem\'{a}tica Aplicada IV, Edificio C-3, Campus Norte
   UPC,\\
   C/ Jordi Girona 1, E-08034 Barcelona, Spain}

\noindent
  {\sc Modesto  Salgado}\ ,
   {\sc Silvia Vilari\~no} \\
   (modesto@zmat.usc.es\ , svfernan@usc.es)\\
   {\it Departamento de Xeometr\'{\i}a e Topolox\'{\i}a,
   Facultade de Matem\'{a}ticas,\\
   Universidade de Santia\-go de Compostela,
   15782-Santia\-go de Compostela, Spain }
\end{center}

\begin{center}\today\end{center}


\begin{abstract}
The $k$-cosymplectic Lagrangian and Hamiltonian formalisms of
first-order field theories are reviewed and completed. In
particular, they are stated for singular and almost-regular systems.
Subsequently, several alternative formulations for $k$-cosymplectic
first-order field theories are developed: First, generalizing the
construction of Tulczyjew for mechanics, we give a new
interpretation of the classical field equations in terms of certain
submanifolds of the tangent bundle of the $k^1$-velocities of a
manifold. Second, the Lagrangian and Hamiltonian formalisms are
unified by giving an extension of the Skinner-Rusk formulation on
classical mechanics. Finally, both formalisms are formulated in
terms of Lie algebroids.
\end{abstract}

\noindent {\sl M.S. Classification (2000)}: 70S05, 53D05, 53Z05

\markright{\sc A.M. Rey {\it et al\/},
     \sl $k$-cosymplectic field theories \ldots}


\renewcommand{\baselinestretch}{1.5}
\renewcommand{\arraystretch}{0.66}

\tableofcontents


\section{Introduction}\label{intro}

G\"{u}nther's ( $k$-symplectic) formalism \cite{aw,aw3,gun,fam} is the
generalization to  first order classical field theories of the
standard symplectic formalism in mechanics, which is the geometric
framework for describing autonomous dynamical systems. In this
sense, the $k$-symplectic formalism is used to give a geometric
description of certain kinds of field theories: in a local
description, those whose Lagrangian does not depend on the
coordinates in the basis (in many of them, the space-time
coordinates); that is, it is only valid for Lagrangians
$L(q^i,v^i_A)$  and  Hamiltonians $H(q^i,p^A_i)$ that depend on the
field coordinates $q^i$ and  on the partial derivatives of the field
$v^i_A$.

The $k$-cosymplectic formalism is the generalization to field
theories of the standard cosymplectic formalism in mechanics, which
is the geometric framework for describing non-autonomous dyna\-mical
systems \cite{mod1,mod2}. This formalism describes field theories
involving the coordinates in the basis $(t^1,\ldots ,t^k)$ on the
Lagrangian $L(t^A,q^i,v^i_A)$ and on the Hamiltonian
$H(t^A,q^i,p^A_i)$.
The $k$-cosymplectic formalism  was introduced in
\cite{mod1,mod2}. One of the advantages of this formalism, and of
G\"unther's ($k$-symplectic) formalism, is that only the
tangent and cotangent bundle of a manifold are required to develop
it. In addition, there are also other polysymplectic formalisms for
describing field theories, such as those developed by G.
Sardanashvily {\it et al} \cite{Sarda2,Sarda1,Sd-95}, and by I.
Kanatchikov \cite{Kana}, as well as the $n$-symplectic formalism of
L. K. Norris \cite{McN,No2}. Let us remark that the multisymplectic
formalism is the most ambitious program to develop the Classical
Field Theory (see for example \cite{CCI,EMR-96,EMR-00,GIM1,GIM2,kt},
and references quoted therein).
In \cite{ANMS-07,LMNRS-2002} the equivalence
between the multisimplectic and $k$-cosymplectic description is shown, when
the theories with trivial configuration bundles are considered.

In this paper, we  first review the $k$-cosymplectic
formalism for singular field theories, improving previous
developments on this topic \cite{mod2}. After this,
 the main aims are:

1. \emph{To introduce certain submanifolds of  $TM\oplus\stackrel{k}{\ldots}\oplus TM$,
  the Whitney sum of $k$ copies of $TM$, which allows us
to describe the Euler-Lagrange and Hamilton equations. }

 This part
of the paper is inspired in  \cite{T1,T2}, where Tulczyjew
formulates Hamiltonian dynamics in terms of Lagrangian submanifolds
of a symplectic manifold.  M. de Le\'{o}n et al. generalize
Tulczyjew's paper to higher-order Lagrangian systems in \cite{LL-89}
and to classical field equations in terms of submanifolds of
multysymplectic manifolds in \cite{LLR-91,LMS-03}. In
\cite{LMM-2005}, Tulczjew's construction is extended to Lagrangian
Mechanics on Lie algebroids in terms of Lagrangian Lie
subalgebroids, which are submanifolds of a Lie algebroid $\tau:E\to Q$.

2. \emph{To extend the {\sl Skinner-Rusk  formalism} to $k$-cosymplectic
classical field theories.}

 In \cite{skinner2}, the authors  developed the {\sl Skinner-Rusk
formalism} in order to give a geometrical unified formalism for
describing mechanical systems. It incorporates all the
characteristics of Lagrangian and Hamiltonian descriptions of these
systems (including dynamical equations and solutions, constraints,
Legendre map, evolution operators, equivalence, etc.). This
formalism has been generalized to time-dependent mechanical systems
\cite{CMC}, to the multisymplectic description of first-order field
theories \cite{bar4,LMM-2002}, and also to the $k$-symplectic
formulation of field theories \cite{RRS}.

We extend this unified framework to the
$k$-cosymplectic description of first-order classical field theories
\cite{mod1,mod2}, and to show how this description comprises the
main features of the Lagrangian and Hamiltonian formalisms, both for
the regular and singular cases.

3. \emph{To develop the $k$-cosymplectic formalism in terms of Lie algebroids.}

In \cite{MAR-01}, a theory of Lagrangian Mechanics is developed in a
similar  way to the formulation of the standard Lagrangian
Mechanics. This approach differs from that followed by A. Weinstein
\cite{WEI-96}. A good survey on this subject is \cite{LMM-2005}. The
multisymplectic formalism for classical field theories is extended
to the setting of Lie algebroids in \cite{MAR-04,MAR-05}, and in
\cite{VC-06} a geometric frameword for discrete Classical Field
theories on Lie groupoids is presented.

The organization of the paper is as follows:

Section 2 is devoted to reviewing the main features of the $k$-cosymplectic formalism
 of Lagrangian and Hamiltonian field theories, and
to stating these formalisms for singular systems. First, the field
theoretic phase space for the Hamiltonian approach space is  $\rk \times
(T^1_k)^{\;*}Q$, where $(T^1_k)^{\;*}Q=T^{\;*}Q\oplus
\stackrel{k}{\dots}\oplus T^{\;*}Q$ is the Whitney sum of $k$-copies
of the cotangent bundle $T^{\;*}Q$ of a manifold $Q$. This space is
the canonical example of a $k$-cosymplectic manifold.
The field phase space for the Lagrangian description is $ \rk \times T^1_k
Q$, where $ T^1_k Q= TQ\oplus \stackrel{k}{\dots} \oplus TQ $ is
the Whitney sum of $k$-copies of the tangent bundle $TQ$.
 $T^1_kQ$ has the canonical $k$-tangent
structure, given by $k$ canonical tensor fields of type $(1,1)$
satisfying certain properties. This structure on $ T^1_k
Q$ can be lifted to $\rk \times T^1_k Q$. Using the extended tensor
fields or the Legendre map and a Lagrangian function, we can construct  a $k$- cosymplectic
(or $k$-precosymplectic)
structure on $\rk \times T^1_k Q$ whose $1$-forms and $2$-forms
enable us to develop the Lagrangian formalism.

In Section 3 we develop the first aim of the paper. We give a new
interpretation of the classical field equations in terms of certain
submanifolds of $T^1_k(\rk\times (T^1_k)^*Q)$. In order to do this,
we introduce $2k$ derivations $\imath_{T_A}$ and $d_{T_A}$, $1\leq A\leq
k $ from $\bigwedge M$ to $\bigwedge T^1_kM$, for a differentiable manifold $M$.
 These derivations are
the main tools for developing the rest of the section and there is a
generalization of the derivations introduced by Tulczyjew in
\cite{T1,T2}.

In Section 4 we develop the unified formalism for field theories
(second aim), which is based on the use of the Whitney sum
${\mathcal M} = \left( \r^k \times T^1_kQ \right) \oplus_{\r^k
\times Q} \left( \r^k \times (T^1_k)^{\;*}Q \right)$. There are
canonical ``precosymplectic" forms on ${\mathcal M}$ (the pull-back of the
canonical cosymplectic forms on each $\r \times T^{\;*}Q$) and a
natural coupling function, which is defined by the contraction
between vectors and covectors. Then, given a Lagrangian $L\in
C^{\infty}(\r^k \times T^1_kQ)$, we can state a field equation on
${\mathcal M}$. This equation has solution only on a submanifold
$M_L$, which is the graph of the Legendre map. Then we prove that if
${\bf Z}=(Z_1, \ldots , Z_k)$ is an integrable $k$-vector field,
which is a solution to this equation and tangent to $M_L$, then the
projection onto the first factor $\rk\times T^1_kQ$  of the integral
sections of ${\bf Z}$ are solutions to the Euler-Lagrange field
equations. If $L$ is regular, the converse also holds. Furthermore,
we establish the relationship between ${\bf Z}$ and the Hamiltonian
and the Lagrangian $k$-vector fields of the $k$-cosymplectic
formalism, ${\bf X}_H$ and ${\bf X}_L$.

In Section \ref{algebroids} we present some basic facts on Lie
algebroids, including results form differential calculus, morphisms
and prolongations of Lie algebroids. In this section we also
introduce a bundle $\mathcal{T}^E_kP=(\rk\times\stackrel{k}{\oplus}
E)\times_{\rk\times T^1_kQ} T^1_k(\rk\times P)\to P$ which is said
to be the $k$-prolongation of a Lie algebroid $\tau:E\to Q$ over a
fibration $\pi:P\to Q$. This space is a generalization of the
prolongation of Lie algebroids and it is the fundamental geometric
element to develop the Lagrangian and Hamiltonian $k$-cosymplectic
field theory on Lie algebroids.

Section \ref{ kcosym alg} is devoted to developing a Lagrangian and
Hamiltonian $k$-cosymplectic description of field theories on Lie
algebroids. In particular, in section \ref{lag form al} we develop
the $k$-cosymplectic Lagrangian formalism on Lie algebroids. The
fundamental point of this development is to consider the manifold
$\mathcal{T}^E_kP$ with $P=\stackrel{k}{\oplus}E$ and the geometric
objects defined on $\mathcal{T}^E_k(\stackrel{k}{\oplus}E)$. Given a
Lagrangian function $L:\rk\times\stackrel{k}{\oplus} E\to \r$,
solving certains equations we obtain a section $\xi_L$ of
$\mathcal{T}^E_k(\stackrel{k}{\oplus}E)$ such that its integral
sections are solutions to the Euler-Lagrange equations for $L$.
Finally in section \ref{615} we recover  the standard Lagrangian
$k$-cosymplectic formalism described in section \ref{lang form} as a
particular case of the section \ref{ kcosym alg} when $E=TQ$.
Section \ref{ham form al} is devoted to developing a Hamiltonian
$k$-cosymplectic description of  field theory on Lie algebroids.
This description is similar to the Lagrangian case; now we consider
the vector bundle $\mathcal{T}^E_k(\stackrel{k}{\oplus}E^*)$, where
$E^*\to Q$ is the dual bundle of $E$. Given a Hamiltonian function
$H:\rk\times\stackrel{k}{\oplus} E^*\to \r$, solving certains
equations we obtain a section $\xi_H$ of
$\mathcal{T}^E_k(\stackrel{k}{\oplus}E^*)$ such that its integral
sections are solutions to the Hamilton equations for $H$.
Taking $E=TQ$, the results in sections \ref{lag form al} and
\ref{ham form al} coincide with the results of the standard
$k$-cosymplectic formalism described in section 2. Thus the standard
$k$-cosymplectic formalism can be recovered as a particular case of
the description on Lie algebroids.

Manifolds are real, paracompact, connected and $C^\infty$. Maps are
$C^\infty$. Sum over crossed repeated indices is understood.

\section{The standard k-cosymplectic formalism in field theory}\label{k-cosymp form}

\subsection{Hamiltonian formalism\cite{mod1}}
\label{ham form}

\subsubsection{Geometric elements}\label{211}

Let $Q$ be a differentiable manifold, $\dim Q = n$, and
$\tau^{\;*}_Q: T^{\;*}Q \to Q$ its cotangent bundle.
Denote by $(T^1_k)^{\;*}Q= T^{\;*}Q \oplus \stackrel{k}{\dots}
\oplus T^{\;*}Q$, the Whitney sum of $k$ copies of $T^{\;*}Q$. The
manifold $(T^1_k)^{\;*}Q$ can be identified with the manifold
$J^1(Q,\rk)_0$ of 1-jets of mappings from $Q$ to $\rk$ with target
at $0\in \rk$,
\[
\begin{array}{ccc}
J^1(Q,\r^k)_0 & \equiv & T^{\;*}Q \oplus \stackrel{k}{\dots} \oplus T^{\;*}Q \\
j^1_{q,0}\sigma  & \equiv & (d\sigma^1(q), \dots ,d\sigma^k(q))\ ,
\end{array}
\]
where $\sigma^A= \pi^A \circ \sigma:Q \longrightarrow \r$ is the
$A^{th}$ component of $\sigma$, and  $\pi^A:\r^k \to \r$ is the
canonical projection onto the $A^{th}$ component, for $1\leq A \leq
k$. $(T^1_k)^{\;*}Q$ is called {\sl the cotangent bundle of
$k^1$-covelocities of the manifold $Q$}.

The manifold $J^1\pi_{Q}$ of  1-jets of sections of the trivial
bundle $\pi_{Q}:\rk \times Q \to Q$ is diffeomorphic to $\rk \times
(T^1_k)^{\;*}Q$, via the diffeomorphism given by
\[
\begin{array}{rcl}
J^1\pi_{Q} & \to & \rk   \times (T^1_k)^{\;*}Q \\
\noalign{\medskip} j^1_q\phi= j^1_q(\phi_Q,Id_{Q})  & \mapsto &
(\phi_Q(q), \alpha^1_q, \ldots ,\alpha^k_q) \ ,
\end{array}
\]
where $\phi_Q: Q \stackrel{\phi}{\to}  \rkq
\stackrel{\pi_{\rk}}{\to}\rk $ , $ \alpha^A_q=d(\phi_Q)^A(q)$,
$1\leq A \leq k$, and $(\phi_Q)^A:Q\stackrel{\phi_Q}{\to} \rk
\stackrel{\pi^A}{\to} \r$.

Throughout the paper we use the following notation for the canonical projections
$$
\rk \times (T^1_k)^{\;*}Q \stackrel{(\pi_{Q})_{1,0}}{\longrightarrow} \rk
\times Q \stackrel{\pi_{Q}}{\longrightarrow} Q \ ,
$$
 and
$(\pi_{Q})_{1}=\pi_{Q}\circ (\pi_{Q})_{1,0}$, where
$$\pi_Q(t,q)=q, \quad (\pi_Q)_{1,0}(t,\alpha^1_q, \ldots
,\alpha^k_q)=(t,q), \quad (\pi_Q)_1(t,\alpha^1_q, \ldots
,\alpha^k_q)=q \, ,$$ with $t\in \rk $, $q\in Q$ and $(\alpha^1_q,
\ldots ,\alpha^k_q)\in (T^1_k)^{\;*}Q$.

If $(q^i)$ are local coordinates on $U \subseteq Q$, then the
induced local coordinates $(q^i , p_i)$, $1\leq i \leq n$, on
$(\tau^{\;*}_Q)^{-1}(U)=T^{\;*}U\subset T^{\;*}Q$, are given by
$$
q^i(\alpha_q)=q^i(q), \quad p_i(\alpha_q)= \alpha_q \left(
\frac{\partial}{\partial q^i}\Big\vert_q \right)\, ,
$$
and the induced local coordinates  $(t^A,q^i ,
p^A_i)$ on $[(\pi_Q)_1]^{-1}(U)=\rk \times (T^1_k)^{\;*}U$ are given
by
$$
t^A(j^1_q\phi) = (\phi_Q(q))^A ,\quad q^i(j^1_q\phi) = q^i(q)\, ,
\quad   p^A_i(j^1_q\phi) =
d(\phi_Q)^A(q)\left(\ds\frac{\partial}{\partial
q^i}\Big\vert_q\right) ,
$$
  or equivalently, for $1\leq i\leq n$ and $1\leq A\leq k$,
$$
t^A(t,\alpha^1_q, \ldots ,\alpha^k_q) = t^A\, , \quad
q^i(t,\alpha^1_q, \ldots ,\alpha^k_q) = q^i(q)\, , \quad
p_A^i(t,\alpha^1_q, \ldots ,\alpha^k_q) =
\alpha^A_q\left(\ds\frac{\partial}{\partial q^i}\Big\vert_q
\right)\, .
$$

On $\rk\times (T^1_k)^{\;*}Q$, we define the differential forms
$$
\eta^A=(\pi^A_1)^{\;*}dt^A\, , \quad \theta_0^A=
(\pi^A_2)^{\;*}\theta_0\, , \quad \omega_0^A=
(\pi^A_2)^{\;*}\omega_0\, ,
$$
where $\pi^A_1:\rk \times (T^1_k)^{\;*}Q \rightarrow \r$ and
$\pi^A_2:\rk \times (T^1_k)^{\;*}Q \rightarrow T^{\;*}Q$ are the
projections defined by
$$
\pi^A_1(t,(\alpha^1_q, \ldots ,\alpha^k_q))=t^A \,,\quad
\pi^A_2(t,(\alpha^1_q, \ldots ,\alpha^k_q))=\alpha^A_q\, ,
$$
$\omega=-d\theta=dq^i \wedge dp_i$ is the canonical symplectic form
on $T^{\;*}Q$ and $\theta=p_i \, dq^i$ is the Liouville $1$-form on
$T^{\;*}Q$. Obviously $\omega^A = -d\theta^A$.

In local coordinates we have
\begin{equation}\label{locexp}
  \eta^A=dt^A \, , \quad \theta^A = \displaystyle \, p^A_i
dq^i \,  ,  \quad \omega^A = \displaystyle  dq^i \wedge dp^A_i\, .
\end{equation}
Finally, consider the vertical distribution of the bundle
$(\pi_{Q})_{1,0}:\rk\times (T^1_k)^{\;*}Q \to \rkq$,
 $$
 V^{\;*}=ker\, \left( \, (\pi_Q)_{1,0}
\right)_*=\left\langle\frac{\displaystyle\partial}
{\displaystyle\partial p^1_i}, \dots,
\frac{\displaystyle\partial}{\displaystyle\partial
p^k_i}\right\rangle_{i=1,\ldots , n}
$$

A simple inspection of the expressions in local coordinates
(\ref{locexp}) shows that the forms $\eta^A$ and $\omega^A$ are
closed, and the following relations hold
\begin{enumerate}
\item $\eta^1\wedge\dots\wedge \eta^k\neq 0$,\quad
$(\eta^A)\vert_{V^{\;*}}=0,\quad (\omega^A)\vert_{V^{\;*}\times
V^{\;*}}=0,$

\item $(\ds {\cap_{A=1}^{k}} \ker \eta^A) \cap (\ds
{\cap_{A=1}^{k}} \ker \omega^A)=\{0\}$, \quad $dim(\ds
{\cap_{A=1}^{k}} \ker \omega^A)=k,$
\end{enumerate}

Then, from the above  geometrical model, we have the following definition \cite{mod1}:

\begin{definition}\label{deest}
Let $M$ be a differentiable manifold  of dimension $k(n+1)+n$. A
{\rm $k$--cosymplectic structure} on $M$ is a family
$(\eta^A,\omega^A,V)$, where each $\eta^A$ is a closed
$1$-form, each $\omega^A$ is a closed $2$-form and $V$ is an
integrable  $nk$-dimensional distribution on $M$, satisfying $1$ and
$2$. In this case, $M$ is said to be an {\rm $k$--cosymplectic
manifold}.
\end{definition}

\begin{theorem}   \cite{mod1} (Darboux Theorem)
If $M$ is   a  $k$--cosymplectic manifold, then around each
  point of $M$ there exist local coordinates
$(x^A,y^i,z^A_i)$; $1\leq A\leq k, 1\leq i \leq n$, such that
$$
\eta^A=dx^A,\quad \omega^A=dy^i\wedge dz^A_i, \quad
V=\left\langle\frac{\displaystyle\partial} {\displaystyle\partial
z^1_i}, \dots, \frac{\displaystyle\partial}{\displaystyle\partial
z^k_i}\right\rangle_{i=1,\ldots , n}.
$$
\end{theorem}

The canonical model for these geometrical structures is   $(\rk
\times (T^1_k)^{\;*}Q,\eta^A,\omega^A,V^{\;*})$.

For every $k$-cosymplectic structure  $(\eta_A ,\omega_A,V)$ on $M$,
there exists a family of $k$ vector fields $\{R_A\}$ characterized by the following conditions
$$
\imath_{R_A}\eta^B=\delta^B_A,\qquad \imath_{R_A}\omega^B=0, \quad
1\leq A,B \leq k
$$
They are called the \textit{Reeb vector fields} associated to the
$k$--cosymplectic structure. In the canonical model $R_A=
\partial/\partial t^A$. Observe that the vector
fields $\{\partial/\partial t^A$ are defined
intrinsically in $\rk\times (T^1_k)^{\;*}Q$, and span locally the
vertical distribution with respect to the projection
$\rk\times (T^1_k)^{\;*}Q\to (T^1_k)^{\;*}Q$.

\subsubsection{k-vector fields and integral sections}\label{212}

Let $M$ be an arbitrary manifold, $T^{1}_{k}M$ the Whitney sum
$TM\oplus \stackrel{k}{\dots} \oplus TM$ of $k$ copies of $TM$, and
$\tau_M : T^{1}_{k}M \longrightarrow M$ its  canonical projection.
$\tau_M : T^{1}_{k}M \longrightarrow M$ is usually called the
tangent bundle of $k^1$-velocities of $M$, the reason for this name
will be explained later in Section \ref{221}

\begin{definition}  \label{kvector}
A  {\rm $k$-vector field} on $M$ is a section ${\bf X} : M \longrightarrow T^1_kM$ of the projection
$\tau_M$.
\end{definition}

Since $T^{1}_{k}M$ is  the Whitney sum $TM\oplus
\stackrel{k}{\dots}\oplus TM$ of $k$ copies of $TM$, we deduce that
giving a $k$-vector field ${\bf X}$ is equivalent to giving a
family of $k$ vector fields $X_{1}, \dots, X_{k}$ on $M$ by
projecting ${\bf X}$ onto every factor. For this reason we will
denote a $k$-vector field by $(X_1, \ldots, X_k)$.

\begin{definition} \label{integsect}
An {\rm integral section} of the $k$-vector field  \, $(X_{1},
\dots, X_{k})$ \, passing through a point $x\in M$  is a map
$\phi:U_0\subset \r^k \rightarrow M$, defined on some neighborhood
$U_0$ of $0\in \rk$, such that
$$
\phi(0)=x, \, \,
\phi_{*}(t)\left(\displaystyle\frac{\displaystyle\partial}{\displaystyle\partial
t^A}\Big\vert_t\right) = X_{A}(\phi (t)) \ , \quad \mbox{for}
\quad t\in U_0\, .
$$
We say that  a $k$-vector field $(X_1,\ldots , X_k)$ on $M$ is {\rm
integrable} if there is an integral section passing through each
point of $M$.
\end{definition}

Observe that, if $k=1$, this definition coincides with the
definition of integral curve of a vector field. In the
$k$-cosymplectic formalism, the solutions to the field equations are
described as the integral sections of some $k$-vector fields.

\subsubsection{$k$-cosymplectic Hamiltonian formalism}\label{213}

Let $H\colon \rk\times (T^1_k)^{\;*}Q\to \r$  be a Hamiltonian
function. Let ${\bf X}=(X_1,\dots,X_k)$ be a $k$-vector field on
$\rk\times (T^1_k)^{\;*}Q$, which is a solution to the following
equations
\begin{equation}
\label{geonah}
dt^A(X_B)=\delta^A_B \quad ,  \quad
\displaystyle \sum_{i=1}^k \, \imath_{X_A}\omega^A
= dH-\displaystyle\sum_{A=1}^k \ds\frac{\partial H}{\partial
t^A}dt^A \ .
\end{equation}

If   ${\bf X}=(X_1,\dots,X_k)$ is an integrable $k$-vector field,
locally given by
$$
X_A=(X_A)^B\frac{\partial}{\ds\partial t^B} +
(X_A)^i\frac{\partial}{\ds\partial q^i}
+(X_A)^B_i\frac{\partial}{\ds\partial p^B_i}
$$
then
\begin{equation}
\label{111} (X_A)^B=\delta_A^B, \quad \quad \ds\frac{\partial
H}{\partial p^A_i}=(X_A)^i, \quad \ds\frac{\partial H}{\partial
q^i}= -\ds\sum_{A=1}^k(X_A)^A_i,
\end{equation}
   and if
$\psi:\rk\to \rk\times (T^1_k)^{\;*}Q$, locally given by
$\psi(t)=(\psi^A(t),\psi^i(t),\psi^A_i(t))$, is an integral section
of ${\bf X}$ , then
$$
\ds\frac{\partial \psi^A}{\partial t^B}=\delta_{AB} ,\quad
\ds\frac{\partial \psi^i}{\partial t^B}=(X_B)^i, \quad
\ds\frac{\partial \psi^A_i}{\partial t^B}=(X_B)^A_i \,  .
$$
Therefore, from (\ref{111}) we obtain that $\phi(t)$ is a solution
to the Hamiltonian field equations
\begin{equation}\label{he}
\frac{\ds\partial H}{\ds\partial q^i}=
-\ds\sum_{A=1}^k\frac{\ds\partial\psi^A_i} {\ds\partial t^A} \quad ,
\quad \frac{\ds\partial H} {\ds\partial p^A_i}=
\frac{\ds\partial\psi^i}{\ds\partial t^A} \, .
\end{equation}
So, equations (\ref{geonah}) can be considered as a {\sl geometric
version} of the Hamiltonian field equations.

\begin{remark}\label{remuni}
{\rm
The above Darboux theorem allows us to write the Hamiltonian
formalism in an arbitrary $k$-cosymplectic manifold
$(M,\eta^A,\omega^A,V)$, substituing the equations (\ref{geonah}) by

\[\eta^A(X_B)=\delta^A_B \quad , \quad
\ds\sum_{A=1}^k\imath_{X_A}\omega^A=dH + \ds\sum_{A=1}^k
(1-{\mathcal R}_A(H))\eta^A\;.\]

If $(M,\eta^A,\omega^A,V)$ is a $k$-cosymplectic manifold, we
can define the vector bundle morphism
$$
\begin{array}{rccl}
\Omega^{\sharp}: & T^1_kM & \longrightarrow & T^{\;*}M  \\
\noalign{\medskip} & (X_1,\dots,X_k) & \rightarrow &
\Omega^{\sharp}(X_1,\dots,X_k) =  \displaystyle \sum_{A=1}^k \,
\imath_{X_A}\omega^A+ \eta^A(X_A)\eta^A
\end{array}
$$
and denoting by $ {\mathcal M}_k(C^\infty(M))$ the space of matrices
of order $k$ whose entries are functions on $M$, we can also define
the vector bundle morphism
$$
\begin{array}{rccl}
\eta^{\sharp}: & T^1_kM & \longrightarrow & {\mathcal M}_k(C^\infty(M))  \\
\noalign{\medskip} & (X_1,\dots,X_k) & \rightarrow &
\eta^{\sharp}(X_1,\dots,X_k) =  (\eta^A(X_B))\, .
\end{array}
$$
Then, the solutions to (\ref{geonah}) are given by
$(X_1,\dots,X_k)+(\ker\Omega^{\sharp}\cap\ker\eta^{\sharp})$, where
$(X_1,\dots,X_k)$ is a particular solution.
}
\end{remark}

\subsection{Lagrangian formalism \cite{mod2}}\label{lang form}

\subsubsection{Geometric elements}
\protect\label{221}

\paragraph{The tangent bundle of $k^1$-velocities of a manifold}\

Let $\tau_Q : TQ \to Q$ be the tangent bundle of $Q$. Let us denote
by $T^1_kQ$ the Whitney sum $TQ \oplus \stackrel{k}{\dots} \oplus
TQ$ of $k$ copies of $TQ$, with projection $\tau^k_Q : T^1_kQ \to
Q$, $\tau^k_Q ({v_1}_\mathbf{q},\ldots ,{v_k}_\mathbf{q})=\mathbf{q}$.
$T^1_kQ$ can be identified with the manifold $J^1_0(\r^k,Q)$ of the
{\it $k^1$-velocities    of $Q$}, that is,  $1$-jets of maps
$\sigma:\rk\to Q$  with source at $0\in \r^k$, say
\[
\begin{array}{ccc}
J^1_0(\r^k,Q) & \equiv & TQ \oplus \stackrel{k}{\dots} \oplus TQ \\
j^1_{0,\mathbf{q}}\sigma & \equiv & ({v_1}_\mathbf{q},\ldots ,
{v_k}_\mathbf{q}) \ ,
\end{array}
\]
where $\mathbf{q}=\sigma (0)$,  and ${v_A}_\mathbf{q}=
\sigma_*(0)(\ds\frac{\partial}{\partial t^A}\Big\vert_{0})$.
$T^1_kQ$ is the {\it tangent bundle of $k^1$-velocities of $Q$}
\cite{mor}.

If $(q^i)$ are local coordinates on $U \subseteq Q$, the induced
coordinates $(q^i , v^i)$ on $TU=\tau_Q^{-1}(U)$ are
$$
 q^i(v_\mathbf{q})=q^i(\mathbf{q}),\qquad
  v^i(v_\mathbf{q})=v_\mathbf{q}(q^i) \ ,
  $$
and  the induced coordinates $(q^i , v_A^i)$, $1\leq i \leq
n,\, 1\leq A \leq k$, on $T^1_kU=(\tau^k_Q)^{-1}(U)$ are given by
$$ q^i({v_1}_\mathbf{q},\ldots , {v_k}_\mathbf{q})=q^i(\mathbf{q}),\qquad
  v_A^i({v_1}_\mathbf{q},\ldots , {v_k}_\mathbf{q})={v_A}_\mathbf{q}(q^i) \, .$$

In general, if $F:M \to N$ is a differentiable map, then the induced
map $T^1_k(F):T^1_kM \to  T^1_kN$ defined by
$T^1_k(F)(j^1_0g)=j^1_0(F \circ g)$ is given by
\begin{equation} \label{tf} T^1_k(F)({v_1}_q,\ldots ,
{v_k}_q)=(F_*(q){v_1}_q,\ldots ,F_*(q){v_k}_q) \quad ,\end{equation}
where ${v_1}_q,\ldots , {v_k}_q\in T_qQ$, $q\in Q$ , and
$F_*(q):T_qM \to T_{F(q)}N$ is the induced map.

\paragraph{The manifold $T^1_k(\rk\times P)$}\

 Let $\pi:P\to Q$ be an arbitrary fibration.
 We use the tangent bundle of $k^1$-velocities
 of the manifold $\rk\times P$, then we consider the
vector bundle $\tau^k_{\rk\times P}:T^1_k(\rk\times P)\to \rk\times P$.

 An element $W_{(\mathbf{t},\mathbf{p})}\in(\tau^k_{\rk\times
P})^{-1}(\mathbf{t},\mathbf{p})$ is given by
$W_{(\mathbf{t},\mathbf{p})}= ({v_1}_{(\mathbf{t},\mathbf{p})},
\ldots ,{v_k}_{(\mathbf{t},\mathbf{p})})$.

Since  $(t^A,q^i, u^\vartheta )$ are the local coordinates on
$\rk\times P$, we write each vector
$(v_A)_{(\mathbf{t},\mathbf{p})}$ as follows
$$ (v_A)_{(\mathbf{t},\mathbf{p})}= (v_A)_B \ds\frac{\partial}{\partial
t^B}\Big\vert_{(\mathbf{t},\mathbf{p})}  +(v_A)^i
\ds\frac{\partial}{\partial q^i}\Big\vert_{(\mathbf{t},\mathbf{p})}
+ (v_A)^\vartheta \ds\frac{\partial}{\partial
 u^\vartheta }\Big\vert_{(\mathbf{t},\mathbf{p})} \  .
$$
Thus the local coordinates $(t^A,q^i, u^\vartheta )$ induce the
local coordinates
$(t^A,q^i, u^\vartheta ,(v_A)_B,(v_A)^i,(v_A)^\vartheta)$ in
$T^1_k(\rk\times P)$.

Taking into account the identification $T^1_k(\rk\times P)\equiv
T^1_k\rk\times T^1_k (P)$ given by
$$
\begin{array}{l}({v_1}_{(\mathbf{t},\mathbf{p})}, \ldots
,{v_k}_{(\mathbf{t},\mathbf{p})})\equiv \\
\noalign{\medskip}\left(\left((v_1)_B \ds\frac{\partial}{\partial
t^B}\Big\vert_{{\bf t}}, \ldots , (v_k)_B
\ds\frac{\partial}{\partial t^B}\Big\vert_{{\bf t}}\right);\left(
(v_1)^i \ds\frac{\partial}{\partial q^i}\Big\vert_{{\bf p}} +
(v_1)^\vartheta \ds\frac{\partial}{\partial
 u^\vartheta }\Big\vert_{{\bf p}}, \ldots ,(v_k)^i
\ds\frac{\partial}{\partial q^i}\Big\vert_{{\bf p}} +
(v_k)^\vartheta \ds\frac{\partial}{\partial  u^\vartheta
}\Big\vert_{{\bf p}} \right)\right) \ , \end{array}
$$
 we consider the map
$$
F=\tau^k_{\rk}\times
T^1_k(\pi):T^1_k(\rk\times P)\equiv T^1_k\rk\times T^1_k P \to
\rk\times T^1_kQ
$$
 which is given by
\begin{equation}
\label{defF}F(W_{(\mathbf{t},\mathbf{p})})=F({v_1}_{(\mathbf{t},\mathbf{p})}, \ldots
,{v_k}_{({\bf t,p})})=({\bf t}, (v_1)^i\ds\frac{
\partial}{\partial q^i}\Big\vert_{\mathbf{q}},\ldots , (v_k)^i\ds\frac{
\partial}{\partial q^i}\Big\vert_{\mathbf{q}}) \ ,
\end{equation}
since $T^1_k(\pi)=T\pi \times \ldots \times T\pi$, and
$$T_{\bf p}\pi\left((v_A)^i \ds\frac{\partial}{\partial
q^i}\Big\vert_{{\bf p}} + (v_A)^\vartheta
\ds\frac{\partial}{\partial u^\vartheta }\Big\vert_{{\bf
p}}\right)=(v_A)^i\ds\frac{
\partial}{\partial q^i}\Big\vert_{\mathbf{q}}  \qquad 1\leq  A \leq k \, .$$
Thus
\begin{equation}\label{locF}
F=\tau^k_{\rk}\times T^1_k\pi:(t^A,q^i, u^\vartheta
,(v_A)^B,(v_A)^i,(v_A)^\vartheta)
  \to (t^A,q^i,(v_A)^i)\end{equation}

This map $F$ will be used in the description of the Tulczyjew's
Lagrangian formalism and in the definition of the fiber bundle
$\mathcal{T}^E_kP$ (see section \ref{k-prol}), which is  the
fundamental geometric element to develop the $k$-cosymplectic
formalism on Lie algebroids (see section \ref{ kcosym alg}).

\paragraph{ The manifold $\rk \times T^1_kQ$}\

Next we see that the manifold $\rk \times
T^1_kQ$ is a cosymplectic manifold when a regular Lagrangian $L:\rk
\times T^1_kQ \to \r$ is given.

The manifold $J^1\pi_{\rk}$ of 1-jets of sections of the trivial
bundle $\pi_{\rk}:\rk \times Q \to \rk$ is diffeomorphic to $\rk
\times  T^1_kQ$, via the diffeomorphism given by
$$
\begin{array}{rcl}
J^1\pi_{\rk} & \to & \rk   \times T^1_kQ \\
\noalign{\medskip} j^1_t\phi= j^1_t(Id_{\rk},\phi_Q) & \to & (
t,v_1, \ldots ,v_k) \ ,
\end{array}
$$
where $\phi_Q: \rk \stackrel{\phi}{\to}  \rkq \stackrel{\pi_Q}{\to}Q
$, and $\ds v_A=(\phi_Q)_*(t)\left(\ds\frac{\partial}{\partial
t^A}\Big\vert_t\right)$.

Denote by $p_Q:\rk\times  T^{1}_{k} Q \to Q$ the canonical
projection, that is $p_Q(t,{v_1}_q,\ldots , {v_k}_q)=q$. If $(q^i)$
are local coordinates on $U \subseteq Q$, then the induced local
coordinates $(q^i , v_i)$, $1\leq i \leq n$, on
$\tau_Q^{-1}(U)=TU\subset TQ$, are given by
$$
q^i(v_q)=q^i(q), \quad v_i(v_q)= v_q(q^i)\, ,
$$
and  then the induced local coordinates  $(t^A,q^i , v^i_A)$ on
$p_Q^{-1}(U)=\rk \times T^1_kU$ are given by
$$
t^A(j^1_t\phi)=t^A\, , \quad q^i(j^1_t\phi)=q^i(\phi_Q(t)) \, ,
\quad v^i_A(j^1_t\phi)=\ds\frac{\partial (q^i\circ\phi_Q)}{\partial
t^A}(t)
$$
or equivalently,
$$t^A(t,{v_1}_q,\ldots , {v_k}_q)   =   t^A; \quad
q^i(t,{v_1}_q,\ldots , {v_k}_q) =q^i(q); \quad
v_A^i(t,{v_1}_q,\ldots , {v_k}_q)   = {v_A}_q(q^i)\, , $$

Throughout the paper we use the following notation for the canonical
projections
$$\rk \times (T^1_k)Q \stackrel{(\pi_{\rk})_{1,0}}{\longrightarrow}
\rk \times Q \stackrel{ \pi_{\rk} }{\longrightarrow} \rk$$ and
$(\pi_{\rk})_{1}=\pi_{\rk}\circ (\pi_{\rk})_{1,0}$,
 where, for $t\in \rk $, $q\in Q$ and $({v_1}_q,\ldots , {v_k}_q)\in
T^1_kQ$,
$$\pi_{\rk}(t,q)=t, \quad (\pi_{\rk})_{1,0}(t,{v_1}_q,\ldots ,
{v_k}_q)=(t,q), \quad (\pi_{\rk})_1(t,{v_1}_q,\ldots , {v_k}_q)=t  \ .$$

\paragraph{Canonical vector fields and tensor fields on $\rk \times T^1_kQ$.
Poincar\'{e}-Cartan forms}\

Denote by $\Delta$ the {\sl canonical vector field (Liouville vector
field)} of the vector bundle $(\pi_{\rk})_{1,0}: \r^{k}\times T^1_kQ
\rightarrow \rk\times Q$. This vector field $\Delta$ is the
infinitesimal generator of  the following flow
$$
\begin{array}{ccc}
\r \times (\r^{k}\times T^1_kQ) & \longrightarrow & \r^{k}\times
T^1_kQ  \\ \noalign{\medskip} (s,(t,{v_1}_q,\ldots , {v_k}_q)) &
\longrightarrow & (t, e^s{v_1}_q, \ldots,e^s{v_k}_q)
\end{array} \, ,
$$
and in local coordinates it has the form
$$
\Delta =   \displaystyle\sum_{i,A} v^i_A
\frac{\displaystyle\partial}{\displaystyle\partial v^i_A}\, .
$$
$\Delta$ can be written as the sum $\Delta=\ds\sum_{A=1}^k\Delta_A$,
where each vector field $\Delta_A$ is the infinitesimal generator of
the following flow
$$
\begin{array}{ccl}
\r \times (\r^{k}\times T^1_kQ) & \longrightarrow & \r^{k}\times
T^1_kQ  \\ \noalign{\medskip} (s,(t,{v_1}_q,\ldots , {v_k}_q )) &
\longrightarrow & (t, {v_1}_q, \ldots, {v_{A-1}}_q , e^s {v_A}_q,
{v_{A+1}}_q, \ldots , {v_k}_q)\, .
\end{array}
$$

\begin{definition}
For a vector  $X_q$ at $Q$, and for $A=1,\ldots, k$, we define its
{\rm vertical $A$-lift} \, $(X_q)^A$ as the local vector field  on
${\tau_Q}^{-1}(q)\subset T_k^1Q$ given by
$$
(X_q)^A(w_q) = \displaystyle\frac{d}{ds} \Big\vert_{s=0} \left(w_q+
(0,\ldots,0,s\,\stackrel{A}{X_q} ,0, \ldots,0 ) \right)
\quad , \quad \mbox{\rm for $w_q=({v_1}_q, \ldots ,{v_k}_q) \in T^1_kQ$} \ .
$$
\end{definition}

In local coordinates, for a vector $X_q = a^i
\,\ds\frac{\partial}{\partial q^i}$ we have
\begin{equation}\label{xa}
(X_q)^A =  a^i \displaystyle\frac{\partial}{\partial v^i_A} \quad .
\end{equation}

The {\sl canonical $k$-tangent structure} on $T^1_kQ$ is the set
$(S^1,\ldots,S^k)$ of tensor fields  of type $(1,1)$ defined by
$$
S^A(w_q)(Z_{w_q})= (\tau_*(w_q)(Z_{w_q}))^A, \quad \mbox{for all} \,
\, Z_{w_q}\in T_{w_q}(T^1_kQ),\, w_q\in T^1_kQ\,.
$$
 From (\ref{xa}), in local coordinates we have
\begin{equation}\label{localJA}
S^A=\displaystyle\frac {\displaystyle\partial}{\displaystyle\partial
v^i_A} \otimes dq^i
\end{equation}

The tensors $S^A$ can be regarded as the
$(0,\ldots,0,\stackrel{A}{1},0,\ldots,0)$-lift of the identity
tensor on $Q$ to $T^1_kQ$ defined in \cite{mor}.

In an obvious way, we consider the extension of $S^A$ to $\rk\times
T^1_kQ$, which we also denote  by $S^A$, and they have the same
local expressions (\ref{localJA}).

  The $k$-tangent manifolds were introduced as a
generalization of the tangent manifolds in \cite{mt1,mt2}. The
canonical model of these manifolds is $T^1_kQ$ with the structure
given by $(S^1, \ldots , S^k)$.

As in the case of mechanical systems, these tensor fields $S^A$
allow us to introduce the forms $\theta_L^A$ and $\omega_L^A$ on
$\rk\times T^1_kQ$ as follows
$$\theta_L^A=dL \circ  S^A \, \quad ,\quad  \omega_L^A=-d\theta_L^A \, , $$
 with local expressions
\begin{equation}\label{am2}
\theta_L^A=   \ds\frac{\displaystyle\partial
L}{\displaystyle\partial v^i_A} \,  dq^i\, \quad
\omega_L^A=dq^i\wedge d\left(\frac{\displaystyle\partial
L}{\displaystyle\partial v^i_A}\right) \  .
\end{equation}

\paragraph{Second order partial differential equations on $\rk \times T^1_kQ$}\

Now we characterize the integrable
$k$-vector fields on $\r^k \times T^1_kQ$ whose integral
sections are canonical prolongations of maps from $\r^k$ to $Q$.

\begin{definition}\label{sode2}
A $k$-vector field  ${\bf X}=(X_1,\dots,X_k)$ on $\r^k \times
T^1_kQ$ is a {\rm second order partial differential equation} ({\sc
sopde} for short)  if  :
$$
dt^A(X_B)= \delta_B^A \, , \quad {S}^A(X_A)=\Delta_A \, .
$$
\end{definition}

Let $(q^i)$ be a coordinate system on $Q$ and $(t^A,q^i,v^i_A)$  the
induced coordinate system on $\r^k \times T^1_kQ$.  A direct
computation in local coordinates  shows  that the local expression
of a {\sc sopde} $(X_1 ,\ldots,X_k) $ is
\begin{equation}\label{localsode2}
X_A(t,q^i,v^i_B)=\frac{\partial}{\partial
t^A}+v^i_A\frac{\displaystyle
\partial} {\displaystyle
\partial q^i}+
(X_A)^i_B \frac{\displaystyle\partial} {\displaystyle \partial v^i_B}\ ,
\end{equation}
where $(X_A)^i_B $ are functions on $\r^k \times T^1_kQ$. As a
direct consequence of the above local expressions, we deduce that
the family of vector fields $\{X_1, \ldots , X_k\}$ are linearly
independent.

\begin{definition}\label{de652}
Let $\phi:\r^k \rightarrow Q$  be  a map, the {\rm first
prolongation} $\phi^{[1]}$ of $\phi$ is the map
$$
\begin{array}{rcl}
\phi^{[1]}:\r^k & \longrightarrow &   \r^k \times T^1_kQ \\ t &
\longrightarrow &
(t,j^1_0\phi_t)\equiv\left(t,\phi_*(t)\left(\ds\frac{\partial}{\partial
t^1} \Big\vert_t\right), \ldots ,
\phi_*(t)\left(\ds\frac{\partial}{\partial t^k}\Big\vert_t \right)
\right)
\end{array}\ ,
$$ where $\phi_t(s)=\phi(t+s)$. In local coordinates
$$
\phi^{[1]}(t^1, \dots, t^k)=\left(t^1, \dots, t^k, \phi^i (t^1,
\dots, t^k),
\frac{\displaystyle\partial\phi^i}{\displaystyle\partial t^A} (t^1,
\dots, t^k)\right),
$$
\end{definition}

\begin{lemma}\label{lem0}
Let $(X_1 ,\ldots,X_k) $ be a {\sc sopde}. A map $\psi:\rk \to \rk
\times T^1_kQ$, given by
$\psi(t)=(\psi^A(t),\psi^i(t),\psi^i_A(t))$, is an integral section
of $(X_1 ,\ldots,X_k) $ if, and only if,
\begin{equation}\label{nn}
\psi^A(t)=t^A+c^A \, , \quad \psi^i_A(t)=\frac{\ds\partial \psi^i}
{\ds\partial t^A}(t) \, , \quad \frac{\ds\partial^2 \psi^i}
{\ds\partial t^A \ds\partial t^B}(t)=(X_A)^i_B(\psi(t)) \, .
\end{equation}
\end{lemma}
\proof
 Equations (\ref{nn}) follow from Definition \ref{integsect}
and (\ref{localsode2}).
\qed

\begin{remark}\label{rem1}
{\rm
The integral sections of a {\sc sopde} are given by
$\psi(t)=\left(t^A+c^A,\psi^i(t),\ds\frac{\partial \psi^i}{\partial
t^A}(t)\right)$, where the functions $\psi^i(t)$ satisfy the third
equation in (\ref{nn}) and $c^A$ are constants. In the particular
case $c=0$, we have that $\psi= \phi^{[1]}$  where $\phi= p_Q\circ
\psi:\rk \stackrel{\psi}{\to}\r^k \times T^1_kQ
\stackrel{p_Q}{\to}Q$; that is, $\phi(t)=(\psi^i(t))$.

Conversely if $\phi:\rk \to Q$ is any map such that
$$
\frac{\ds\partial^2 \phi^i}{\ds\partial t^A \ds\partial
t^B}(t)=(X_A)^i_B(\phi^{[1]}(t)),$$ then $\phi^{[1]}$ is an integral
section of $(X_1 ,\ldots,X_k) $.

Observe that if $(X_1 ,\ldots,X_k)$ is integrable, from
(\ref{nn}) we deduce that $(X_A)^i_B=(X_B)_A^i$.
}
\end{remark}

\begin{lemma}\label{lem1} Let ${\bf X}=(X_1,\ldots , X_k)$ be
an integrable $k$-vector field on $\r^k \times T^1_kQ$. If  every
integral section of ${\bf X}$ is the  first prolongation
$\phi^{[1]}$ of map $\phi:\rk \to Q$, then ${\bf X}$ is a {\sc
sopde}.
\end{lemma}
\proof Let us suppose that each $X_A$ is locally given by
\begin{equation}\label{localsode22}
X_A(t,q^i,v^i_B)=(X_A)^B\frac{\partial}{\partial
t^B}+(X_A)^i\frac{\displaystyle
\partial} {\displaystyle
\partial q^i}+
(X_A)^i_B \frac{\displaystyle\partial} {\displaystyle
\partial v^i_B}\ .
\end{equation}
Let $\psi=\phi^{[1]}:\rk \to \r^k \times T^1_kQ$ be an integral
section of ${\bf X}$, then from (\ref {localsode2}), (\ref
{localsode22}) and Definitions \ref{integsect} and \ref{de652} we
obtain
$$
(X_A)^B(\phi^{[1]}(t))=\delta_A^B, \quad
(X_A)^i(\phi^{[1]}(t))=\ds\frac{\partial \phi^i}{\partial
t^A}(t)=v^i_A(\phi^{[1]}(t)), \quad (X_A)^i_B(\phi^{[1]}(t))=
\ds\frac{\partial^2 \phi^i}{\partial t^A
\partial t^B }(t)$$
thus $X_A$ is locally given as in (\ref{localsode2}).
\qed

\subsubsection{The Legendre map and the Lagrangian forms}

Given a Lagrangian $L:\rk \times T^1_kQ \to \r$ the  Legendre map
$FL: \rk\times T^1_kQ \longrightarrow \rk\times (T^1_k)^{\;*}Q$ is
defined as follows:
$$
FL(t,{v_1}_q,\ldots , {v_k}_q)=(t,\ldots, [FL(t,{v_1}_q,\ldots ,
{v_k}_q)]^A,\ldots )\ ,
$$ where
$$ [FL(t,{v_1}_q,\ldots , {v_k}_q)]_q^A(u_q)=
\displaystyle\frac{d}{ds}\Big\vert_{s=0}\displaystyle L\left(
t,{v_1}_q, \dots ,{v_A}_q+su_q, \ldots , {v_k}_q \right) \ .
$$
 It is locally given by
\begin{equation}\label{locfl1}
FL:(t^A,q^i,v^i_A)  \longrightarrow  \left(t^A,q^i,
\frac{\displaystyle\partial L}{\displaystyle\partial v^i_A }
\right)\, .
\end{equation}

 From (\ref{am2}) and (\ref{locfl1}) the following identities hold
\begin{equation}\label{nueva}
\theta_L^A=FL^{\;*}\theta^A\, , \quad\omega_L^A=FL^{\;*}\omega^A \ .
\end{equation}

\begin{definition}\label{de811}
A Lagrangian function $L:\rk\times T^1_kQ\longrightarrow \r$ is said
to be {\rm regular}  (resp. {\rm hyperregular}) if the corresponding
Legendre map  $FL$ is a local (resp. global) diffeomorphism.
Elsewhere $L$ is called a {\rm singular} Lagrangian.
\end{definition}

{}From (\ref{locfl1}) we obtain that   $L$ is regular if, and only
if, $\det \left(\frac{\displaystyle\partial^2L}
{\displaystyle\partial v^i_A
\partial v^j_B}\right)\neq 0$, $1 \leq i,j \leq n$, $1\leq A,B \leq k$.
Therefore (see \cite{mod2}):

\begin{proposition}  The following conditions are equivalent: $(i) $ $L$ is
regular, $(ii)$ $FL$  is a local diffeomorphism. $(iii)$
$(dt^A,\omega_L^A,V_*)$ is a $\,k$-cosymplectic structure on $\rk
\times T^1_kQ$ where $V_*=ker \,
\left((\pi_{\rk})_{1,0}\right)_*=\big\langle\frac{\displaystyle\partial}
{\displaystyle\partial v^i_1}, \dots,
\frac{\displaystyle\partial}{\displaystyle\partial
v^i_k}\big\rangle_{i=1,\ldots , n}$ is the vertical distribution of
the bundle $(\pi_{\rk})_{1,0}:\rk\times T^1_kQ \to
\rkq$.\end{proposition}

\begin{definition}
A singular Lagrangian function $L:\rk\times T^1_kQ\longrightarrow
\r$ is called {\rm almost-regular} if $\mathcal{P}:= FL(T^1_kQ)$ is
a closed submanifold of $\rk\times(T^1_k)^{\;*}Q$ (we will denote
the natural imbedding by $\jmath_0:\mathcal{P}\hookrightarrow
\rk\times(T^1_k)^{\;*}Q$), $FL$ is a submersion onto its image, and
the fibres $FL^{-1}(FL(t,w_q))$, for every $(t,w_q)\in \rk\times
T^1_kQ$, are connected submanifolds of $\rk\times T^1_kQ$.
\end{definition}

Observe that the vector fields
$\frac{\displaystyle\partial}{\displaystyle\partial t^A}$ are
tangent to $\mathcal{P}$.

\subsubsection{$k$-cosymplectic Lagrangian formalism}

Suppose that a physical system is described by $n$
functions $\psi^i(t^1,\ldots,t^k)$. Associated with this system there is a
Lagrangian $L(t^A,\psi^i,\psi^i_A)$, with
$\ds \psi^i_A=\ds\frac{\partial \psi^i }{\partial t^A}$;
then the Euler-Lagrange equations are
$$
\ds\sum_{A=1}^k \left(\ds\frac{\partial^2L}{\partial t^A \partial
\psi^i_A} \, + \, \ds\frac{\partial^2L}{\partial q^j \partial
\psi^i_A} \,   \ds\frac{\partial \psi^j}{\partial t^A} \, +
\,\ds\frac{\partial^2L}{\partial \psi^j_B \partial \psi^i_A} \,
\ds\frac{\partial^2 \psi^j}{\partial t^A \partial t^B}\,\right) = \,
\ds\frac{\partial L}{ \partial q^i}
$$
and we can consider that the Lagrangian $L$ is defined on $\rk\times
T^1_kQ$, that is  $L=L(t^A,q^i,v^i_A)$,   and we can write the
Euler-Lagrange   equations as
\begin{equation}\label{e-l-2}
\displaystyle \sum_{A=1}^k\ds\frac{\partial}{\partial
t^A}\Big\vert_t  \left(\frac{\displaystyle\partial L}{\displaystyle
\partial v^i_A}(\psi(t)) \right)= \frac{\displaystyle \partial
L}{\displaystyle
\partial q^i}(\psi(t)) \, ,\quad v^i_A(\psi(t))=\ds\frac{\partial( q^i
\circ \psi )}{\partial t^A}(t)\, ,
\end{equation}
where each solution $\psi:U_0\subset \rk \to \rk \times T^1_kQ$ is
given by $\psi(t)=\left(t,\psi^i(t),\ds\frac{\partial   \psi^i
}{\partial t^A}(t)\right)$.

Thus each solution $\psi(t)$ of the Euler-Lagrange equations
(\ref{e-l-2}) is a first prolongation of a map $\phi:U_0\subset
\r^k\to Q$ given by $\phi(t)=(\psi^i(t))$.

Next we give a geometrical description of these equations.
Let us consider the equations
\begin{equation}\label{lageq0}
dt^A ((X_L)_B)  =   \delta^A_B \quad , \quad
\ds\sum_{A=1}^k \, i_{(X_L)_A} \omega_L^A   =
dE_L + \,\ds\sum_{A=1}^k\ds\frac{\partial L}{\partial t^A}dt^A\, ,
\end{equation}
where $E_L=\Delta(L)-L$. If $(X_L)_A$ is locally given by
$$
(X_L)_A=((X_L)_A)^B\frac{\partial}{\ds\partial t^B} +
((X_L)_A)^i\frac{\partial}{\ds\partial q^i}
+((X_L)_A)^i_B\frac{\partial}{\ds\partial v_B^i} \ ,
$$
we obtain that (\ref{lageq0}) is equivalent to the equations
$$
((X_L)_A)^B=\delta_A^B \quad , \quad ((X_L)_B)^i  \,
\ds\frac{\partial^2 L}{\partial t^A \partial v^i_B}= v^i_B \,
\ds\frac{\partial^2 L}{\partial t^A \partial v^i_B}
$$
\begin{equation}\label{lagloc2}
((X_L)_C)^j \, \ds\frac{\partial^2 L}{\partial v^i_B \partial
v^j_C}= v^j_C\,  \ds\frac{\partial^2 L}{\partial v^i_B \partial v^j_C}
\end{equation}
\begin{equation}\label{lagloc22}
\ds\frac{\partial^2 L}{\partial q^j \partial v^i_B}\left(v^i_B
-((X_L)_B)^i\right)+ \ds\frac{\partial^2 L}{\partial t^B \partial
v^i_B}+v^k_B \ds\frac{\partial^2 L}{\partial q^k \partial v^i_B}+
((X_L)_B)_C^k \ds\frac{\partial^2 L}{\partial v^k_C \partial v^i_B}=
\ds\frac{\partial  L}{\partial q^i}\ .
\end{equation}

When $L$ is regular, from (\ref{lagloc2}) we obtain that  this last
equation can be written as follows
\begin{equation}\label{lagloc3}
\ds\frac{\partial^2 L}{\partial t^B \partial v^i_B}+v^k_B
\ds\frac{\partial^2 L}{\partial q^k \partial v^i_B}+ ((X_L)_B)_C^k
\ds\frac{\partial^2 L}{\partial v^k_C \partial v^i_B}=
\ds\frac{\partial  L}{\partial q^i} \, ,
\end{equation}
and then $(X_L)_A$ is locally given by
$$
(X_L)_A= \ds\frac{\partial}{\partial t^A}+ v^i_B
\ds\frac{\partial}{\partial q^i}+ ((X_L)_A)_B^i \ds\frac{\partial
}{\partial v^i_B}\ ;
$$
that is, $((X_L)_1 , \ldots , (X_L)_k)$ is a {\sc sopde}.

\begin{theorem}\label{rel} Let $L$ be a   Lagrangian and
${\bf X}_L=((X_L)_1,\dots,(X_L)_k)$ a $k$-vector field such that
\begin{equation}\label{eqthel11}
dt^A ((X_L)_B) \, = \, \delta^A_B, \quad , \quad
 \ds\sum_{A=1}^k \, i_{(X_L)_A} \omega_L^A =\,
dE_L + \,\ds\sum_{A=1}^k\ds\frac{\partial L}{\partial t^A}dt^A \, .
\end{equation}
\begin{enumerate}
\item If $L$ is regular, then ${\bf X}_L=((X_L)_1,\dots,(X_L)_k)$
is a {\sc sopde}. If $\psi:\rk \to \rk\times T^1_kQ$ is an integral
section of ${\bf X}_L$, then $\phi:\rk \stackrel{\psi}{\to}\rk
\times T^1_kQ \stackrel{p_Q}{\to}Q$ is a solution to the
Euler-Lagrange equations (\ref{e-l-2}).
\item If $((X_L)_1,\dots,(X_L)_k)$ is integrable,
and $\phi^{[1]}:\rk \to \rk\times T^1_kQ$ is an integral section,
then $\phi:\rk \to Q$ is a solution to the Euler-Lagrange equations
(\ref{e-l-2}).
\end{enumerate}
\end{theorem}
\proof ${\it 1}$ is a consequence of  (\ref{lagloc2}) and
(\ref{lagloc3}). If $\phi^{[1]}$ is an integral section, from
 (\ref{lagloc22})  and the local expression of $\phi^{[1]}$
we have that $\phi$ is a solution to the Euler-Lagrange equations
(\ref{e-l-2}).
 \qed

\begin{remark}
{\rm
If $L:\rk\times T^1_kQ\longrightarrow \r$ is a regular Lagrangian,
then $(dt^A,\omega_L^A,V_*)$ is a $k$-cosymplectic structure on $\rk
\times T^1_kQ$. The Reeb vector fields $(R_L)_A$ corresponding to
this $k$-cosymplectic structure  are characterized by
$$
i_{(R_L)_A} \, dt^B=\delta_A^B \quad , \quad i_{(R_L)_A} \,
\omega^B_L=0 \,  ,
$$
and satisfy  $(R_L)_A(E_L)=-\partial L /\partial t^A$.
}
\end{remark}

If the Lagrangian $L$ is hyper-regular, that is, $FL$ is a
diffeomorphism, then we can define a Hamiltonian function $H:\rk
\times (T^1_k)^*Q \to \r$ by $H=E_L \circ FL^{-1}$ where $FL^{-1}$
is the inverse map of $FL$. Then we have the following:

\begin{theorem}\label{te421}
\begin{enumerate}
\item If $\;{\bf X}_L=((X_L)_1,\dots,(X_L)_k)$ is a solution to
(\ref{lageq0}), then ${\bf X}_H=((X_H)_1, \ldots, (X_H)_k)$, where
$(X_H)_A=FL_*((X_L)_A)$ is a solution to the
equations (\ref{geonah}) in $\rk \times (T^1_k)^*Q$, with
$\eta^A=\eta_0^A$, $\omega^A=\omega^A_0$, and $H=E_L\circ FL^{-1}$.
\item If ${\bf X}_L=((X_L)_1,\dots,(X_L)_k)$ is integrable and
$\phi^{[1]}$ is an integral section, then $\varphi=FL \circ
\phi^{[1]}$ is an integral section of \, ${\bf X}_H=((X_H)_1,
\ldots, (X_H)_k)$ and thus it is a solution to the Hamilton field
equations (\ref{he}) for $H=E_L\circ FL^{-1}$.
\end{enumerate}
\end{theorem}
\proof
\begin{enumerate}
\item It is an immediate consequence of (\ref{geonah}) and
(\ref{lageq0}) using that $FL^*\eta_0^A=dt^A$,
$FL^*\omega^A_0=\omega^A_L$, and $E_L=H\circ FL^{-1}$. \item It is
an immediate consequence of Definition \ref {integsect} of integral
section of a $k$-vector field.
\qed
\end{enumerate}

\begin{remark}
{\rm
If we rewrite the equations (\ref{eqthel11}) for the case $k=1$, we
have
$$
dt(X_L)=1 \quad, \quad i_{X_L}\omega_L = dE_L + \ds\frac{\partial L}{\partial t }dt\ ,
$$
which are equivalent to the dynamical equations
$$
dt(X_L)=1 \quad, \quad i_{X_L}\Omega_L =0
$$
where $\Omega_L=\omega_L+dE_L\wedge dt$ is the Poincar\'{e}-Cartan
$2$-form associated to the Lagrangian $L$ (see \cite{mt0}). This
describes the non-autonomous Lagrangian mechanics. Then, applying
theorem \ref{te421} the non-autonomous Hamiltonian mechanics is
obtained.
}
\end{remark}

If the Lagrangian $L$ is singular, then the existence of solutions
to the equations (\ref{lageq0}) is not assured except, perhaps, in a
submanifold of $\rk \times T^1_kQ$ (see \cite{mod2}). Furthermore,
when these solutions exist, they are not {\sc sopde}, in general.
Thus, in order to recover the Euler-Lagrange equations
(\ref{e-l-2}), the following condition must be added to the
equations (\ref{lageq0}) (see definition \ref{sode2}):
$$
{S}^A(X_A)=\Delta_A \ .
$$

If the Lagrangian is almost-regular, then there exists
$H_0\in\Cinfty({\mathcal P})$ such that $(FL_0)^{\;*}H_0=E_L$, where
$FL_0\colon \rk \times T^1_kQ\to{\mathcal P}$ is defined by
$\jmath_0\circ FL_0=FL$. The Hamiltonian field equation analogous to
(\ref{geonah}) should be
$$
\label{geonah0} \jmath_0^{\;*}(\eta^A)((X_0)_B)=\delta^A_B\, ,\quad
\displaystyle \sum_{i=1}^k \,
\imath_{(X_0)_A}(\jmath_0^{\;*}(\omega^A_0)) =
dH_0-\displaystyle\sum_{A=1}^k \ds\frac{\partial H_0}{\partial t^A}j_0^{\;*}(\eta^A) \, .
$$
where ${\bf X}_0=((X_0)_1,\ldots,(X_0)_k)$ (if it exists) is a
$k$-vector field on ${\mathcal P}$. The existence of $k$-vector
fields in ${\mathcal P}$ solution to the above equations
is not assured except, perhaps, in a submanifold of ${\mathcal P}$.

\section{Tulczyjew's submanifolds for $k$-cosymplectic field theories}\label{Tulczyjew}

This approach is inspired in papers \cite{T1,T2} and
\cite{LLR-91,LMS-03}.

\subsection{Derivations on $T^1_kM$}\label{31}

 Let us denote with $\bigwedge N$   the algebra
of the exterior differential forms on an arbitrary manifold $N$. In
\cite{T1,T2}, a derivation $i_{\ds T} $ of degree $-1$ from
$\bigwedge M$ on $\bigwedge TM$ over $\tau_M:TM\to M$ was defined in
an arbitrary manifold $M$ by $i_{T}\mu=0$ if $\mu$ is a function on
$M$, and by
$$
i_{\ds T}\mu(v_x)  (Z^1_{v_x},\ldots ,Z^l_{v_x})=\mu
(p)(v_x,(\tau_M)_*(v_x)(Z^1_{v_x}),\ldots
,(\tau_M)_*(v_x)(Z^l_{v_x})),
$$
if $\mu$ is a $l+1$-form, where $x\in M$, $Z^i_{v_x}\in T_{v_x}(TM)$.

A derivation $d_T$ of degree $0$ from $\bigwedge (M)$ on $\bigwedge
TM$ over $\tau$ is defined by $d_T\mu=i_Td\mu +di_T\mu$, where $d$
is the exterior derivative. We have $dd_T=d_Td$.

We extend the above definition as follows: for every $A=1 ,
\ldots , k$ we  define a derivation $i_{\ds T_A} $ of degree $-1$
from $\bigwedge (M)$ on $\bigwedge T ^1_kM$ over $\tau:T^1_kM\to M$
by $i_{T_A}\mu=0$ if $\mu$ is a function on $M$, and by
\begin{equation}\label{itmua}
i_{\ds T_A}\mu(w_x)  (\widetilde{Z}^1_{w_x},\ldots
,\widetilde{Z}^l_{w_x})=\mu
(p)(\tau_A(w_x),\tau_*(w_x)(\widetilde{Z}^1_{w_x}),\ldots
,\tau_*(w_x)(\widetilde{Z}^l_{w_x})),
\end{equation}
if $\mu$ is an $l+1$-form, where $\tau_A:T^1_kM\to M$ is the
projection on the $A^{th}$-component of $T^1_kM$, $w_x\in T^1_kM$
and $\widetilde{Z}^i_{w_x}\in T_{w_x}(T^1_kM), \quad 1\leq i \leq
l,$.

A derivation $d_{T_A}$ of degree $0$ from $\bigwedge (M)$ on
$\bigwedge T^1_kM$ over $\tau$ is defined by $d_{T_A}\mu=i_{T_A}
d\mu +di_{T_A}\mu$, where $d$ is the exterior derivative. We have
$d\, d_{T_A}=d_{T_A}d$.

\subsection{Tulczyjew's Hamiltonian formulation}\label{tulzch}
In this subsection it is important to consider the paragraph ``{\it
The manifold $T^1_k(\rk\times P)$}" of the Section \ref{221}. Here
we consider the particular case $P=(T^1_k)^*Q$.

Let $W_{(t,\alpha)}=(v_{1_{(t,\alpha)}}, \ldots ,
v_{k_{(t,\alpha)}})$ be a point in $T^1_k(\rk\times (T^1_k)^{\;*}Q)$
where $(t,\alpha)\in\rk\times (T^1_k)^{\;*}Q $, that is
$v_{A_{(t,\alpha)}}\in T_{(t,\alpha)}(\rk\times (T^1_k)^{\;*}Q)$,
$1\leq A \leq k$.  We write
$$
v_{A_{\ds(t,\alpha)}}= (v_A)_B \ds\frac{\partial}{\partial
t^B}\Big\vert_{(t,\alpha)}+(v_A)^i \ds\frac{\partial}{\partial
q^i}\Big\vert_{(t,\alpha)} + (v_A)^B_i
\ds\frac{\partial}{\partial p^B_i}\Big\vert_{(t,\alpha)}
$$
 and we introduce the canonical coordinates
$(t^A,q^i,p^A_i,(v_A)_B,(v_A)^i,(v_A)^B_i)$ on $T^1_k(\rk\times
(T^1_k)^{\;*}Q)$.

Let $\omega^A$ be the canonical $2$-forms on $\rk\times (T^1_k)^{\;*}Q$
introduced in the subsection $2.1.1$.
 From (\ref{itmua}) we introduce the  $1$-forms $i_{T_A}\omega^A$ on
$T^1_k(\rk\times (T^1_k)^{\;*}Q)$, which are locally given by
\begin{equation}\label{itw1}
i_{T_A}\omega^A=\ds\sum_{i=1}^k ((v_A)^i\, \,  dp^A_i - (v_A)^A_i dq^i )\ .
\end{equation}

Let $H:\rk\times (T^1_k)^{\;*}Q$ be a Hamiltonian function.
Associated with $H$ we define the    submanifold
$$\begin{array}{lcl} D_H&=&\Big\{
W_{(t,\alpha)}\in T^1_k(\rk\times (T^1_k)^{\;*}Q) \, :
dt^B(v_{A_{(t,\alpha)}})=\delta^B_A\,,\\\noalign{\medskip}&& \,
\Big(\ds\sum_{A=1}^k i_{T_A}\omega^A \Big)(W_{(t,\alpha)}) =
\Big({\tau_{\rk\times (T^1_k)^{\;*}Q}}^{\;*}(dH- \ds\frac{\partial
H}{\partial t^A}dt^A)\Big) (W_{(t,\alpha)})\Big \}\;,\end{array}
$$where ${\tau_{\rk\times (T^1_k)^{\;*}Q}}:T^1_k(\rk\times
(T^1_k)^{\;*}Q)\to \rk\times (T^1_k)^{\;*}Q$ is the canonical
projection of the tangent bundle of $k^1$-velocities of $\rk\times
(T^1_k)^{\;*}Q$.

 From (\ref{itw1})    we deduce
that  $D_H$ is locally defined by the constraints
\begin{equation}\label{dhe2}
\begin{array}{c}
(v_A)^B(W_{(t,\alpha)})=\delta_A^B\; ,\quad (v_A)^i(W_{(t,\alpha)})=
\ds\frac{\partial H}{\partial p^A_i}\Big\vert_{\tau_{\rk\times
(T^1_k)^{\;*}Q}(W_{(t,\alpha)})} \; , \;\\\noalign{\medskip}
   -\;\ds\sum_{A=1}^k(v_A)^A_i(W_{(t,\alpha)})=
\ds\frac{\partial H}{\partial q^i}\Big\vert_{\tau_{\rk\times
(T^1_k)^{\;*}Q}(W_{(t,\alpha)})}  \; .\end{array}
\end{equation}
and thus $\dim D_H=k+(nk)+k(nk)$.

\begin{proposition}
Let ${\bf X}=(X_1,\ldots , X_k)$ be an integrable $k$-vector field
on $\rk\times (T^1_k)^{\;*}Q$ such that $Im \,{\bf X}\subset D_H$.
Then its integral sections are solutions of the
HDW-equations.\end{proposition} \proof  If
$$X_A=(X_A)^B  \ds\frac{\partial} {\partial
t^B}+(X_A)^i  \ds\frac{\partial} {\partial q^i}+ (X_A)^B_i
\ds\frac{\partial}{\partial  p^B_i}$$
 then from (\ref{dhe2}) and definition of $D_H$ we have that
\begin{equation}\label{xh1}
\begin{array}{c}
(X_A)^B(W_{(t,\alpha)})=\delta_A^B\; ,\quad (X_A)^i(W_{(t,\alpha)})=
\ds\frac{\partial H}{\partial p^A_i}\Big\vert_{\tau_{\rk\times
(T^1_k)^{\;*}Q}(W_{(t,\alpha)})} \; , \;\\\noalign{\medskip}
 -\;  \ds\sum_{A=1}^k(X_A)^A_i(W_{(t,\alpha)})=
\ds\frac{\partial H}{\partial q^i}\Big\vert_{\tau_{\rk\times
(T^1_k)^{\;*}Q}(W_{(t,\alpha)})}  \; .\end{array}
\end{equation}
If ${\bf X}$ is integrable and    if   $\psi:\rk \to \rk\times
(T^1_k)^{\;*}Q$, given by
$\psi(t)=(\psi^A(t),\psi^i(t),\psi^A_i(t))$, is an integral section
of ${\bf X}$ , then   we obtain
\begin{equation}\label{ll}
\ds\frac{\partial \psi^B}{\partial
t^A}\Big\vert_{t}=(X_A)^B(\psi(t)), \quad \ds\frac{\partial
\psi^i}{\partial t^B}\Big\vert_{t}=(X_B)^i(\psi(t)), \quad
\ds\frac{\partial \psi^A_i}{\partial
t^B}\Big\vert_{t}=(X_B)^A_i(\psi(t)) \, .
\end{equation} Therefore, from  (\ref{xh1}) and (\ref{ll})
we obtain that $\psi(t)$ is a solution to the Hamilton field equations
(\ref{he})  where  $\psi(t)=(t+cte,\psi^i(t),\psi^A_i(t))$. \qed

\subsection{Tulczyjew's Lagrangian formulation}\label{33}

Let $\theta^A$ be the canonical $1$-forms on
$\rk\times(T^1_k)^{\;*}Q$ introduced in subsection $2.1.1$. Then
using that $d_{T_A}\theta^A=-i_{T_A}\omega^A+di_{T_A}\theta^A_0, \,$
where $ \omega^A= -d\theta^A$,  we obtain from (\ref{itmua}) that
the $1$-forms $d_{T_A}\theta^A$ on $T^1_k(\rk\times (T^1_k)^{\;*}Q)$
are locally given by
\begin{equation}\label{sa}
d_{T_A}\theta^A=  (v_A)^A_i\, \,   dq^i + p^A_i d(v_A)^i\in
\bigwedge^1T^1_k(\rk\times(T^1_k)^{\;*}Q) \quad .
\end{equation}

Let $L(t^A,q^i,v_A^i)$ be a function on $\rk\times T^1_kQ$ and let
$D_L$ be the submanifold
 {\small$$D_L= \{ W_{(t,\alpha)}\in
T^1_k(\rk\times (T^1_k)^{\;*}Q)  \, :
dt^B(v_{A_{(t,\alpha)}})=\delta^B_A\,, \, \ds\sum_{A=1}^k d_{T_A}
\theta^A (W_{(t,\alpha)}) = F^{\;*}(dL - \frac{\partial L}{\partial
t^A}dt^A)  (W_{(t,\alpha)}) \} \ ,
$$}
$F =  (\tau_{\r^k}\times  T^1_k(\pi^k_Q):T^1_k(\rk\times
(T^1_k)^*Q)\to \rk\times T^1_kQ$ being the map defined by (\ref{locF}),
which, in this particular case (i.e., with $P=(T^1_k)^*Q$ and
$\pi=\pi^k_Q$) is locally given by
$$F(t^A,q^i,p^A_i,(v_A)_B,(v_A)^i,(v_A)^B_i)=(t^A,q^i,(v_A)^i)\, .$$

 From   (\ref{sa}) we deduce that $D_{\mathcal{L}}$ is characterized
by the equations

\begin{equation}\label{L2}
(v_A)_B(W_{(t,p)})=\delta_A^B\quad , \quad  p^A_i(W_{(t,p)}) =
\ds\frac{\partial L}{\partial v^i_A}\Big\vert_{ F(w_{(t,p)})} \quad
, \quad \ds\sum_{A=1}^k(v_A)^A_i(W_{(t,p)})= \ds\frac{\partial
L}{\partial q^i}\Big\vert_{F(W_{(t,p)})}  \;\; .
\end{equation}
and thus $D_L$ has dimension $ k+nk+nk^2$.
\bigskip

\begin{proposition}Let $\Psi:\rk\to \rk\times(T^1_k)^{\;*}Q$ be a
section of $\rk\times(T^1_k)^{\;*}Q \to \rk$. If $$
\Psi^{(1)}(t)=(\Psi_*(t)(\ds\frac{\partial}{\partial t^1}(t)),
\ldots ,\Psi_*(t)(\ds\frac{\partial}{\partial t^k}(t)))\in D_L$$
then $\psi =(\pi_Q)_1\circ\Psi:\rk \to Q$ is a solution to the
Euler-Lagrange equations.
\end{proposition}
\proof  If the local expression of $\Psi$ is given by
$\Psi(t)=(t^A,\Psi^i(t),\Psi^A_i(t))$, then
$$\Psi_*(t)(\ds\frac{\partial }{\partial t^A}(t))=
\ds\frac{\partial}{\partial t^A}_{\ds |\Psi(t)}+ \ds\frac{\partial
\Psi^i}{\partial t^A}(t)\ds\frac{\partial} {\partial q^i}_{\ds
|\Psi(t)} +\ds\frac{\partial \Psi^B_i}{\partial t^A
}(t)\ds\frac{\partial}{\partial p^B_i}_{\ds |\Psi(t)}
$$
and thus, we deduce from (\ref{L2}) that
\beann
1)&\qquad& (v_A)_B(\Psi^{(1)}(t))=\delta^B_A\quad ,\\
2)&\qquad& p^A_i(\Psi^{(1)}(t))=p^A_i(\Psi(t))=\Psi^A_i(t)=\ds\frac{\partial
L}{\partial v^i_A}\Big\vert_{F(\Psi^{(1)}(t))}\quad , \\
3)&\qquad&\ds\sum_{A=1}^k(v_A)^A_i( \Psi^{(1)}(t))=
\ds\sum_{A=1}^k\ds\frac{\partial \Psi^A_i}{\partial t^A }(t)=
\ds\frac{\partial L}{\partial q^i}\Big\vert_{F( \Psi^{(1)}(t))}
\quad .
\eeann
\ From $2)$ and $3)$ we obtain
$$\ds\sum_{A=1}^k \ds\frac{\partial  }{\partial t^A}
\big(\ds\frac{\partial L  }{\partial v^i_A}\Big\vert_{F(
\Psi^{(1)}(t))}\big)=\ds\frac{\partial L}{\partial q^i}\Big\vert_{F(
\Psi^{(1)}(t))}\;.
$$
Now, we consider the map $\psi =(\pi_Q)_1\circ\Psi$ locally given by
$\psi(t)=(\Psi^i(t))$. It is easy to show that  $F\circ
\Psi^{(1)}(t)=(t,\psi_*(t)(\ds\frac{\partial}{\partial t^1}(t)),
\ldots , \psi_*(t)(\ds\frac{\partial}{\partial t^k}(t)))
=\psi^{[1]}(t)$, where $\psi^{[1]}$ is defined in Definition
\ref{de652}. So, we obtain
$$
\ds\sum_{A=1}^k \ds\frac{\partial  }{\partial t^A}
\big(\ds\frac{\partial {\mathcal{L}}  }{\partial v^i_A}_{\ds
|\psi^{[1]} (t)}\big)=\ds\frac{\partial {\mathcal{L}}}{\partial
q^i}_{\ds |\psi^{[1]}(t)}\;,
$$ that is, $\psi$ is a solution to the Euler-Lagrange equations.
\qed

\section{Skinner-Rusk formulation}

\subsection{The Skinner-Rusk formalism for $k$-cosymplectic field theories}

Let us consider the {\sl Whitney sum} ${\mathcal M}=\left( \r^k
\times T^1_kQ \right) \oplus_{\r^k \times Q} \left( \r^k \times
(T^1_k)^{\;*}Q \right)$, with natural coordinates
$(t^A,q^i,v^i_A,p^A_i)$. It has natural bundle structures over $\r^k
\times T^1_kQ$ and $\r^k \times (T^1_k)^{\;*}Q$. Let us denote by
$pr_1:{\mathcal M} \to \r^k \times T^1_kQ$ the projection into the
first factor, $pr_1(t^A,q^i,v^i_A,p^A_i)=(t^A,q^i,v^i_A)$ and by
$pr_2:{\mathcal M} \to \r^k \times(T^1_k)^{\;*}Q$ the projection
into the second factor, $pr_2(t^A,q^i,v^i_A,p^A_i)=(t^A,q^i,p^A_i)$.

Let $(\eta^1, \ldots, \eta^k,\omega^1, \ldots , \omega^k)$ be the
canonical forms of the canonical $k$-cosymplectic structure on $\r^k
\times (T^1_k)^{\;*}Q$. We denote
$$
\vartheta^A=(pr_2)^{\;*} dt^A=d t^A\, , \quad
\Omega^A=(pr_2)^{\;*}\omega^A \, ,
$$
and so we have the family $(\vartheta^1, \ldots , \vartheta^k,
\Omega^1, \ldots ,\Omega^k)$ in ${\mathcal M}$.

Now, taking the $k$-vector field $\left(\derpar{}{t^1}, \ldots,
\derpar{}{t^k}\right)$ in $\r^k \times (T^1_k)^{\;*}Q$, we can
define a family of $k$-vector fields $(\xi_1, \ldots, \xi_k)$ in
${\mathcal M}$ such that
$$
(pr_2)_*\xi_A=\derpar{}{t^A} \ .
$$
These $k$-vector fields $(\xi_1, \ldots, \xi_k)$ satisfy that, for
$1 \leq A,B \leq k$, \beann \imath_{\xi_A}\vartheta^B &=&
\imath_{\xi_A} (pr_2^{\;*}dt^B)=
pr_2^{\;*}(\imath_{\derpar{}{t^A}}dt^B)=\delta_A^B
\\
\imath_{\xi_A}\Omega^B &=& \imath_{\xi_A}(pr_2^{\;*}\omega^B_0)=
pr_2^{\;*}(\imath_{\derpar{}{t^A}}\omega^B)=0 \ , \eeann
 and they are locally given by
\begin{equation}\label{37}
\xi_A=\ds\frac{\partial}{\partial t^A} + (\xi_A)^i_B
\ds\frac{\partial}{\partial v^i_B} \, ,
\end{equation}
where $(\xi_A)^i_B$ are arbitrary local functions in ${\mathcal M}$.
Hence, this $k$-vector field is not unique.

Finally, the {\sl coupling function} in ${\mathcal M}$, denoted by
${\mathcal C}$, is defined as follows:
$$
\begin{array}{ccccl}
{\mathcal C} & : & {\mathcal M}=\left( \r^k \times T^1_kQ \right)
\oplus_{\r^k \times Q} \left( \r^k \times
(T^1_k)^{\;*}Q \right) & \longrightarrow & \r \\
\noalign{\medskip} & & (t, {v_1}_q,\ldots ,{v_k}_q,\alpha^1_q,\ldots
,\alpha^k_q) & \mapsto & \ds\sum_{A=1}^k \alpha^A_q({v_A}_q)
\end{array}
$$

Given a Lagrangian $L\in\Cinfty \left(\r^k \times T^1_kQ \right)$,
we define the {\sl Hamiltonian function}
$\mathcal{H}\in\Cinfty\left({\mathcal M}\right)$ as
\begin{equation}\label{s0}
\mathcal{H}={\mathcal C}-pr_1^{\;*}L \ ,
\end{equation}
which, in coordinates, is given by
\begin{equation}\label{s1}
\mathcal{H}=  p^A_i \, v^i_A-L(t^A,q^i,v^i_A) \quad .
\end{equation}
Then, in this formalism, we have the following problem:

\begin{state}
\label{problema0}   Suppose that an integrable
$k$-vector field ${\bf Z}=(Z_1,\ldots,Z_k)$ exists in ${\mathcal M}$, such
that
\begin{equation}\label{s3}
\vartheta^A(Z_B)=\delta^A_B \quad , \quad  \ds\sum_{A=1}^k
\imath_{Z_A}\Omega^A=d
\mathcal{H}-\ds\sum_{A=1}^k\xi_A(\mathcal{H})\, \vartheta^A \, .
\end{equation}
The problem is to find the integral sections $\psi\colon\rk\to
{\mathcal M}$ of ${\bf Z}=(Z_1,\ldots,Z_k)$.
\end{state}

Equations (\ref{s3}) give different kinds of information. In fact,
writing locally each $Z_A$ as
$$
Z_A= (Z_A)^B \ds\frac{\partial}{\partial t^B} + (Z_A)^i
\ds\frac{\partial}{\partial q^i} +
(Z_A)^i_B\ds\frac{\partial}{\partial v^i_B}+ (Z_A)^B_i
\ds\frac{\partial}{\partial p^B_i} \, , \label{zeta}
$$
from (\ref{locexp}), (\ref{s1}) and (\ref{s3}) we obtain
\begin{eqnarray}
\label{s4b} (Z_A)^B=\delta^B_A \\
\label{s6}
p^A_i= \ds\frac{\partial L}{\partial v^i_A}\circ pr_1 \\
\label{s4} (Z_A)^i=v^i_A \\
\label{s5} \ds\sum_{A=1}^k (Z_A)^A_i=\ds\frac{\partial L}{\partial
q^i}\circ pr_1 \ .
\end{eqnarray}
Then the vector
fields $Z_A$ are locally given by
\begin{equation}\label{s7}
Z_A=  \ds\frac{\partial}{\partial t^A} + v^i_A
\ds\frac{\partial}{\partial q^i} +
(Z_A)^i_B\ds\frac{\partial}{\partial v^i_B}+ (Z_A)^B_i
\ds\frac{\partial}{\partial p^B_i} \quad ,
\end{equation}
where the coefficients $(Z_A)^B_i$ are related by the equations
(\ref{s5}). Observe that these equations do not depend on the
arbitrary functions $(\xi_A)^i_B$, that is, on the family of vector
fields $\left\{\xi_A\right\}$ that we have chosen to extend the
vector fields $\left\{\derpar{}{t^A}\right\}$.

So, in particular, we have obtained information on four different classes:
\begin{enumerate}
\item
The constraint equations (\ref{s6}), which are algebraic (not
differential) equations defining a submanifold $M_L$ of ${\mathcal
M}$ where the equation (\ref{s3}) has solution. Observe that this
submanifold is just the graph of the Legendre map $FL$ defined by
the Lagrangian $L$.
\item Let us observe that, as a consequence of  (\ref{s6}), the
$k$-vector field ${\bf Z}=(Z_1,\ldots,Z_k)$, $Z_A\in
\mathfrak{X}({\mathcal M})$, satisfies equation  (\ref{s3})  only on
$M_L$ .
\item Equations (\ref{s4}), called the {\sl {\sc sopde}
condition}, will be used in the following subsection (see Theorem
\ref{ELH-eq}), to show that the integral sections of ${\bf
Z}=(Z_1,\ldots,Z_k)$ can be obtained from first prolongations
$\phi^{[1]}$ of maps $\phi:\rk \to Q$.
\item
Equations (\ref{s5}), which, taking into account (\ref{s4b}),
(\ref{s6}) and (\ref{s4}), will give the classical Euler-Lagrange
equations for the integral sections of ${\bf Z}$ (see Theorem
\ref{ELH-eq}).
\item From (\ref{s4b}), (\ref{s6}), (\ref{s4}) and (\ref{s5}) we
deduce that the solutions of equations (\ref{s3}) do not depend on
the $k$-vector field $(\xi_1,\ldots,\xi_k)$ chosen.
\end{enumerate}

If ${\bf Z}=(Z_1,\ldots,Z_k)$ is a solution to (\ref{s3}), then each
$Z_A$ is tangent to the submanifold $M_L$ if, and only if, the
functions $Z_A\left(p^B_j -\ds\frac{\partial L}{\partial v^j_B}\circ
pr_1\right)$ vanish at the points of $M_L$. Then from (\ref{s7}) we deduce that
this is equivalent to the following equations
\begin{equation}\label{s8}
(Z_A)^B_j= \ds\frac{\partial^2 L}{\partial t^A\partial v^j_B} +
v^i_A \, \ds\frac{\partial^2 L}{\partial q^i\partial v^j_B} +
(Z_A)^i_C \, \ds\frac{\partial^2 L}{\partial v^i_C\partial v^j_B}
\quad ,
\end{equation}
which are conditions for the coefficients $(Z_A)^i_C$.

Taking into account that the $k$-vector fields ${\bf Z}$ must be
tangent to the submanifold $M_L$, the above problem can be stated in
$M_L$, instead of in ${\mathcal M}$. First observe that the family
composed of the $k$ vector fields $(\xi_1, \ldots,\xi_k)$ on ${\mathcal
M}$ are tangent to $M_L$ if and only if
$$
\ds\frac{\partial^2 L}{\partial t^A \partial v^i_B}\circ pr_1 +
(\xi_A)^j_C \ds\frac{\partial^2 L}{\partial v^j_C \partial
v^i_B}\circ pr_1 = 0\, ,
$$
since the constraint function defining $M_L$ is
$p^A_i-\ds\frac{\partial L}{\partial v^i_A}\circ pr_1$. Thus taking
into account $3$, we have

\begin{state} We denote by $\jmath\colon M_L \to {\mathcal M}$ the natural
imbedding. The problem is to find  the integral sections $\psi\colon\rk\to
M_L\subset {\mathcal M}$ of integrable $k$-vector fields ${\bf
Z}_L=((Z_L)_1,\ldots,(Z_L)_k)$ on $M_L$ solution to the following
equations
\begin{equation}\label{s3bis}
(\jmath^{\;*}\vartheta^A)((Z_L)_B)=\delta^A_B \quad , \quad
\ds\sum_{A=1}^k
\imath_{(Z_L)_A}(\jmath^{\;*}\Omega^A)=d(\jmath^{\;*}\mathcal{H})-
\jmath^{\;*}\left[\ds\sum_{A=1}^k  \xi_A(\mathcal{H})\right]
\,(\jmath^{\;*}\vartheta^A)\, ,
\end{equation}
Of course, $\jmath_*(Z_L)_A=Z_A\vert_{M_L}$, where ${\bf
Z}=(Z_1,\ldots,Z_k)$ is the $k$-vector field on ${\mathcal M}$
solution to (\ref{s3}). \end{state}

It is interesting to remark that:
\begin{enumerate}
\item
In general, equations (\ref{s3}) (or, what is equivalent, equations
(\ref{s3bis})) do not have a unique solution. Solutions to
(\ref{s3}) are given by $(Z_1,\ldots,Z_k)+\left( \ker\Omega^{\sharp}
\cap \ker\vartheta^{\sharp} \right)$, where $(Z_1,\dots,Z_k)$ is a
particular solution, and $\Omega^{\sharp}$ is the morphism defined by
$$
\begin{array}{rccl}
\Omega^{\sharp}: & T^1_k{\mathcal M} & \longrightarrow & T^{\;*}{\mathcal M}  \\
\noalign{\medskip} & (Y_1,\dots,Y_k) & \rightarrow &
\Omega^{\sharp}(Y_1,\dots,Y_k) =  \displaystyle \sum_{A=1}^k \,
\imath_{Y_A}\Omega^A+ \vartheta^A(Y_A)\, \vartheta^A \, ,
\end{array}
$$
and, denoting by ${\mathcal M}_k(C^\infty({\mathcal M}))$ the space
of matrices of order $k$ whose entries are functions on ${\mathcal
M}$, the vector bundle morphism $\vartheta^{\sharp}$ is defined by
$$
\begin{array}{rccl}
\vartheta^{\sharp}: & T^1_k {\mathcal M} & \longrightarrow &
{\mathcal M}_k(C^\infty({\mathcal M}))
\\
\noalign{\medskip} & (Y_1,\ldots,Y_k) & \rightarrow &
\vartheta^{\sharp}(Y_1,\ldots,Y_k) =  (\vartheta^A(Y_B))\, .
\end{array}
$$
\item
If $L$ is regular, then taking into account (\ref{s4b}), (\ref{s4})
and (\ref{s5}) we can define a local $k$-vector field $(Z_1 ,\ldots
, Z_k)$ on a neighborhood of each point in $M_L$ which is a solution
to (\ref{s3}). Each $Z_A$ is locally given by
$$
(Z_A)^B=\delta^B_A\quad ,\quad (Z_A)^i=v^i_A\quad ,\quad
(Z_A)^B_i=\ds\frac{1}{k}  \, \ds\frac{\partial L}{\partial q^i} \,
\delta_A^B \, ,
$$
with $(Z_A)^i_B$ satisfying (\ref{s8}). Now, by using a partition of
the unity, one can construct a global $k$-vector field which is a
solution to (\ref{s3}).
\end{enumerate}

When the Lagrangian function $L$ is singular, we cannot ensure the
existence of solutions to the equations (\ref{s3}) or (\ref{s3bis}).
Thus we must develop a constraint algorithm for obtaining a
constraint submanifold (if it exists) where these solutions exist.
Next, we outline this procedure (see also \cite{LMM-2002}, where a
similar algorithm is sketched in the multisymplectic formulation).

Assuming that the Lagrangian is almost-regular, we start with
$P_0=M_L$. Then, let $P_1$ be the subset of $P_0$ composed of those
points where a solution to (\ref{s3bis}) exists, that is,
$$
P_1= \{ z \in P_0 \ |\ \exists ((Z_L)_1, \ldots, (Z_L)_k)\in
(T^1_k)_z P_0 \mbox{ solution to } (\ref{s3bis}) \} \ .
$$
If $P_1$ is a submanifold of $P_0$, there exists a section of
the canonical projection $\tau_{P_0}:T^1_k P_0 \to P_0$ defined on
$P_1$ which is a solution to (\ref{s3bis}), but which does not
define a $k$-vector field on $P_1$, in general. In order to find
solutions taking values into $T^1_k P_1$, we define a new subset
$P_2$ of $P_1$ as
$$
P_2= \{ z \in P_1 \ |\ \exists ((Z_L)_1, \ldots, (Z_L)_k) \in
(T^1_k)_z P_1 \mbox{ solution to } (\ref{s3bis}) \}\ .
$$
If $P_2$ is a submanifold of $P_1$, then there exists a section of
the canonical projection $\tau_{P_1}:T^1_k P_1 \to P_1$ defined on
$P_2$ which is a solution to (\ref{s3bis}), but which does not
define, in general, a $k$-vector field on $P_2$. Procceding further,
we get a family of constraint manifolds
$$
\ldots \, \hookrightarrow \, P_2 \, \hookrightarrow \, P_1 \,
\hookrightarrow \,  P_0=M_L \, \hookrightarrow \, {\mathcal M}\ .
$$
If there is a natural number $f$ such that $P_{f+1}=P_f$ and
$dim \, P_f >k$, then we call $P_f$ the {\sl final constraint
submanifold} on which we can find solutions to equation
(\ref{s3bis}). The solutions are not unique (even in
the regular case) and, in general, they are not integrable. In order
to find integrable solutions to equation (\ref{s3bis}), a constraint
algorithm based on the same idea must be developed.

\subsection{The field equations for sections}

Consider the following restrictions of the projections $pr_1$ and $pr_2$,
$$
pr_1^0\colon M_L \to \r^k \times T^1_kQ\quad , \quad pr_2^0\colon
M_L \to \r^k \times (T^1_k)^{\;*}Q \ .
$$

\begin{remark}
{\rm
Observe that, as $M_L$ is the graph of $FL$, it is diffeomorphic to
$\r^k \times T^1_kQ$, and this means that $pr_1^0$ is really a diffeomorphism.
}
\end{remark}

Let ${\bf Z}=(Z_1,\ldots , Z_k)$ be an integrable $k$-vector field
solution to (\ref{s3}). Every integral section $\psi\colon t\in\rk
\to (\psi^A(t),\psi^i(t),\psi^i_A(t),\psi^A_i(t))\in {\mathcal M}$
of ${\bf Z}$   is of the form $\psi=(\psi_L,\psi_H)$, with
$\psi_L=pr_1\circ\psi\colon\rk\to \r^k \times T^1_kQ$, and if $\psi$
takes values in $M_L$ then $\psi_H=FL\circ\psi_L$. In fact, from
(\ref{s6}),
$$
\psi_H(t)=(pr_2 \circ \psi)(t)=(\psi^A(t),\psi^i(t),\psi^A_i(t))
=\left(\psi^A(t),\psi^i(t), \ds\frac{\partial L}{\partial
v^i_A}(\psi_L(t))\right) =(FL \circ \psi_L)(t) \, .
$$
In this way, every constraint, differential equation, etc. in the
unified formalism can be translated into the non autonomous Lagrangian
or Hamiltonian formalism by restriction to the first or second
factors of the product bundle. In particular, conditions (\ref{s6})
generate, by $pr_2$-projection, the primary constraints of the
Hamiltonian formalism for singular Lagrangians (i.e., the image of
the Legendre transformation, $FL(\r^k \times T^1_kQ)\subset \r^k
\times (T^1_k)^{\;*}Q$), and they can be called {\sl the primary
Hamiltonian constraints}.

Hence the main result in this subsection is the following:

\begin{theorem}\label{ELH-eq}
Let ${\bf Z}=(Z_1,\ldots, Z_k)$ be an integrable $k$-vector field in
${\mathcal M}$ solution to (\ref{s3}), and let $\psi\colon\rk\to
M_L\subset {\mathcal M}$ be an integral section of ${\bf
Z}=(Z_1,\ldots,Z_k)$, with
$\psi=(\psi_L,\psi_H)=(\psi_L,FL\circ\psi_L)$. Then $\psi_L$ is the
canonical lift $\phi^{[1]}$ of the projected section $\phi= p_Q\circ
pr^0_1\circ\psi:\rk
\stackrel{\psi}{\to}M_L\stackrel{pr^0_1}{\approx} \r^k \times T^1_kQ
\stackrel{p_Q}{\to}Q $, and $\phi$ is a solution to the
Euler-Lagrange field equations (\ref{e-l-2}).
\end{theorem}
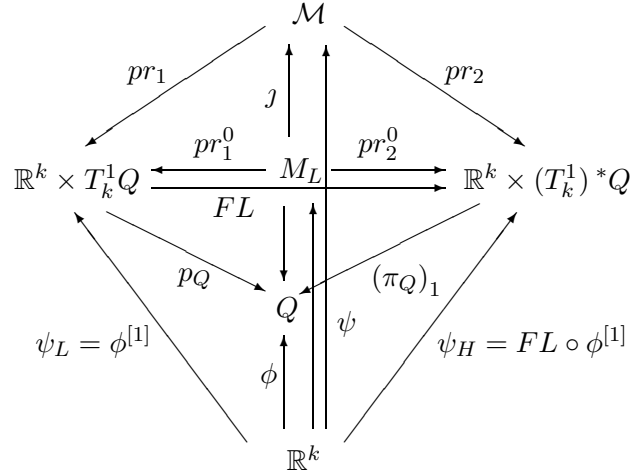
$$
\begin{array}{ccc}
 & \begin{picture}(135,50)(0,0) \put(70,36){\mbox{${\mathcal M}$}}
\put(8,15){\mbox{$pr_1$}} \put(128,15){\mbox{$pr_2$}}
\put(60,35){\vector(-3,-2){68}} \put(90,35){\vector(3,-2){68}}
\put(60,6){\mbox{$\jmath$}} \put(69,-7){\vector(0,1){35}}
\end{picture} &
\\
\begin{picture}(25,20)(0,0)
\put(0,-4){\mbox{$\r^k \times T^1_kQ$}}
\end{picture}
&
\begin{picture}(135,20)(0,0)
\put(33,10){\mbox{$pr^0_1$}} \put(60,3){\vector(-1,0){43}}
\put(65,0){\mbox{$M_L$}} \put(95,10){\mbox{$pr^0_2$}}
\put(85,3){\vector(1,0){43}}
\end{picture}
&
\begin{picture}(80,20)(0,0)
\put(-10,-4){\mbox{$\r^k \times (T^1_k)^{\;*}Q$}}
\end{picture}
\\ &
\begin{picture}(135,105)(0,0)
\put(17,104){\vector(1,0){111}} \put(40,93){\mbox{$FL$}}
\put(27,69){\mbox{$p_Q$}} \put(100,67){\mbox{${(\pi_Q)}_1$}}
\put(-27,42){\mbox{$\psi_L=\phi^{[1]}$}}
\put(125,42){\mbox{$\psi_H=FL\circ\phi^{[1]}$}}
\put(87,50){\mbox{$\psi$}} \put(58,30){\mbox{$\phi$}}
  \put(64,55){\mbox{$Q$}} \put(68,-3){\mbox{$\rk$}}
\put(67,97){\vector(0,-1){28}} \put(0,95){\vector(2,-1){60}}
\put(141,98){\vector(-2,-1){68}} \put(53,8){\vector(-3,4){65}}
\put(90,8){\vector(3,4){65}} \put(67,13){\vector(0,1){35}}
\put(78,13){\vector(0,1){85}} \put(83,13){\vector(0,1){145}}
\end{picture} &
\end{array}
$$
\proof
If \
$\ds \psi(t)=\left(\psi^A(t),\psi^i(t),\psi^i_A(t),\psi^A_i(t)=\frac{\partial
L}{\partial v^i_A}(\psi_L(t)) \right)$\
is an integral section of ${\bf Z}$, then
\begin{equation}
Z_A(\psi(t))=\ds\frac{\partial\psi^B }{\partial t^A}(t)
\ds\frac{\partial}{\partial t^B}\Big\vert_{\psi(t)} +
\ds\frac{\partial\psi^i }{\partial t^A}(t)
\ds\frac{\partial}{\partial q^i}\Big\vert_{\psi(t)} +
\ds\frac{\partial \psi^B_i}{\partial t^A}(t)
\ds\frac{\partial}{\partial p^B_i}\Big\vert_{\psi(t)} +
\ds\frac{\partial\psi^i_B}{\partial t^A}(t) \ds\frac{\partial
}{\partial v^i_B}\Big\vert_{\psi(t)} \label{aux0}\ .
\end{equation}
From (\ref{s4b}), (\ref{s6}), (\ref{s4})   and (\ref{aux0}) we
obtain
\begin{eqnarray}
\ds\frac{\partial \psi^B}{\partial t^A}(t) & = &
(Z_A)^B(\psi(t))=\delta^B_A \label{aux0b}
\\
\psi^A_i(t)& = & p^A_i(\psi(t))= \left(\frac{\partial L}{\partial
v^i_A}\circ pr_1\right)(\psi(t))= \frac{\partial L}{\partial
v^i_A}(\psi_L(t)) \label{aux2}
\\
\psi^i_A(t) & = & v^i_A(\psi(t)) =
(Z_A)^i(\psi(t))=\ds\frac{\partial \psi^i }{\partial t^A}(t)
\label{aux1}
\\
\ds\frac{\partial \psi^B_i }{\partial t^A}(t) & = &
(Z_A)^B_i(\psi(t)) \ . \label{aux4}
\end{eqnarray}
Therefore from (\ref{s5}), (\ref{aux2})  and (\ref{aux4}) we obtain
$$
\ds\frac{\partial L}{\partial q^i}(\psi_L(t))   = \ds\sum_{A=1}^k
(Z_A)^A_i(\psi(t))   = \ds\sum_{A=1}^k\ds\frac{\partial
\psi^A_i}{\partial t^A}(t)= \ds\sum_{A=1}^k\ds\frac{\partial}
{\partial t^A} \left(\ds\frac{\partial L}{\partial
v^i_A}(\psi_L(t))\right) \ ,
$$
and from (\ref{aux0b}) we obtain $\psi^A(t)=t^A+c^A$. Taking
$c^A=0$, from (\ref{aux1}) we have
$$
\psi_L(t)=\left(t,\psi^i(t),\ds\frac{\partial \psi^i }{\partial t^A}
(t)\right) \quad ,
$$
and from the last two equations we deduce that $\psi_L=\phi^{[1]}$
and $\phi= p_Q\circ pr^0_1\circ \psi:\rk \stackrel{\psi}{\to}M_L
\stackrel{pr^0_1}{\approx} \r^k \times T^1_kQ \stackrel{p_Q}{\to}Q
$, is a solution to the Euler-Lagrange field equations
(\ref{e-l-2}), where $\phi(t)=(\psi^i(t))$.
\qed

\begin{proposition}
According to the hypothesis of Theorem \ref{ELH-eq}, if $L$ is
regular then $\psi_H=FL\circ\psi_L$ is a solution to the Hamilton
field equations (\ref{he}), where the Hamiltonian $H$ is given by $H
\circ FL = E_L$. \label{ELH-eqbis}
\end{proposition}
\proof
 Since $L$ is regular, $FL$ is a local diffeomorphism, and
thus we can choose for each point in $\r^k \times T^1_kQ$ an open
neighborhood $U \subset \r^k \times T^1_k Q$ such that $FL\vert_{
U}:U \to FL(U)$ is a diffeomorphism. So we can define $H_U:FL(U) \to
\r$ as $H_U= (E_L)\vert_{U} \circ (FL\vert_{U})^{-1}$.

Denoting by $H\equiv H_U$, $E_L\equiv(E_L)\vert_{U}$ and $FL\equiv
FL\vert_{U}$, we have $E_L=H\circ FL$, which provides the identities
\begin{equation}\label{sh0}
\ds\frac{\partial H}{\partial p^A_i}\circ FL =v^i_A \quad , \quad
\ds\frac{\partial H}{\partial q^i} \circ FL= \, - \,
\ds\frac{\partial L}{\partial q^i} \quad .
\end{equation}
Now considering the open subset $V=\psi_L^{-1} (U) \subset \r^k$,
 we have $\psi\vert_{V} : V \subset \r^k \to U \oplus
FL(U) \subset M_L$, where $(\psi_L)\vert_{V} : V \subset \r^k \to U
\subset \r^k \times T^1_k Q$ and $(\psi_H)\vert_{V} = FL \circ
(\psi_L)\vert_{V} : V \subset \r^k \to FL(U) \subset \r^k \times
(T^1_k)^{\;*} Q$.
Therefore from (\ref{s5}), (\ref{aux1}), (\ref{aux4}) and
(\ref{sh0}), for every $t \in V \subset \r^k$ we obtain
\beann
\ds\frac{\partial H}{\partial p^A_i}(\ds\psi_H(t))&=&
\left(\ds\frac{\partial H}{\partial p^A_i}\circ
FL\right)(\psi_L(t))= v^i_A(\psi_L(t)) =
\ds\frac{\partial\psi^i}{\partial t^A}(t)\\
\ds\frac{\partial H}{\partial q^i}(\ds\psi_H(t))&=& \left(
\ds\frac{\partial L}{\partial q^i} \circ FL \right)(\psi_L(t))= -
\ds\frac{\partial L}{\partial q^i}(\psi_L(t)) = -
(Z_A)^A_i(\psi(t))= - \ds\frac{\partial \psi^A_i}{\partial t^A}(t) \ ,
\eeann
from which we deduce that $(\psi_H)\vert_{V}$ is a solution to the
Hamilton field equations (\ref{he}).
\qed

Conversely, we can state:

\begin{proposition}\label{33b}
If $L$ is regular and ${\bf X}=(X_1, \ldots, X_k)$ is a  solution to
(\ref{lageq0}) then:
\begin{enumerate} \item The $k$-vector field
${\bf Z}=(Z_1, \ldots, Z_k)$ given by $Z_A=(Id_{\r^k \times T^1_kQ}
\oplus FL)_*(X_A)$ is a solution to
(\ref{s3}).
\item If $\psi_L:  \rk \to   \r^k \times T^1_kQ$ is an integral
section of \, ${\bf X}=(X_1, \ldots, X_k)$ (and thus, from Remark
\ref{rem1}  and  from Theorem \ref{rel}, $\phi= \rho \circ
\psi_L:\rk \stackrel{\psi_L}{\to}\r^k \times T^1_kQ
\stackrel{\rho}{\to}Q$ is a solution to the Euler-Lagrange field
equations) then $\psi=(\psi_L, FL \circ \psi_L): \rk \to M_L \subset
{\mathcal M}$ is an integral section of ${\bf Z}=(Z_1, \ldots,
Z_k)$.
\end{enumerate}
\end{proposition}
\proof Now from the local expression (\ref{locfl1})of $FL$ we deduce
that
\begin{enumerate}
\item If $L$ is regular and ${\bf X}=(X_1,
\ldots, X_k)$ is a  solution to (\ref{lageq0}), then from Theorem
\ref{rel} we know that $X_A$ is a {\sc sopde}, and thus $X_A$ is
locally given by
\begin{equation}\label{xi11}
X_A = \ds\frac{\partial}{\partial t^A} + v^i_A \,
\ds\frac{\partial}{\partial q^i}+ (X_A)^i_B
\ds\frac{\partial}{\partial v^i_B} \ ,
\end{equation}
where $(X_A)^i_B$ satisfy
\begin{equation}\label{lleq1}
\ds\frac{\partial^2 L}{\partial t^A \partial v^i_A} + v^j_A \,
\ds\frac{\partial^2 L}{\partial q^j \partial v^i_A} + (X_A)^j_B \,
\ds\frac{\partial^2 L}{\partial v^j_B \partial v^i_A} \, = \,
\ds\frac{\partial L}{\partial q^i} \ .
\end{equation}
Now from the local expression (\ref{locfl1}) of $FL$ we deduce that
\begin{equation}\label{idea}
\begin{array}{ccl}
Z_A &=& (Id_{\r^k \times T^1_kQ} \oplus FL)_*(X_A) =
\ds\frac{\partial}{\partial t^A} \, + \, v^i_A \,
\ds\frac{\partial}{\partial q^i} \, + \,
(X_A)^i_B \, \ds\frac{\partial}{\partial v^i_B}  \\
\noalign{\medskip}
  &  & + \, \left( \ds\frac{\partial^2 L}{\partial t^A \partial v^j_C} \, + \,
v^i_A \, \ds\frac{\partial^2 L }{\partial q^i \partial v^j_C} \, +
\, (X_A)^i_B \, \ds\frac{\partial^2 L }{\partial v^i_B
\partial v^j_C} \right) \, \ds\frac{\partial}{\partial p^C_j}
\end{array} \ .
\end{equation}
Then from   (\ref{lleq1}) and (\ref{idea}) we deduce that ${\bf
Z}=(Z_1,\ldots , Z_k)$ satisfy equations (\ref{s4b}), (\ref{s4}),
(\ref{s5}) and $Z_A \left(p^B_k - \ds\frac{\partial L}{\partial
v^k_B}\right)=0 $, that is, the $k$-vector field ${\bf
Z}=(Z_1,\ldots, Z_k)$ is a solution to (\ref{s3}) and each $Z_A$ is
tangent to $M_L$ .
\item
It is an immediate consequence of the definition  of  ${\bf Z}$ and the
definition of integral section.
 \qed
\end{enumerate}

\begin{remark}
{\rm
The last result really holds for regular and almost-regular
Lagrangians. In the almost-regular case, assuming as additional
hypothesis that ${\bf X}_L$ is a {\sc sopde}, the proof is the same,
but the sections $\psi$, $\psi_L$ and $\psi_H$ take values not on
$M_L$, $\r^k \times T^1_kQ$ and $\r^k \times (T^1_k)^{\;*}Q$, but in
the final constraint submanifold $P_f$ and on the projection
submanifolds $pr_1(P_f)\subset \r^k \times T^1_kQ$ and
$pr_2(P_f)\subset \r^k \times (T^1_k)^{\;*}Q$, respectively.
}
\end{remark}

\subsection{The field equations for $k$-vector fields}

Next we establish the relationship between
$k$-vector fields that are solutions to (\ref{lageq0}) and
$k$-vector fields that are solutions to (\ref{s3}) or, what is
equivalent, solutions to (\ref{s3bis}).
First, observe that:

\begin{lemma}
We have that
\begin{equation}\label{jomega}
\jmath^{\;*}\vartheta^A = (pr^0_1)^{\;*} dt^A \quad , \quad
\jmath^{\;*}\Omega^A = (pr^0_1)^{\;*} \omega^A_L \quad .
\end{equation}
\end{lemma}
\proof It is immediate from (\ref{nueva}),taking into account  that
$FL \circ pr^0_1=pr_2 \circ j$.
\qed

\begin{theorem}\label{ELH-kvf}
a)  Let $L: \r^k \times T^1_k Q \to \r$ be a Lagrangian and let
${\bf Z}_L=((Z_L)_1,\ldots, (Z_L)_k)$ be a $k$-vector field on $M_L$
solution to (\ref{s3bis}). Then the $k$-vector field ${\bf
X}_L=((X_L)_1,\ldots, (X_L)_k)$ on $\r^k \times T^1_kQ$ defined by
\begin{equation} \label{xlz}
{\bf X}_L\circ pr^0_1=T^1_k (pr^0_1)\circ{\bf Z}_L
\end{equation}
is a $k$-vector field solution to (\ref{lageq0}), where $T^1_k
(pr^0_1)\colon T^1_k(M_L) \to T^1_k(\r^k \times T^1_kQ)$ is the
natural extension of $pr^0_1$, introduced in (\ref{tf}).

Conversely, every $k$-vector field ${\bf X}_L$ solution to
(\ref{lageq0}) can be recovered in this way from a $k$-vector field
${\bf Z}_L$ in $M_L$ solution to (\ref{s3bis}).

b) The $k$-vector field ${\bf Z}_L$ is integrable if, and only if,
the $k$-vector field ${\bf X}_L$ is an integrable {\sc sopde}.
\end{theorem}
\proof a) Since $pr^0_1:M_L \to \r^k \times T^1_kQ$ is a
diffeomorphism, then the $k$-vector field ${\bf X}_L$ on $\r^k
\times T^1_kQ$ defined by (\ref{xlz}) is given by
\begin{equation}\label{xaz}
(X_L)_A \, = \, \left( (pr^0_1)^{-1} \right)^{\;*} \, (Z_L)_A  \, .
\end{equation}
Furthermore, we obtain that
\begin{equation}\label{jesth}
\jmath^{\;*}\mathcal{H} = \jmath^{\;*} ( \mathcal{C} - (pr_1)^{\;*}L
) = \jmath^{\;*}\mathcal{C} - \jmath^{\;*} (pr_1)^{\;*}L =
(pr^0_1)^{\;*} (C(L))- (pr^0_1)^{\;*}L =  (pr^0_1)^{\;*}E_L \, .
\end{equation}
From (\ref{jomega}) and (\ref{xaz}) we deduce that
\begin{equation}\label{sumas0}
\jmath^{\;*} \vartheta^A ((Z_L)_B) \, = \, \left( (pr^0_1)^{\;*}
dt^A \right) \left( (pr^0_1)^{\;*} (X_L)_B \right) \, = \,
(pr^0_1)^{\;*} \left( dt^A \left( (X_L)_B \right) \right) \ ,
\end{equation}
and from (\ref{37}), (\ref{s0}),   (\ref{s1}) and (\ref{s6})
\bea\label{xir} \jmath^{\;*}\left[ \xi_A(\mathcal{H})\right] &=&
\jmath^{\;*}\left[  \left(\ds\frac{\partial}{\partial t^A} +
(\xi_A)^i_B \ds\frac{\partial}{\partial v^i_B}\right)( p^C_j\,
v^j_C-(pr_1^{\;*}L))\right] \nonumber \\ &=&
\jmath^{\;*}\left[(\xi_A)^i_B\left(p^B_i- \ds\frac{\partial
L}{\partial v^i_B}\circ pr_1\right)-
pr_1^{\;*}\left(\ds\frac{\partial L}{\partial t^A}\right)\right]= -
(pr^0_1)^{\;*}\left(\ds\frac{\partial L}{\partial t^A}\right) \ . \eea
Therefore from  (\ref{s3bis}), (\ref{jomega}), (\ref{xaz}),
(\ref{jesth}) and (\ref{xir}), we obtain
\begin{equation}\label{dif}
\begin{array}{l}
\ds\sum_{A=1}^k \, \imath_{(Z_L)_A}\jmath^{\;*}\Omega^A\,
-d(\jmath^{\;*}\mathcal{H})+   \jmath^{\;*}\left[\ds\sum_{A=1}^k
\xi_A(\mathcal{H})\right]
\,(\jmath^{\;*}\vartheta^A)\,   \\
\noalign{\medskip} = \ds\sum_{A=1}^k \,
\imath_{(pr^0_1)^{\;*}(X_L)_A} (pr^0_1)^{\;*} \omega^A_L \, -\,
d((pr^0_1)^{\;*} E_L)- \ds\sum_{A=1}^k
(pr^0_1)^{\;*}\left(\ds\frac{\partial L}{\partial  t^A}\right)\,(pr^0_1)^{\;*}dt^A \\
\noalign{\medskip} = (pr^0_1)^{\;*} \left(\ds\sum_{A=1}^k \,
\imath_{ (X_L)_A} \omega^A_L - dE_L - \ds\sum_{A=1}^k
\ds\frac{\partial L}{\partial  t^A} \, dt^A \right) \, .
\end{array}
\end{equation}
Since $pr^0_1$ is a diffeomorphism, from (\ref{sumas0}) and
(\ref{dif}) we deduce that the $k$-vector field ${\bf Z}_L$ is a
solution to (\ref{s3bis}) if, and only if, the $k$-vector field
${\bf X}_L$ is a solution to (\ref{lageq0}).

b) Suppose now that the $k$-vector field ${\bf Z}_L$ is integrable.
Let $\varphi:\rk \to \rk\times T^1_kQ$ be an integral section of
${\bf X}_L$, that is,
$(X_L)_A(\varphi(t))=\varphi_*(t)\left(\ds\frac{\partial}{\partial
t^A}\Big\vert_t\right)$. Thus
$$\begin{array}{lcl}
(Z_L)_A((pr^0_1)^{-1}\circ\,\varphi(t))&=&((pr^0_1)^{-1})^{\;*}(X_L)_A
((pr^0_1)^{-1}\circ\,\varphi(t))=((pr^0_1)^{-1})_*(\varphi(t))((X_L)_A(\varphi(t)))
\\
\noalign{\medskip} &=&
((pr^0_1)^{-1})_*(\varphi(t))\left(\varphi_*(t)\left(\ds\frac{\partial}{\partial
t^A}\Big\vert_t\right)\right)
=((pr^0_1)^{-1}\circ\varphi(t))_*\left(\ds\frac{\partial}{\partial
t^A}\Big\vert_t\right)\, ,
\end{array}$$
which means $\psi= (pr^0_1)^{-1}\circ\,\varphi: \rk \to M_L$ is an
integral section of ${\bf Z}_L$.

   Since $\psi:\rk \to M_L$, then we know that the integral section $j
\circ \psi: \rk \to {\mathcal M}$ is given by $((j\circ \psi)_L,FL
\circ (j\circ \psi)_L)$, and from  Theorem \ref{ELH-eq}, we know
that $(j \circ \psi)_L=\phi^{[1]}$, where $\phi= p_Q \circ \psi:\rk
\stackrel{\psi}{\to}M_L\approx\r^k \times T^1_kQ
\stackrel{p_Q}{\to}Q $. Then we  have
$$\phi^{[1]}=(j\circ \psi)_L=pr_1 \circ j \circ \psi=pr^0_1 \circ
\psi=\varphi \ .$$ Since every integral section $\varphi$ of ${\bf X}_L$
is a first prolongation $\phi^{[1]}$ of a map $\phi:\rk \to Q$ space,
we deduce from Lema \ref{lem1} that ${\bf X}_L$ is a {\sc sopde}.

If $m$ is an arbitrary point of $\rk \times T^1_kQ $, we consider
the integral section $\psi$ of ${\bf Z}_L$ passing through
$(pr_0^1)^{-1}(m)$, then $pr_1^0 \circ \psi$ is an integral section
of ${\bf X}_L$ passing through $m$. Thus, ${\bf X}_L$ is integrable.

Conversely, let ${\bf X}_L$ be an integrable {\sc sopde}. If $m$ is
an arbitrary point of $M_L $, we consider the integral section
$\varphi$ of ${\bf X}_L$ passing through $(pr_0^1)(m)$ then
$(pr^0_1)^{-1}\circ\,\varphi$ is an integral section of ${\bf Z}_L$
passing through $m$. Thus, ${\bf Z}_L$ is integrable.
\qed

If $L$ is regular, in a neighborhood of each point of $\r^k \times
T^1_kQ$ there exists a local solution ${\bf X}_L=((X_L)_1, \ldots ,
(X_L)_k)$ to (\ref{lageq0}). As $L$ is regular, $FL$ is a local
diffeomorphism, so this open neighborhood can be chosen in such a
way that  $FL$ is a diffeomorphism onto its image. Thus in a
neighborhood of each point of $FL(\r^k \times T^1_kQ)$ we can define
$(X_H)_A=[(FL)^{-1}]^{\;*}(X_L)_A$, or equivalently, in terms of $k$-vector fields
$T^1_k(FL) \circ X_L \, = \, X_H$.

\begin{proposition}\label{ELH-kvfbis}
\begin{enumerate}
\item
The local $k$-vector field ${\bf X}_H=((X_H)_1,\ldots ,(X_H)_k)$ is
a solution to (\ref{geonah}), where the Hamiltonian $H$ is locally
given by $H \circ FL = E_L$. (In other words, the local $k$-vector
fields ${\bf X}_L$ and ${\bf X}_H$ solution to (\ref{lageq0}) and
(\ref{he}), respectively, are $FL$-related).
\item
Every local integrable $k$-vector field solution to (\ref{he}) can
be recovered in this way from a local integrable $k$-vector field
${\bf Z}$ in ${\mathcal M}$ solution to (\ref{s3}).
\end{enumerate}
\end{proposition}
\proof
\begin{enumerate}
\item
This is the local version of Theorem \ref{te421}.
\item
Furthermore, if ${\bf X}_H$ is a local integrable $k$-vector field
solution to (\ref{geonah}), then we can obtain the $FL$-related
local integrable $k$-vector field ${\bf X}_L$ solution to
(\ref{lageq0}). By Theorem \ref{ELH-kvf}, we recover ${\bf X}_L$ by
a local integrable $k$-vector field ${\bf Z}_L$ solution to
(\ref{s3bis}).
 \qed
\end{enumerate}

It is interesting to point out
that the Skinner-Rusk formalism developed in \cite{BEMMR-2008} and \cite{CMC} for the
time-dependent mechanics is just a particular case of the
Skinner-Rusk formalism which we present here for the
$k$-cosymplectic formulation of first-order field theories.

\section{Lie algebroids  and associated spaces}\label{algebroids}

In this section we present some basic facts on Lie algebroids that are
necessary for further developments. We refer to the reader to
\cite{CW-1999,HM-1990,Mack-1987,Mack-1995} for details about Lie
groupoids, Lie algebroids and their role in differential geometry.

\subsection{Lie algebroids}\label{51}

Let $E$ be a vector bundle of rank $m$ over a manifold $Q$ of
dimension $n$, and let $\tau:E\to Q$ be the vector bundle projection.
Denote by ${\it Sec}(E)$ the $C^\infty(Q)$-module of sections of
$\tau:E\to Q$. A {\sl Lie algebroid structure}
$([\cdot,\cdot]_E,\rho)$ on $E$ is a Lie bracket
$[\cdot,\cdot]_E:{\it Sec}(E)\times {\it Sec}(E)\to {\it Sec}(E)$ on
the space ${\it Sec}(E)$, together with  a bundle  map $\rho:E\to
TQ$, called {\sl the anchor map}, such that if we denote by
$\rho:{\it Sec}(E)\to \vf{(Q)}$ the homomorphism of the
$C^\infty(Q)$-module induced by the anchor map, then they satisfy
the {\sl compatibility condition}
\[[\sigma_1,f\sigma_2]_E=
f[\sigma_1,\sigma_2]+(\rho(\sigma_1)f)\sigma_2\,.\]

Here $f$ is a smooth function on $Q$; $\,\sigma_1,\sigma_2$ are
sections of $E$, and we denote by $\rho(\sigma_1)$ the vector
field on $Q$ given by $\rho(\sigma_1)(\mathbf{q})=\rho(\sigma_1({\bf
q}))$. The triple $(E,[\cdot,\cdot]_E,\rho)$ is called a {\sl Lie
algebroid over $Q$}. From the compatibility condition and the Jacobi
identity, it follows that the anchor map $\rho:{\it
Sec}(E)\to\vf{(Q)}$ is a homomorphism between the Lie algebras
$({\it Sec}(E),[\cdot,\cdot]_E)$ and $(\vf{(Q)},[\cdot,\cdot])$.

 In this paper, we
consider a Lie algebroid as a substitute of the tangent bundle of
$Q$. In this way, one regards an element $a$ of $E$ as a generalized
velocity, and the actual velocity $v$ is obtained when applying the
anchor map to $a$, i.e. $v=\rho(a)$.

Let $(q^i)_{i=1}^n$ be local coordinates on $Q$ and
$(e_\alpha)_{\alpha=1}^m$ be a local basis of sections of $\tau$.
Given ${\bf a}\in E$ such that $\tau(a)=\mathbf{q}$, we can write
$a=y^\alpha(a)e_\alpha(\mathbf{q})\in E_\mathbf{q}$, thus the
coordinates of $a$ are $(q^i(a),y^\alpha(a))$. Therefore, each
section $\sigma$ is locally given by $\sigma\Big\vert_{U}=y^\alpha
e_\alpha$.

In local form, the Lie algebroid structure is determined by the
local functions  $\rho^i_\alpha,\; \mathcal{C}^\gamma_{\alpha\beta}$
on $Q$. Both are determined by the relations
\begin{equation}\label{structure} \rho(e_\alpha)=\rho^i_\alpha
\ds\frac{\partial}{\partial q^i} ,\quad
[e_\alpha,e_\beta]_E=\mathcal{C}^\gamma_{\alpha\, \beta}e_\gamma\, .
\end{equation}
 The functions $\rho^i_\alpha$ and
$\mathcal{C}^\gamma_{\alpha\beta}$ are said to be the {\sl structure
functions} of the Lie algebroid in the above coordinate system. They
satisfy the following relations (as a consequence of the compability
condition and Jacobi's identity)
\begin{equation}\label{ecest}\ds\sum_{cyclic(\alpha,\beta,\gamma)}\left(\rho^i_\alpha\ds\frac{\partial
\mathcal{C}^\nu_{\beta\gamma}}{\partial
q^i}+\mathcal{C}^\nu_{\alpha\mu}\mathcal{C}^\mu_{\beta\gamma}\right)=0
\quad , \quad  \rho^j_\alpha\ds\frac{\partial \rho^i_\beta}{\partial
q^j}- \rho^j_\beta\ds\frac{\partial \rho^i_\alpha}{\partial
q^j}=\rho^i_\gamma \mathcal{C}^\gamma_{\alpha\beta}\;,
\end{equation}
which are usually called {\sl the structure equations} of the Lie algebroid.

\paragraph{Exterior differential}\

 The structure of the Lie algebroid on
$E$ allows us to define {\sl the exterior differential of $E$},
$d^E:{\it Sec}(\bigwedge^l E^*)\to {\it Sec}(\bigwedge^{l+1} E^*)$,
as follows
$$\begin{array}{lcl}
d^E\mu(\sigma_1,\ldots, \sigma_{l+1})&=&
\ds\sum_{i=1}^{l+1}(-1)^{i+1}\rho(\sigma_i)\mu(\sigma_1,\ldots,
\widehat{\sigma_i},\ldots, \sigma_{l+1})\\\noalign{\medskip} &+&
\ds\sum_{i<j}(-1)^{i+j}\mu([\sigma_i,\sigma_j]_E,\sigma_1,\ldots,
\widehat{\sigma_i},\ldots, \widehat{\sigma_j},\ldots
\sigma_{l+1})\;,
\end{array}$$ for $\mu\in {\it Sec}(\bigwedge^lE^*)$ and
$\sigma:1,\ldots,\sigma_{l+1}\in {\it Sec}(E)$. It follows that $d$
is a cohomology operator, that is, $d^2=0$.

In particular, if $f:Q\to\r$ is a real smooth function then
$df(\sigma)=\rho(\sigma)f$, for $\sigma\in Sec(E)$. Locally, the
exterior differential is determined by
\[dq^i=\rho^i_\alpha e^\alpha\quad \makebox{and}\quad
de^\gamma=-\ds\frac{1}{2}\mathcal{C}^\gamma_{\alpha\beta}e^\alpha\wedge
e^\beta\,,\]where $\{e^\alpha\}$ is the dual basis of
$\{e_\alpha\}$.

The usual Cartan calculus extends to the case of Lie algebroids: for
every section $\sigma$ of $E$ we  have a derivation $\imath_\sigma$
(contraction) of degree $-1$ and a derivation
$\mathcal{L}_\sigma=\imath_\sigma\circ d + d\circ \imath_\sigma$
(Lie derivative) of degree $0$, (for more details, see
\cite{Mack-1987,Mack-1995}).

\paragraph{Morphisms}\

 Let $(E,[\cdot,\cdot]_E,\rho)$ and
$(E',[\cdot,\cdot]_E',\rho')$ be two Lie algebroids over $Q$ and
$Q'$ respectively. Suppose that
$\Phi=(\overline{\Phi},\underline{\Phi})$ is a vector bundle map,
that is $\overline{\Phi}:E\to E'$ is a fiberwise linear map over
$\underline{\Phi}:Q\to Q'$. The pair
$(\overline{\Phi},\underline{\Phi})$ is said to be a {\sl Lie
algebroid morphism} if
 \begin{equation}\label{lie morph}d^E
(\Phi^*\sigma')=\Phi^*(d^{E'}\sigma ')\,,\quad\makebox{\rm for all }
\sigma '\in Sec(\bigwedge^l (E')^*)\makebox{\rm and for all
l.}\end{equation}
 Here $\Phi^*\sigma '$ is the section of the vector
bundle $\bigwedge^k E^*\to Q$ defined for $l>0$ by
\[(\Phi^*\sigma ')_q(a_1,\ldots,a_l) = \sigma_{\underline{\Phi}
(\mathbf{q})}' (\overline{\Phi}(a_1),\ldots,
\overline{\Phi}(a_l))\,,\]
 for $q\in Q$ and $a_1,\ldots, a_l\in
E_q$.  In the particular case when $Q=Q'$ and
$\underline{\Phi}=id_Q$ then (\ref{lie morph}) holds if, and only if,
\[[\overline{\Phi}\circ\sigma_1,\overline{\Phi}\circ\sigma_2]_{E'} =
\overline{\Phi}[\sigma_1,\sigma_2]_E,\quad
\rho'(\overline{\Phi}\circ\sigma)=\rho(\sigma),\quad \makebox{for }
\sigma,\sigma_1,\sigma_2\in Sec(E)\,.\]

Let $(q^i)$ be a local coordinate system on $Q$ and $(q^i{'})$ a
local coordinate system on $Q'$. Let $\{e_\alpha\}$ and
$\{e_{\alpha}'\}$ be local bases of section of $E$ and $E'$,
respectively, and $\{e^\alpha\}$ and $\{e^{'\alpha}\}$ the dual
bases. The vector bundle map $\Phi$ is determined by the relations
$\Phi^*{q^i{'}}=\phi^i(\mathbf{q})$ and $\Phi^* e^{'\alpha}=
\phi^\alpha_\beta e^\beta$ for certain local functions $\phi^i$ and
$\phi^\alpha_\beta$ on $Q$. Then
$\Phi=(\overline{\Phi},\underline{\Phi})$ is a morphism of Lie
algebroids if, and only if,
\begin{equation}\label{morp cond}
 \rho^j_\alpha \ds\frac{\partial \phi^i}{\partial q^j}=\rho^{'i}_
 \beta\phi^\beta_\alpha\quad,\quad
\phi^\beta_\gamma\mathcal{C}^\gamma_{\alpha\delta}
=\left(\rho^i_\alpha\ds\frac{\partial \phi^\beta_\delta}{\partial
q^i} - \rho^i_\delta \ds\frac{\partial \phi^\beta_\alpha}{\partial
q^i}\right) +
\mathcal{C}^{'\beta}_{\theta\sigma}\phi^\theta_\alpha\phi^\sigma_\delta
\,.\end{equation} In these expressions $\rho^i_\alpha,
\mathcal{C}^\alpha_{\beta\gamma}$ are the structure functions on $E$
and $\rho^{'i}_\alpha, \mathcal{C}{'\alpha}_{\beta\gamma}$ are the
structure functions on $E'$.

\paragraph{The prolongation of a Lie algebroid over a fibration}\

(See \cite{CLMMM-2006,HM-1990,LMM-2005,Mart-2001}). Let
$(E,[\cdot,\cdot]_E,\rho)$ be a Lie algebroid over a manifold $Q$
and $\pi:P\to Q$ a fibration. Consider the subset of $E\times TP$
\[\mathcal{T}^E_pP=\{(b,v)\in E_q\times T_pP\, |\,
\rho(b)=T_p\pi(v)\}\] where $T\pi:TP\to TQ$ is the tangent map to
$\pi\,,\,p\in P_q $, and $\pi(p)=q$.   $\,\mathcal{T}^EP=\cup_{p\in
P}\mathcal{T}^E_pP$ is a vector bundle over $P$ and the vector
bundle projection is $\tau^E_P:\mathcal{T}^EP\to P$.  We can
consider a structure of Lie algebroid on $\mathcal{T}^EP$  given by
the anchor $\rho^\pi:\mathcal{T}^EP\to TP,\;\rho^\pi(b,v_p)=v_p$.

 This Lie algebroid will be used in Section \ref{624}, when we introduce
the solutions of the Hamilton field equations on Lie algebroids.

\subsection{The manifold $\stackrel{k}{\oplus} E$}\label{52}

The standard $k$-cosymplectic Lagrangian formalism is developed in the
bundle $\rk\times T^1_kQ$. When we consider a Lie algebroid $E$ as a substitute of
the tangent bundle, it is natural, in this situation,  to consider that the
analog of the bundle of $k^1$-velocities $T^1_kQ$ is the Whitney sum
of $k$ copies of the algebroid $E$, and thus the analog of the
manifold $\rk\times T^1_kQ$ is $\rk\times\stackrel{k}{\oplus}E$.

We denote by $\stackrel{k}{\oplus} E=E\oplus \stackrel{k}{\ldots}
\oplus E$, the Whitney sum of $k$ copies of the vector bundle $E$,
with projection map $\widetilde{\tau}:\stackrel{k}{\oplus} E\to Q,$
given by $
\widetilde{\tau}({a_1}_\mathbf{q},\ldots,{a_k}_\mathbf{q})= q$. The
Lie structure of $E$ allows us to introduce the following maps:
\[\xymatrix{\stackrel{k}{\oplus}E\equiv E\oplus
\stackrel{k}{\ldots} \oplus E\ar[rr]^-{\widetilde{\rho}=(
\rho,\,\stackrel{k}{\ldots} , \rho) }\ar[dr]_-{\widetilde{\tau}} &&
TQ\oplus \stackrel{k}{\ldots} \oplus TQ\equiv
T^1_kQ\ar[dl]^-{\tau^k_Q} \\ &Q&
 }\]where $\widetilde{\rho}
({a_1}_\mathbf{q},\ldots,{a_k}_\mathbf{q})=(\rho({a_1}_{\bf
q}),\ldots,\rho({a_k}_\mathbf{q}))$, and $\rho:E\to TQ$ is the
anchor map of $E$.

\paragraph{Local basis of sections of
$\widetilde{\tau}:\stackrel{k}{\oplus}E \to Q$}\

A local basis $\{e_\alpha\}_{\alpha=1}^m$ of $Sec(E)$ induces  {\sl
a local basis of sections of the bundle
$\widetilde{\tau}:\stackrel{k}{\oplus}E \to Q$}. In fact, let
$a_\mathbf{q}=({a_1}_\mathbf{q},\ldots,{a_k}_\mathbf{q})$ be an
arbitrary point of $ \stackrel{k}{\oplus}E$, then for each
$A\;(A=1,\ldots, k),\,$ $a_{A_\mathbf{q}}\in E$ and since
$\{e_\alpha\}$ is a local basis of sections of $E$ we have
$a_{A_\mathbf{q}}=y^\alpha(a_{A_\mathbf{q}}) e_\alpha(\mathbf{q})$.
Therefore
$$\begin{array}{lcl} a_\mathbf{q}&=&(y^\alpha({a_1}_\mathbf{q})
e_\alpha(\mathbf{q}),\ldots, y^\alpha({a_k}_\mathbf{q})
e_\alpha(\mathbf{q}))\\\noalign{\medskip}&=&y^\alpha({a_1}_\mathbf{q})
(e_\alpha(\mathbf{q}),0,\ldots, 0) + \ldots+y^\alpha({a_k}_{\bf
q})(a_\mathbf{q}) (0,\ldots, 0, e_\alpha({\bf
q}))=y^\alpha(a_{a_\mathbf{q}})\,
\widetilde{e}_\alpha^A(\mathbf{q})\;,
\end{array}$$ where $\widetilde{e}_\alpha^A(\mathbf{q})=(0,\ldots,\stackrel{A}{\widetilde{e_\alpha(\mathbf{q})}},\ldots,
0)\;.$ Thus a local basis $\{e_\alpha\}$ of $Sec(E)$ induces a local
basis $\{\widetilde{e}^A_\alpha\}$  of $Sec(\stackrel{k}{\oplus}E)$
defined by
$\widetilde{e}_\alpha^A(\mathbf{q}):
=(0,\ldots,\stackrel{A}{\widetilde{e_\alpha(\mathbf{q})}},\ldots,0)$,
 where $\stackrel{A}{\widetilde{\qquad}}$ indicates the
$A^{th}$ position of $\widetilde{e}_\alpha^A(\mathbf{q})$.

If $(q^i,y^\alpha)$ are local coordinates on
${\tau}^{-1}(U)\subseteq E$, then the induced local coordinates
$(q^i,y^\alpha_A)$ on $\widetilde{\tau}^{-1}(U)\subseteq
\stackrel{k}{\oplus} E$ are given by
$$
q^i({a_1}_\mathbf{q},\ldots,{a_k}_\mathbf{q})=q^i(\mathbf{q})\,,\quad
y^\alpha_A({a_1}_\mathbf{q},\ldots,{a_k}_\mathbf{q})=y^\alpha(a_{A_{\bf
q}})\;.
$$

\subsection{The $k$-prolongation of a Lie algebroid over a
fibration}\label{k-prol}

 Let $\pi:P\to Q$ be a bundle. Now we
define a vector bundle which generalizes the concept of
prolongation of a Lie algebroid over a fibration $\pi:P\to Q$. We
denote this bundle by $\mathcal{T}^E_kP$ and it is called the
{\sl $k$-prolongation of $P$ with respect to a Lie algebroid $E$}.

Throughout this paper we consider two particular cases of
$k$-prolongations. The first corresponds to the case
$P=\stackrel{k}{\oplus}E$, and the bundle
$\mathcal{T}^E_k(\stackrel{k}{\oplus}E)$ will allow us to develop
the Lagrangian formalism on Lie algebroids, (see section \ref{lag
form al}). This bundle plays the role of the bundle $T^1_k(\rk\times
T^1_kQ)\to \rk\times T^1_kQ$ in the Lagrangian k-cosymplectic
formalism. The second particular case is
$P=\stackrel{k}{\oplus}E^*$, where we will develop the Hamiltonian
formalism (see section \ref{ham form al}).

The total space of the $k$-prolongation of $P$ with respect to $E$
$$\mathcal{T}^E_kP=(\rk\times \stackrel{k}{\oplus}
E)\times_{\rk\times T^1_kQ} T^1_k(\rk\times P)$$ is the total space
of the pull-back of the map $F=\tau^k_{\rk}\times
T^1_k\pi:T^1_k(\rk\times P)\to \rk\times T^1_kQ$, (locally defined
by (\ref{locF})), by the map $Id_{\rk}\times\widetilde{\rho}\equiv
Id_{\rk}\times\rho\oplus\stackrel{k}{ \ldots} \oplus \rho:\rk\times
\stackrel{k}{\oplus} E\to  \rk\times T^1_kQ\,,$
$$
\mathcal{T}^E_kP=\{ \left({\bf s},a_{\mathbf{q}})
,W_{(\mathbf{t},\mathbf{p})}\right)\in (\rk\times
\stackrel{k}{\oplus} E)\times T^1_k(\rk\times P)\, :
Id_{\rk}\times\widetilde{\rho}({\bf s},a_{{\bf q}})=
F(W_{(\mathbf{t},\mathbf{p})}) \}\;,
$$where
$W_{(\mathbf{t},\mathbf{p})}=((v_1)_{(\mathbf{t},\mathbf{p})},\ldots,(v_k)_{(\mathbf{t},\mathbf{p})})\in
T^1_k(\rk\times P).$

Let us observe that $({\bf s},a_{\mathbf{q}}
,(v_1)_{(\mathbf{t},\mathbf{p})},\ldots,(v_k)_{(\mathbf{t},\mathbf{p})})\in\mathcal{T}^E_kP$
means that ${\bf s}={\bf t}$ and $\mathbf{q}=\pi ({\bf p})$.
Therefore, an element $({\bf s},a_{\mathbf{q}}
,(v_1)_{(\mathbf{t},\mathbf{p})},\ldots,(v_k)_{(\mathbf{t},\mathbf{p})})$
of $\mathcal{T}^E_kP$ can be identified with a family
$(a_{\pi(\mathbf{p})},$ $
(v_1)_{(\mathbf{t},\mathbf{p})},\ldots,(v_k)_{(\mathbf{t},\mathbf{p})})$.

\begin{remark}{\rm Throughout the paper we shall denote the elements of
$\mathcal{T}^E_kP$ as $(a_\mathbf{q},W_{(\mathbf{t},\mathbf{p})})$
where $a_\mathbf{q}\in \stackrel{k}{\oplus} E$,
$W_{(\mathbf{t},\mathbf{p})}\in T^1_k(\rk\times P)$ and
$\pi(\mathbf{p})=\mathbf{q}$.}\end{remark}

The {\sl $k$-prolongacion of a Lie algebroid $E$ over a fibration
$\pi:P\to Q$} is the space $\mathcal{T}^k_EP$ fibrered over
$\rk\times P$ with the projection
\[\begin{array}{rcl} \widetilde{\tau}_{\rk\times P}:\mathcal{T}^E_kP&\to&\rk\times P\\\noalign{\medskip} (a_\mathbf{q},W_{(\mathbf{t},\mathbf{p})}
)&\mapsto &\widetilde{\tau}_{\rk\times P}(a_{\bf
q},W_{(\mathbf{t},\mathbf{p})} )=\tau^k_{\rk\times
P}(W_{(\mathbf{t},\mathbf{p})}) =(\mathbf{t},\mathbf{p})
\end{array}\]
where $\tau^k_{\rk\times P}:T^1_k(\rk\times P) \to \rk\times P$ is
the projection of the bundle of $k^1$-velocities of $P$.

If $\mathbf{q}\in Q,\;a_\mathbf{q}\in \stackrel{k}{\oplus} E,\;
W_{(\mathbf{t},\mathbf{p})}
=({v_1}_{(\mathbf{t},\mathbf{p})},\ldots,
{v_k}_{(\mathbf{t},\mathbf{p})})\in T^1_k(\rk\times P)$ and $(a_{\bf
q},W_{(\mathbf{t},\mathbf{p})})\in \mathcal{T}^E_kP$,
 we have the following natural projections
$$
\begin{array}{lclclcl}
\widetilde{\tau}_1(a_{\bf
q},W_{(\mathbf{t},\mathbf{p})})&=&(\mathbf{t},a_\mathbf{q}) & , &
\widetilde{\tau}_2(a_{\bf
q},W_{(\mathbf{t},\mathbf{p})})&=&W_{(\mathbf{t},\mathbf{p})}
\\\noalign{\medskip}
\widetilde{\tau}_{1\,,\stackrel{k}{\oplus} E}(a_{\bf
q},W_{(\mathbf{t},\mathbf{p})}) &=& a_\mathbf{q}  & , &
\widetilde{\tau}_2^A(a_\mathbf{q},W_{(\mathbf{t},\mathbf{p})})&=&
v_A{_{(\mathbf{t},\mathbf{p})}}
\\\noalign{\medskip}
\widetilde{\tau}_{\rk\times P}(a_{\bf
q},W_{(\mathbf{t},\mathbf{p})})&=& (\mathbf{t},\mathbf{p}) & , &
\tau^{k,A}_{\rk\times
P}(W_{(\mathbf{t},\mathbf{p})})&=&v_A{_{(\mathbf{t},\mathbf{p})}}\quad ,
\end{array}
$$
and we have the following diagram
\[\xymatrix{& \mathcal{T}^E_kP \ar@(ur,ul)@/^{12mm}/[rrrr]^-{\widetilde{\tau}_2^A}
\ar[dr]^-{\widetilde{\tau}_{\rk\times
P}}\ar[rr]^-{\widetilde{\tau}_2}
\ar[dd]^-{\widetilde{\tau}_1}\ar[ddl]_-{\widetilde{\tau}_{1\,,\stackrel{k}{\oplus}
E}}& & T^1_k(\rk\times P)\ar[rr]^-{\tau^{k,A}_{\rk\times
P}}\ar[dl]_-{{\tau}^k_{\rk\times P}}\ar[dd]^-{F=\tau^k_{\rk}\times
T^1_k\pi}
  &&
  T(\rk\times P) \\
   & & \rk\times P & && \\
   \stackrel{k}{\oplus} E & \rk\times \stackrel{k}{\oplus} E \ar[l]_-{\pi_2}
   \ar[rr]^-{Id_{\rk}\times \widetilde{\rho}}
&& \rk\times T^1_kQ && }\]

\paragraph{Local coordinates on $\mathcal{T}^E_kP$}\

Let $(a_\mathbf{q},W_{(\mathbf{t},\mathbf{p})})$ be an arbitrary
point of $\mathcal{T}^E_kP$, then
$a_\mathbf{q}=(a_{1_{\mathbf{q}}},\ldots,a_{k_{\mathbf{q}}})\in
\stackrel{k}{\oplus} E$ and
$W_{(\mathbf{t},\mathbf{p})}=((v_1)_{(\mathbf{t},\mathbf{p})},\ldots,
(v_k)_{(\mathbf{t},\mathbf{p})})\in T^1_k(\rk\times P)\,,$ satisfy
\begin{equation}\label{cond k-prol}Id_{\rk}\times\widetilde{\rho}({\bf t},a_{{\bf q}})=
\tau^k_{\rk}\times T^1_k\pi({v_1}_{(\mathbf{t},\mathbf{p})},\ldots,
{v_k}_{(\mathbf{t},\mathbf{p})})\,,
\end{equation}
Given local coordinates $(q^i,u^\vartheta),\;1\leq i\leq
dim\,Q,\;1\leq\vartheta\leq s$ on $P$, and a local basis
$\{e_{\alpha}\}_{\alpha=1}^m$ of sections of $E$, we have that
\begin{equation}\label{coord}
a_\mathbf{q}=y^\alpha_A(a_\mathbf{q})\widetilde{e}_\alpha^A({\bf
q})\;,\quad (v_A)_{(\mathbf{t},\mathbf{p})}=(v_A)^B
\ds\frac{\partial}{\partial
t^B}\Big\vert_{{(\mathbf{t},\mathbf{p})}}  +(v_A)^i
\ds\frac{\partial}{\partial
q^i}\Big\vert_{{(\mathbf{t},\mathbf{p})}} + (v_A)^\vartheta
\ds\frac{\partial}{\partial
u^\vartheta}\Big\vert_{{(\mathbf{t},\mathbf{p})}}\;.
\end{equation}
As
\[\rho(a_{A_\mathbf{q}})=
\rho(y^\alpha_A(a_\mathbf{q})e_\alpha (\mathbf{q}))=
y^\alpha_A(a_\mathbf{q})\rho(e_\alpha (\mathbf{q}))=
y^\alpha_A(a_\mathbf{q})\rho^i_\alpha(\mathbf{q})\ds\frac{
\partial}{\partial
q^i}\Big\vert_\mathbf{q}\,,\]
and from (\ref{defF}) we have that the
condition (\ref{cond k-prol}) is equivalent to
\begin{equation}\label{fund}
(v_A)^i=y^\alpha_A(a_\mathbf{q})\rho^i_\alpha(\mathbf{q})\, .
\end{equation}

 Taking into account (\ref{fund}) we
introduce the local coordinates
$(t^A,q^i,u^\vartheta,z^\alpha_A,v_A^B,(v_A)^\vartheta )$ on
$\mathcal{T}^E_kP$ given by {\small\begin{equation}\label{local
coord k-prol}\begin{array}{lll}
t^A(a_\mathbf{q},W_{(\mathbf{t},\mathbf{p})})=t^A(t) & q^i(a_{\bf
q},W_{(\mathbf{t},\mathbf{p})})= q^i(\mathbf{q}) \;,\quad &
u^\vartheta(a_\mathbf{q},W_{(\mathbf{t},\mathbf{p})})=u^\vartheta(p)\;,
\\\noalign{\medskip}
z^\alpha_A(a_\mathbf{q},W_{(\mathbf{t},\mathbf{p})})=y_A^\alpha(a_{\bf
q}) &
  v_A^B(a_\mathbf{q},W_{(\mathbf{t},\mathbf{p})})=v_{A_{(\mathbf{t},\mathbf{p})}}(t^B)\;,\quad &
(v_A)^\vartheta (a_\mathbf{q},W_{(\mathbf{t},\mathbf{p})})=
v_A{_{(\mathbf{t},\mathbf{p})}}(u^\vartheta)\;. \\
\end{array}
\end{equation}}

\paragraph{Local basis of sections of the bundle
$\widetilde{\tau}_{\rk\times P}:\mathcal{T}^E_kP\to \rk\times P$}\

 Given local coordinates $(t^A,q^i,u^\vartheta)$ on
$\rk\times P$ and a local basis $\{e_{\alpha}\}$ of sections of $E$,
we can define a local basis
$\{\mathcal{X}_\alpha^A,\,\mathcal{Y}^A_B,\,\mathcal{V}^A_\vartheta\}$
of sections of $\widetilde{\tau}_{\rk\times P}:\mathcal{T}^E_kP\to
\rk\times P$. In fact, from (\ref{coord}) and (\ref{fund}) we
deduce that for a point $(a_\mathbf{q},W_{(\mathbf{t},\mathbf{p})})$
one has
\begin{equation}\label{expr z}
\begin{array}{lcl}
(a_\mathbf{q},W_{(\mathbf{t},\mathbf{p})})& = & y^\alpha_A(a_{\bf
q})(\widetilde{e}^A_\alpha(\mathbf{q}),
(0,\ldots,\stackrel{A}{\widetilde{\rho^i_\alpha(x)\ds\frac{\partial
}{\partial q^i}\Big\vert_{(\mathbf{t},\mathbf{p})}}},\ldots, 0))
\\\noalign{\medskip}&+& v_A^B(\mathbf{0}_q,(0,\ldots,
\stackrel{A}{\widetilde{\ds\frac{\partial}{\partial
t^B}\Big\vert_{(\mathbf{t},\mathbf{p})}}},\ldots, 0))
+(v_A)^\vartheta (\mathbf{0}_q,(0,\ldots,
\stackrel{A}{\widetilde{\ds\frac{\partial} {\partial
u^\vartheta}\Big\vert_{(\mathbf{t},\mathbf{p})}}},\ldots, 0))  \,.
\end{array}\end{equation}

Thus the set
$\{\mathcal{X}_\alpha^A,\,\mathcal{Y}^A_B,\,\mathcal{V}^A_\vartheta\}$with
$1\leq A,B\leq k\,;\, 1\leq \alpha\leq m\,;\, 1\leq \vartheta\leq
s$, defined by
\begin{equation}\label{base k-prol}
\begin{array}{lcl}
\mathcal{X}_\alpha^A(\mathbf{t},\mathbf{p})\; &=&
(\widetilde{e}^A_\alpha(\mathbf{q}),
(0,\ldots,\stackrel{A}{\widetilde{\rho^i_\alpha({\bf
q})\ds\frac{\partial }{\partial
q^i}\Big\vert_{(\mathbf{t},\mathbf{p})}}},\ldots, 0))
\\\noalign{\medskip}
\mathcal{Y}^A_B(\mathbf{t},\mathbf{p})& = &( \mathbf{0}_q,(0,\ldots,
\stackrel{A}{\widetilde{\ds\frac{\partial}{\partial
t^B}\Big\vert_{(\mathbf{t},\mathbf{p})}}},\ldots,0))
\\\noalign{\medskip}
\mathcal{V}^A_\vartheta(\mathbf{t},\mathbf{p}) &=&
(\mathbf{0}_q,(0,\ldots,
\stackrel{A}{\widetilde{\ds\frac{\partial}{\partial
u^\vartheta}\Big\vert_{(\mathbf{t},\mathbf{p})}}},\ldots, 0))
\end{array}
\end{equation}
 is a local basis of
sections of the vector bundle $\widetilde{\tau}_{\rk\times
P}:\mathcal{T}^E_kP\to \rk\times P$.

\begin{remark}
{\rm
Throughout this paper, a section of $\mathcal{T}^E_kP$ means
a section of $\widetilde{\tau}_{\rk\times P}:\mathcal{T}^E_kP\to\rk\times P$.
}
\end{remark}

\paragraph{$k$-vector field on $\rk\times P$
associated to a section of $\mathcal{T}^E_kP$}\

Every section $\sigma$ of $\mathcal{T}^E_kP$ has associated a
$k$-vector field on $\rk\times P$ given by
$$\widetilde{\tau}_2(\sigma)=(\widetilde{\tau}_2^1(\sigma),\ldots,
\widetilde{\tau}_2^k(\sigma))\quad .$$

Let $\sigma:\rk\times P\to\mathcal{T}^E_kP$ be an arbitrary section
of $\widetilde{\tau}_{\rk\times P}$, such that $
\sigma(\mathbf{t},\mathbf{p})
=(\mathbf{t},\mathbf{p},\sigma^\alpha_A(\mathbf{t},\mathbf{p}),$ $
\sigma_A^B(\mathbf{t},\mathbf{p}),
\sigma_A^\vartheta(\mathbf{t},\mathbf{p}))\;, $ then  from
(\ref{expr z}) and (\ref{base k-prol}) we have that the expression
of $\sigma$ in terms of the basis
$\{\mathcal{X}_\alpha^A,\,\mathcal{Y}^A_B,\,\mathcal{V}^A_\vartheta\}$
is
$$\sigma=\sigma^\alpha_A\mathcal{X}_\alpha^A+
\sigma_A^B
\mathcal{Y}^A_B+\sigma_A^\vartheta\mathcal{V}^A_\vartheta\;,$$ and
the associated $k$-vector field
$\widetilde{\tau}_2(\sigma)=(\widetilde{\tau}_2^1(\sigma),\ldots,
\widetilde{\tau}_2^k(\sigma))$ is locally given by
\begin{equation}\label{k-vector asoc k-prol}
\widetilde{\tau}_2^A(\sigma)=\sigma_A^B\ds\frac{\partial}{\partial
t^B}+ \rho^i_\alpha \sigma^\alpha_A\ds\frac{\partial}{\partial q^i}+
\sigma_A^\vartheta\ds\frac{\partial}{\partial u^\vartheta}\in
\mathfrak{X}(\rk\times P)\;.
\end{equation}

These $k$-vector fields play an important role in the development of
the Lagrangian and Hamiltonian formalism on Lie algebroids.(See
section \ref{614} and \ref{623})

\paragraph{Lie bracket of section of $\mathcal{T}^E_kP$}\

A Lie bracket associated to the Lie bracket on $Sec(E)$ can be
easily defined in terms of projectable sections as follows: a
section $Z$ of $\mathcal{T}^E_kP$ is said to be {\sl projectable}
if there exists a section $\sigma$ of $\oplus^kE$ such that the
following diagram is commutative
\[\xymatrix{\rk\times P\ar[r]^-Z\ar[d] &\mathcal{T}^E_kP\ar[d]^-{\widetilde{\tau}_{1,\stackrel{k}{\oplus}E}}\\
 Q\ar[r]_-{\sigma} &
\stackrel{k}{\oplus}E} \ .
\]
Equivalently, a section $Z$ is projectable if, and only if, it is of
the form $Z(\mathbf{t},\mathbf{p})=(\sigma(\pi(\mathbf{p})),{\bf
X}(\mathbf{t},\mathbf{p}))$, for some section
$\sigma=(\sigma_1,\ldots, \sigma_k)$ of $\stackrel{k}{\oplus}E$ and
some $k$-vector field ${\bf X}=(X_1,\ldots, X_k)$ on $\rk\times P$.
The Lie bracket of two projectable sections $Z$ and $Z'$ is then
given by
$$
[Z,Z']^\pi(\mathbf{p})
=([\sigma_1,\sigma_1']_E(\mathbf{q}),\ldots,[\sigma_k,\sigma_k']_E(\mathbf{q}),
[X_1,X_1'](\mathbf{t},\mathbf{p}),\ldots,
[X_k,X_k'](\mathbf{t},\mathbf{p})),
$$
 where $(\mathbf{t},\mathbf{p})\in \rk\times P,\,q=\pi(\mathbf{p})\,.$ It is
easy to see that $[Z,Z']^\pi(\mathbf{t},\mathbf{p})$ is an element of
$\mathcal{T}^E_kP$. Since any section of $\mathcal{T}^E_kP$ can be
locally written as a linear combination of the projectable sections
$\{\mathcal{X}_\alpha^A,\,\mathcal{Y}^A_B,\,\mathcal{V}^A_\vartheta\}$,
the definition of a Lie bracket for arbitrary sections of
$\mathcal{T}^E_kP$ follows.

The Lie brackets of the elements of the local basis
$\{\mathcal{X}_\alpha^A,\,\mathcal{Y}^A_B,\,\mathcal{V}^A_\vartheta\}$
are
\begin{equation} \label{lie brack k-prol}
\begin{array}{lll}
[\mathcal{X}_\alpha^A,\mathcal{X}_\beta^B]^\pi=\delta_B^A
\mathcal{C}^\gamma_{\alpha\beta}\mathcal{X}^A_\gamma\quad &
[\mathcal{X}_\alpha^A,\mathcal{Y}^B_C]^\pi=0\quad &
[\mathcal{X}_\alpha^A,\mathcal{V}^C_\vartheta]^\pi=0
\\ \noalign{\medskip}
[\mathcal{Y}^A_B,\mathcal{Y}^C_D]^\pi=0 &
[\mathcal{Y}^A_B,\mathcal{V}^C_\vartheta ]^\pi=0 \quad&
[\mathcal{V}^A_\varphi ,\mathcal{V}^B_\vartheta ]^\pi=0
\end{array}\quad .
\end{equation}

\paragraph{The derivation $d^A$}\

 Now, for each $A$, we define a derivation of degree $1$ on the set
$Sec(\bigwedge (\mathcal{T}^E_kP)^*)$, that is, \ $d^A:Sec(\bigwedge^l
(\mathcal{T}^E_kP)^*)\to Sec(\bigwedge^{l+1} (\mathcal{T}^E_kP)^*)$\
given by
\begin{equation}\label{dif k-prol}\begin{array}{lcl}
d^A\mu(Z_1,\ldots,
Z_{l+1})&=&\ds\sum_{i=1}^l\,(-1)^{i+1}\hat{\rho}_A(Z_i)\mu(Z_1,\ldots,
\widehat{Z_i},\ldots, Z_{l+1})\\\noalign{\medskip} &+&
\ds\sum_{i<j}(-1)^{i+j}\mu([Z_i,Z_j]^\pi,Z_1,\ldots,
\widehat{Z_i},\ldots, \widehat{Z}_j,\ldots, Z_{l+1})\;,
\end{array}
\end{equation}for $\mu\in Sec(\bigwedge^l
(\mathcal{T}^E_kP)^*)$ and $Z_1,\ldots, Z_k\in
Sec(\mathcal{T}^E_kP)$.
 In particular, if $f:\rk\times P\to \r$ is a real
smooth function, then
\ $d^Af:\rk\times P \to (\mathcal{T}^E_kP)^*$\ is defined by
$d^Af(Z)=\widetilde{\tau}_2^A(Z)f$, for $Z\in
Sec(\mathcal{T}^E_kP)$.

  From (\ref {base k-prol}) and
(\ref{k-vector asoc k-prol}) we deduce that
 $d^Af$ is locally given by
\begin{equation}
\label{dif func}d^Af=\ds\frac{\partial f}{\partial t^C}\overline{\mathcal{Y}}_A^C +
\rho^i_\alpha \ds\frac{\partial f}{\partial q^i}\mathcal{X}^\alpha_A
+ \ds\frac{\partial f}{\partial
u^\vartheta}\mathcal{V}_A^\vartheta\;,
\end{equation}
where
$\{\mathcal{X}^\alpha_A,\,\overline{\mathcal{Y}}_A^B,
\,\mathcal{V}_A^\vartheta\}$ is the local basis of sections of
$(\mathcal{T}^E_kP)^*$ dual to the local
basis$\{\mathcal{X}_\alpha^A,\,\mathcal{Y}^A_B,
\,\mathcal{V}^A_\vartheta\}$ of sections of $ \mathcal{T}^E_kP $.

The derivation $d^A$ allows us to introduce the Poincar\'{e}-Cartan
$2$-sections of the Lagrangian formalism on Lie algebroids (see
section \ref{614}),
 and the Liouville $2$-section of the Hamiltonian formalism (see section \ref{623}).

\section{$k$-cosymplectic classical field theory on Lie algebroids}\label{ kcosym alg}

\subsection{Lagrangian formalism}\label{lag form al}

In this subsection we give a description of Lagrangian
$k$-cosymplectic first-order classical field theory on Lie
algebroids. The Lagrangian  field theory on Lie algebroids is
developed in the $k$-prolongation
$\mathcal{T}^E_k(\stackrel{k}{\oplus}E)$ of a Lie algebroid $E$ over
the vector bundle projection
$\widetilde{\tau}:\stackrel{k}{\oplus}E\to Q$.  This vector bundle
$\mathcal{T}^E_k(\stackrel{k}{\oplus}E)$ plays the role of
$\tau^k_{\rktkq}:T^1_k( \rktkq)\to  \rktkq$, the tangent bundle of
$k^1$-velocities of $\rktkq$, in the standard Lagrangian
$k$-cosymplectic formalism.

\subsubsection{Geometric elements}\label{611}

\paragraph{The vector bundle $\mathcal{T}^E_k(\stackrel{k}{\oplus} E)$}\

Consider the $k$-prolongation
$\mathcal{T}^E_k(\stackrel{k}{\oplus}E)$
 of a Lie algebroid $E$ over the fibration
$\widetilde{\tau}:\stackrel{k}{\oplus}E\to Q$ (observe that,
in this case, the fiber $\pi:P\to Q$ is $\widetilde{\tau}:
\stackrel{k}{\oplus} E\to Q$). Taking into account the general
description of the $k$-prolongation (see the previous section), if
$(q^i,y^\alpha_A)$ are local coordinates on $\stackrel{k}{\oplus}E$,
then  we have the local coordinates $(t^A,
q^i,y^\alpha_A,z^\alpha_A,v_A^B,(v_A)^\beta_B)$ on
$\mathcal{T}^E_k(\stackrel{k}{\oplus}E)$ given by (see (\ref{local
coord k-prol}))
{\small \begin{equation}\label{coordfin2}\begin{array}{lll}
t^A(a_\mathbf{q},W_{(\mathbf{t},b_{\mathbf{q}})})=t^A(t)\;, &
q^i(a_{\bf q},W_{(\mathbf{t},b_{\mathbf{q}})})= q^i(\mathbf{q})
\;,\quad &
y^\alpha_A(a_\mathbf{q},W_{(\mathbf{t},b_{\mathbf{q}})})=y^\alpha_A(p)\;,
\\\noalign{\medskip}
z^\alpha_A(a_\mathbf{q},W_{(\mathbf{t},b_{\mathbf{q}})})=y_A^\alpha(a_{\bf
q})\;, &
  v_A^B(a_\mathbf{q},W_{(\mathbf{t},b_{\mathbf{q}})})=
v_{A_{(\mathbf{t},b_{\mathbf{q}})}}(t^B)\;,\quad &
(v_A)^\beta_B (a_\mathbf{q},W_{(\mathbf{t},b_{\mathbf{q}})})=
v_A{_{(\mathbf{t},b_{\mathbf{q}})}}(y^\beta_B)\ ,
\end{array}
\end{equation}}
and the local basis
$\{\mathcal{X}_\alpha^A,\;\mathcal{Y}^A_B,\;(\mathcal{V}^A)^B_\beta\}$
of sections of $\tau_{\rk\times\stackrel{k}{\oplus}E }:
\mathcal{T}^E_k(\stackrel{k}{\oplus}E)\to \rk\times
\stackrel{k}{\oplus}E$, defined in (\ref{base k-prol}), is written
here as
\begin{equation}\label{base}
\begin{array}{lcl}
\mathcal{X}_\alpha^A(\mathbf{t},b_{\mathbf{q}})\; &=&
(\widetilde{e}^A_\alpha(\mathbf{q}),
(0,\ldots,\stackrel{A}{\widetilde{\rho^i_\alpha({\bf
q})\ds\frac{\partial }{\partial
q^i}\Big\vert_{(\mathbf{t},b_{\mathbf{q}})}}},\ldots, 0))
\\\noalign{\medskip}
\mathcal{Y}^A_B(\mathbf{t},b_{\mathbf{q}})& = &(
\mathbf{0}_q,(0,\ldots,
\stackrel{A}{\widetilde{\ds\frac{\partial}{\partial
t^B}\Big\vert_{(\mathbf{t},b_{\mathbf{q}})}}},\ldots,0))
\\\noalign{\medskip}
(\mathcal{V}^A)^B_\beta(\mathbf{t},b_{\mathbf{q}}) &=&
(\mathbf{0}_q,(0,\ldots,
\stackrel{A}{\widetilde{\ds\frac{\partial}{\partial
y^\beta_B}\Big\vert_{(\mathbf{t},b_{\mathbf{q}})}}},\ldots, 0))
\end{array}\end{equation}

From (\ref{k-vector asoc k-prol}), we know that the
   $k$-vector fields  associated to the basis  $\{
\mathcal{X}_\alpha^A,\;\mathcal{Y}^A_B,\;(\mathcal{V}^A)^B_\beta\}$
are
  \begin{equation}\label{anchorLE}
\widetilde{\tau}_2^A(\mathcal{X}^B_\alpha)=\delta_A^B\rho^i_\alpha\ds\frac{\partial}{\partial
q^i} \quad , \quad
  \widetilde{\tau}_2^A(\mathcal{Y}^B_C)= \delta_B^C   \ds\frac{\partial}{\partial
t^C}\quad , \quad
\widetilde{\tau}_2^A((\mathcal{V}^B)^C_\alpha)=\delta_A^B\ds\frac{\partial}{\partial
y^\alpha_C}\,.
\end{equation}

The Lie brackets (\ref{lie brack k-prol}) of the elements of the
local basis  of sections are now
\begin{equation} \label{algebra LE}
\begin{array}{lll}
[\mathcal{X}_\alpha^A,\mathcal{X}_\beta^B]^{\widetilde{\tau}}=\delta_B^A
\mathcal{C}^\gamma_{\alpha\beta}\mathcal{X}^A_\gamma\quad &
[\mathcal{X}_\alpha^A,\mathcal{Y}^B_C]^{\widetilde{\tau}}=0\quad &
[\mathcal{X}_\alpha^A,(\mathcal{V}^C)^B_\beta]^{\widetilde{\tau}}=0
\\ \noalign{\medskip}
[\mathcal{Y}^A_B,\mathcal{Y}^C_D]^{\widetilde{\tau}}=0 &
[\mathcal{Y}^A_B,(\mathcal{V}^C)^D_\beta ]^{\widetilde{\tau}}=0
\quad&
[(\mathcal{V}^A)^B_\alpha,(\mathcal{V}^C)^D_\beta]^{\widetilde{\tau}}=0
\end{array}\quad .
\end{equation}

In $\mathcal{T}^E_k(\stackrel{k}{\oplus}E)$ there are two families
of canonical objects: the Liouville sections and the vertical
endomorphism whose definitions and properties mimic those of the
corresponding canonical objects in $\rktkq$ (see
\cite{LMNRS-2002,mod1,mod2,MSV-2005}).
First, we need to introduce:

\paragraph{Vertical $A$-lifts}\

An element $(a_\mathbf{q},W_{({\bf t},b_\mathbf{q})})$ of
$\mathcal{T}^E_k(\stackrel{k}{\oplus} E) $ is said to be {\sl vertical} if
$$
\widetilde{\tau}_{1,\stackrel{k}{\oplus}E}
(a_\mathbf{q},W_{({\bf t},b_\mathbf{q})})=\mathbf{0}_q\equiv(0_q,\stackrel{k}{\ldots},0_q)\in
\stackrel{k}{\oplus} E$$
 where
$\widetilde{\tau}_{1,\stackrel{k}{\oplus}E}:
\mathcal{T}^E_k(\stackrel{k}{\oplus}E) \to \stackrel{k}{\oplus}E$ is
the projection on the first factor. This condition means that

1) $a_{A_\mathbf{q}}=0_q$.

2) If $Z_{({\bf t},b_\mathbf{q})}=(\mathbf{0}_q,W_{({\bf t},b_{\bf
q})})$, with $W_{({\bf t},b_\mathbf{q})}=(v_{1_{({\bf t},b_{\bf
q})}},\ldots, v_{k_{({\bf t},b_\mathbf{q})}})\in
T^1_k(\rk\times\stackrel{k}{\oplus}E)$, since $Z\in
\mathcal{T}^E_k(\stackrel{k}{\oplus} E) $  from (\ref{fund}) we have
 $(v_A)^i=y^\alpha_A(\mathbf{0}_q)\rho^i_\alpha=0$

Therefore each $v_{A_{({\bf t},b_\mathbf{q})}}$ is locally given by
$$v_{A_{(\mathbf{t},b_\mathbf{q})}}= v_A^B \ds\frac{\partial  }{\partial
t^B}\Big\vert_{(\mathbf{t},b_\mathbf{q})} + (v_A)^\alpha_B
\ds\frac{\partial }{\partial
y^\alpha_B}\Big\vert_{(\mathbf{t},b_{\bf q})}\in T_{({\bf
t},b_\mathbf{q})}(\rk\times \stackrel{k}{\oplus} E)$$ and it is
vertical with respect to the canonical projection
$\rk\times\stackrel{k}{\oplus} E \to Q$.

 For each $A=1,\ldots, k$, we call the following map the {\sl vertical $A^{th}$-lifting map}
\begin{equation}\label{a-lift}
\begin{array}{rcl}
\xi^{V_A}:\rk\times\stackrel{k}{\oplus}
E\times_Q\stackrel{k}{\oplus} E & \longrightarrow &
\mathcal{T}^E_k(\stackrel{k}{\oplus} E)  \\
(t,a_\mathbf{q},b_\mathbf{q}) & \longmapsto & \xi^{V_A}(t,a_{\bf
q},b_\mathbf{q})=(\mathbf{0}_q,(0,\ldots,
\stackrel{A}{\widetilde{(a_\mathbf{q})^{V_A}_{(t,b_\mathbf{q})}}},\ldots,0)) \\
\end{array} \ ,
\end{equation}
 where for an arbitrary function $f$ defined on
$\rk\times\stackrel{k}{\oplus} E$,
$(a_\mathbf{q})^{V_A}$ is given by
 \begin{equation}\label{vertical}
(a_\mathbf{q})^{V_A}_{(t,b_{\bf
q})}f=\ds\frac{d}{ds}\Big\vert_{s=0}f(t,b_{1_q},\ldots, b_{a_{\bf
q}}+s a_{a_\mathbf{q}},\ldots, b_{k_q})\ .
\end{equation}

\,From (\ref{vertical}) we deduce that the local expression of
$\;(a_\mathbf{q})^{V_A}_{(\mathbf{t},b_\mathbf{q})}$ is
\begin{equation}\label{localvert}
(a_\mathbf{q})^{V_A}_{(t,b_{\bf
q})}=y^\alpha_A(a_{q})\ds\frac{\partial}{\partial
y^\alpha_A}\Big\vert_{(\mathbf{t},b_\mathbf{q})}\in T_{(t,b_{\bf
q})}(\rk\times \stackrel{k}{\oplus} E)\; .
\end{equation}

\, From (\ref{base}), (\ref{localvert}) and (\ref{a-lift}) we obtain
\begin{equation}\label{localxia}
\xi^{V_A}(t,a_\mathbf{q},b_\mathbf{q})=(\mathbf{0}_q,(0,\ldots,
\stackrel{A}{\widetilde{y^\alpha_A(a_{q})\ds\frac{\partial}{\partial
y^\alpha_A}\Big\vert_{(\mathbf{t},b_\mathbf{q})}}},\ldots, 0)) =
y^\alpha_A(a_\mathbf{q})(\mathcal{V}^A)^A_\alpha(t,b_{\bf q})\,.
\end{equation}

 The vertical $A^{th}$-lifting map allows us to define vertical lifts of section
of $\widetilde{\tau}:\stackrel{k}{\oplus}E\to Q$ to sections of
$\widetilde{\tau}_{\rk\times\stackrel{k}{\oplus}E}:\mathcal{T}^E_k(\stackrel{k}{\oplus}E)\to
\rk\times \stackrel{k}{\oplus}E$. If $\sigma$ is a section of $
\stackrel{k}{\oplus}E \to Q$, then the section
$\sigma^{V_A}:\rk\times \stackrel{k}{\oplus}E \to
\mathcal{T}^E_k(\stackrel{k}{\oplus}E)$ of
$\widetilde{\tau}_{\rk\times\stackrel{k}{\oplus}E}$ is defined by
$\sigma^{V_A}(t,b_\mathbf{q})=\xi^{V_A}(t,
\sigma(\mathbf{q}),b_\mathbf{q})$, and it will be called the vertical
$A^{th}$-lift of $\sigma$. In particular, from (\ref{localxia}) we
obtain
\begin{equation}\label{vert e}
(\widetilde{e}^B_\alpha)^{V_A}(t,b_\mathbf{q})=
\delta^B_A(\mathcal{V}^A)^A_\alpha(t,b_\mathbf{q})\qquad
1\leq \alpha\leq dim\,E\ .
\end{equation}

\paragraph{The Liouville sections}\

The Liouville $A^{th}$-section $\Delta_A$ is the section of
$\widetilde{\tau}_{\rk\times\stackrel{k}{\oplus}E}:\mathcal{T}^E_k(\stackrel{k}{\oplus}
E) \to \rk\times \stackrel{k}{\oplus} E$ given by
$$
\begin{array}{rcl}
\Delta_A:\rk\times \stackrel{k}{\oplus} E & \to &
\mathcal{T}^E_k(\stackrel{k}{\oplus} E) \\\noalign{\medskip}
(\mathbf{t},b_\mathbf{q})&\mapsto
&\Delta_A(\mathbf{t},b_\mathbf{q})=\xi^{V_A}(t,b_\mathbf{q},b_\mathbf{q})
\end{array}\;.
$$

\, From (\ref{localxia}) we obtain that $\Delta_A$ is locally given
by
\begin{equation}\label{Liouville}
\Delta_A=\ds\sum_{\alpha} y^\alpha_A(\mathcal{V}^A)^A_\alpha\;.
\end{equation}
The  set  $\{\widetilde{\tau}_2(\Delta_A)=
(\widetilde{\tau}_2^1(\Delta_A),\ldots,\widetilde{\tau}_2^k(\Delta_A))\}$,
where each $\widetilde{\tau}_2(\Delta_A)$ is the $k$-vector field
associated to each $\Delta_A$ given by (\ref{k-vector asoc k-prol}),
enables us to introduce the Lagrangian energy function $E_L$

\begin{remark} \label{liovilleA}{\rm
In the standard case, every section $\Delta_A$ translates into the
$k$-vector field on $\rk\times T^1_kQ$ given by
$(0,\ldots,\stackrel{A}{\widetilde{v^i_A\ds\frac{\partial}{\partial
v^i_A}}},\ldots, 0)$, and since $A$ is fixed, $\Delta_A$ is identified
with the canonical vector field, $\Delta_A=
v^i_A\ds\frac{\partial}{\partial v^i_A}$. (See
\cite{LMNRS-2002,mod1,mod2,MSV-2005}).}
\end{remark}

\paragraph{The vertical endomorphism}\

\begin{definition}
The {\rm $A^{th}$-vertical endomorphisms} of
$\mathcal{T}^E_k(\stackrel{k}{\oplus} E) $, for $1\leq A \leq k$,
are defined as
$$\begin{array}{rccl}
\widetilde{S}^A:&\mathcal{T}^E_k(\stackrel{k}{\oplus} E)  & \to &
\mathcal{T}^E_k(\stackrel{k}{\oplus} E) \\\noalign{\medskip}
&(a_\mathbf{q},W_{(\mathbf{t},b_\mathbf{q})})&\mapsto
&\widetilde{S}^A(a_{\bf
q},W_{(\mathbf{t},b_\mathbf{q})})=\xi^{V_A}(t,a_\mathbf{q},b_\mathbf{q})
\end{array}\ .
$$
\end{definition}

Next, we express $\widetilde{S}^A$ using the basis of local
sections $\{\mathcal{X}_\alpha^A,\,\mathcal{Y}^A_B,\,
(\mathcal{V}^A)^B_\beta\}$ of $\mathcal{T}^E_k(\stackrel{k}{\oplus}
E) $ and its dual basis
$\{\mathcal{X}^\alpha_A,\,\overline{\mathcal{Y}}_A^B,\,
(\mathcal{V}_A)_B^\beta\}$ of $(\mathcal{T}^E_k(\stackrel{k}{\oplus}
E) )^{\;*}$. In fact, from (\ref{base}) and (\ref{vert e}) we obtain
\begin{equation}\label{localSA}
\widetilde{S}^A=\ds\sum_{\alpha}(\mathcal{V}^A)^A_\alpha\otimes\mathcal{X}_A^\alpha\ .
\end{equation}

\begin{remark}\label{remSA}{\rm In the standard case, the endomorphism
$\widetilde{S}^A$ coincides with the $A^{th}$ element of the family
of the canonical $k$-tangent structure $(S^1,\ldots, S^A)$.
}\end{remark}

\paragraph{Second order partial differential equations}\

In the standard $k$-cosymplectic Lagrangian formalism, the solutions
to the Euler-Lagrange equations are obtained as integral sections of
second order partial differential equations ({\sc sopde}). Taking
into account the Definition \ref{sode2} and the remarks
\ref{liovilleA} and \ref{remSA} we introduce the following
\begin{definition}
 A section  $\xi:\rk\times\stackrel{k}{\oplus} E\to
\mathcal{T}^E_k(\stackrel{k}{\oplus} E) $  of
$\mathcal{T}^E_k(\stackrel{k}{\oplus} E) $ is called {\rm a second
order partial differential equation ({\sc sopde})} if
$$
\widetilde{S}^A(\xi)=\Delta_A \quad \makebox{and} \quad
\overline{\mathcal{Y}}_A^B(\xi)=\delta^A_B\ .
$$
\end{definition}

It is easy to  deduce that the local expression of a {\sc sopde}
$\xi$ is
$$
\xi=\mathcal{Y}^A_A+y^\alpha_A\mathcal{X}_\alpha^A+
(\xi_A)^\alpha_B(\mathcal{V}^A)^B_\alpha\;,
$$where $(\xi_A)^\alpha_B$ are functions on $\rk\times
\stackrel{k}{\oplus}E$.

From (\ref{k-vector asoc k-prol}), we note that the  $k$-vector field
$\widetilde{\tau}_2(\xi)=(\widetilde{\tau}_2^1(\xi),\ldots,\widetilde{\tau}_2^k(\xi))\,$
on $\rk\times\stackrel{k}{\oplus} E$ associated to $\xi$ is locally
given by
\begin{equation}\label{sopde asso}
\widetilde{\tau}_2^A(\xi)= \ds\frac{\partial }{
\partial t^A}
+\rho^i_\alpha y^\alpha_A \ds\frac{\partial}{\partial
q^i}+(\xi_A)^\alpha_B\ds\frac{\partial}{\partial y^\alpha_B}\ .
\end{equation}

\begin{definition}
A map $\eta:\rk\to \rk\times\stackrel{k}{\oplus} E$ is an
{\rm integral section} of the {\sc sopde} $\xi$, if $\eta$ is an integral
section of the associated $k$-vector field $\widetilde{\tau}_2(\xi)$
, that is,
\begin{equation}\label{int sect}
\widetilde{\tau}_2^A(\xi)(\eta(t))=
\eta_*(t)\left(\ds\frac{\partial}{\partial t^A}\Big\vert_{
t}\right)\ .
\end{equation}\end{definition}

Locally, if $\eta(t)=(\eta^A(t),\eta^i(t),\eta^\alpha_A(t))$, from
(\ref{sopde asso}) we deduce that (\ref{int sect}) is equivalent to
\begin{equation}\label{integral sect}
\ds\frac{\partial \eta^B}{\partial t^A}\Big\vert_{t}=\delta^A_B
\;,\quad \ds\frac{\partial \eta^i}{\partial t^A}\Big\vert_{t}=
\eta^\alpha_A(t)\rho^i_\alpha\;,\quad \ds\frac{\partial
\eta^\beta_B}{\partial t^A}\Big\vert_{t}=(\xi_A)^\beta_B(\eta(t))\;.
\end{equation}

\subsubsection{Lagrangian formalism}\label{614}

In this section, we develop a geometric framework, enabling
us to write the Euler-Lagrange equations associated with the
Lagrangian function $L$ in an intrinsic way.

Let $L:\rk\times\stackrel{k}{\oplus} E\to \r$ be a function which we
call a {\sl  Lagrangian function}.

\paragraph{Poincar\'{e}-Cartan sections and the Lagrangian energy function}\

We introduce {\sl the Poincar\'{e}-Cartan $1$-sections} associated with $L$:
$$
\begin{array}{rcc}
\Theta_L^A:\rk\times \stackrel{k}{\oplus}E & \longrightarrow &
(\mathcal{T}^E_k(\stackrel{k}{\oplus}E))^{\;*}
\\\noalign{\medskip}
(\mathbf{t},b_\mathbf{q}) & \longmapsto &
\Theta_L^A(\mathbf{t},b_\mathbf{q}) \ ,
\end{array}
$$
 where $\Theta_L^A(\mathbf{t},b_\mathbf{q})$ is defined by
$$\begin{array}{rlcl}
\Theta_L^A(\mathbf{t},b_\mathbf{q}): &
(\mathcal{T}^E_k(\stackrel{k}{\oplus}E)
)_{(\mathbf{t},b_\mathbf{q})}& \longrightarrow & \r
\\\noalign{\medskip}
   & Z_{(\mathbf{t},b_\mathbf{q})} & \longmapsto &
(\Theta_L^A)_{(\mathbf{t},b_\mathbf{q})}(Z_{(\mathbf{t},b_\mathbf{q})})=
(d^AL)_{(\mathbf{t},b_\mathbf{q})}((\widetilde{S}^A)_{(\mathbf{t},b_\mathbf{q})}
(Z_{(\mathbf{t},b_\mathbf{q})}))
\end{array}\; .
$$
\; From (\ref{dif k-prol}) we obtain that
$$
(\Theta_L^A)_{(\mathbf{t},b_\mathbf{q})}(Z_{(\mathbf{t},b_\mathbf{q})})=
(d^AL)_{(\mathbf{t},b_\mathbf{q})}
((\widetilde{S}^A)_{(\mathbf{t},b_\mathbf{q})}(Z_{(\mathbf{t},b_\mathbf{q})}))=
\widetilde{\tau}_2^A((\widetilde{S}^A)_{(\mathbf{t},b_\mathbf{q})}(Z_{(\mathbf{t},b_\mathbf{q})}))L\ ,
$$
where ${(\mathbf{t},b_\mathbf{q})}\in
\rk\times\stackrel{k}{\oplus}E$, and
$Z_{(\mathbf{t},b_\mathbf{q})}\in
(\mathcal{T}^E_k(\stackrel{k}{\oplus}E))_{(\mathbf{t},b_\mathbf{q})}$.

 From (\ref{anchorLE}) and (\ref{localSA}) we obtain the  local
expression of $\Theta_L^A$,
\begin{equation}\label{local theta}
\Theta_L^A=\ds\frac{\partial L}{\partial y^\alpha
_A}\mathcal{X}^\alpha_A \ .
\end{equation}

The {\sl Poincar\'{e}-Cartan $2$-sections}
$\Omega_L^A:\rk\times\stackrel{k}{\oplus}E \to
(\mathcal{T}^E_k(\stackrel{k}{\oplus}E))^{\;*}\wedge(\mathcal{T}^E_k(\stackrel{k}{\oplus}E))^{\;*}$
associated with $L$ are given by
$$
\Omega_L^A=-d^A\Theta_L^A\ .
$$

 From (\ref{ecest}), (\ref{dif k-prol}), (\ref{anchorLE}), (\ref{algebra LE}) and
(\ref{local theta}) we obtain
{\small\begin{equation}\label{local omega} \omega_L^A
=\ds\frac{\partial^2 L}{\partial t^B\partial
y^\alpha_A}\mathcal{X}^\alpha_ A \wedge
\overline{\mathcal{Y}}_A^{\,B} + \ds\frac{1}{2} \left(\rho^i_\beta
\ds\frac{\partial ^2 L}{\partial q^i\partial y^\alpha_A} -
\rho^i_\alpha \ds\frac{\partial ^2 L}{\partial q^i\partial
y^\beta_A}+ \mathcal{C}^\gamma_{\alpha\beta}\ds\frac{\partial
L}{\partial y^\gamma_A}\right) \mathcal{X}^\alpha_ A \wedge
\mathcal{X}^\beta_A + \ds\frac{\partial ^2 L}{\partial y^\beta_
B\partial y^\alpha_A}\, \mathcal{X}^\alpha_A \wedge
(\mathcal{V}_A)_B^\beta\;.
\end{equation}}

The {\sl energy function} $E_L:\rk\times\stackrel{k}{\oplus} E \to
\r$ defined by $L$ is
$$
E_L=\ds\sum_{A=1}^k \widetilde{\tau}_2^A(\Delta_A)L-L\;,
$$
and from (\ref{anchorLE}) and (\ref{Liouville}) one deduces that
$E_L$ is locally given by
\begin{equation}\label{local ener}
E_L=\ds\sum_{A=1}^k y^\alpha_A\ds\frac{\partial L}{\partial
y^\alpha_A}- L\;.
\end{equation}

\paragraph{Morphisms}\

Before addressing the Euler-Lagrange equations on Lie algebroids,
we show a new point of view for the solutions to the standard
Euler-Lagrange equations, which allows us to consider a solution as a
morphism of Lie algebroids.

In the standard Lagrangian $k$-cosymplectic  description of first
order classical field theories, a solution to the Euler-Lagrange
equation is a field $\phi:\rk\to Q$ such that its first prolongation
$\phi^{[1]}:\rk\to \rk\times T^1_kQ$ (see Definition \ref{de652})
satisfies the Euler-Lagrange field equations, that is,
$$
\displaystyle \sum_{A=1}^k\ds\frac{\partial}{\partial
t^A}\Big\vert_{\mathbf{t}} \left(\frac{\displaystyle\partial
L}{\displaystyle
\partial v^i_A}\Big\vert_{\phi^{[1]}(\mathbf{t})} \right)= \frac{\displaystyle \partial
L}{\displaystyle
\partial q^i}\Big\vert_{\phi^{[1]}(\mathbf{t})}\,.
$$

The map $\phi$ induces  the following morphism of Lie algebroids
\[\xymatrix
{T\rk  \ar[r]^-{T\phi}\ar[d]_{\tau_{\rk}} & TQ\ar[d]^-{\tau_Q}\\
\rk\ar[r]_-{\phi} & Q}\]

If we consider the canonical basis of section of $\tau_{\rk}$,
$\;\left\{\ds\frac{\partial}{\partial t^1},\ldots,
\ds\frac{\partial}{\partial t^k}\right\}$, then the  first
prolongation $\phi^{[1]}$ of $\phi$, can be written as follows:
\[\phi^{[1]}(t)=(t,T_t\phi(\ds\frac{\partial}{\partial
t^1}\Big\vert_{t}),\ldots,T_t\phi(\ds\frac{\partial}{\partial
t^k}\Big\vert_{t}))\,.\]

Returning to the case of algebroids, the analog of the field
solution to the Euler-Lagrange equations to be considered here  is a
morphism of Lie algebroid $\Phi=(\overline{\Phi},\underline{\Phi})$
\[\xymatrix
{T\rk  \ar[r]^-{\overline{\Phi}}\ar[d]_{\tau_{\rk}} & E\ar[d]^-{\tau}\\
\rk\ar[r]_-{\underline{\Phi}} & Q}\]

Taking a local basis $\{e_A\}_{A=1}^k$ of local sections of $T\rk$,
one can define a section
$\widetilde{\Phi}:\rk\to\rk\times\stackrel{k}{\oplus} E$ associated
to $\Phi$ and given by
\[\begin{array}{rcl}
\widetilde{\Phi}:\rk &\to &\rk\times \stackrel{k}{\oplus}E\equiv
E\oplus\stackrel{k}{\ldots}\oplus E\\
t &\to &(t,\overline{\Phi}(e_1(t)),\ldots,
\overline{\Phi}(e_k(t)))\;.
\end{array}\]

Let $(t^A)$ and $(q^i)$ be a local coordinate system on $\rk$ and
$Q$, respectively. Let $\{e_A\}$ be a local basis of sections of
$\tau_{\rk}$ and $\{e_\alpha\}$ be a local basis of sections of $E$;
we denote by $\{e^A\}$ and $\{e^\alpha\}$ the dual basis. Then
$\Phi$ is determined by the relations
$\underline{\Phi}(t)=(\phi^i(t))$ and $\Phi^* e^\alpha=\phi^\alpha_A
e^A$ for certain local functions $\phi^i$ and $\phi^\alpha_A$ on
$\rk$. Thus, the associated map $\widetilde{\Phi}$ is locally given
by $\widetilde{\Phi}(t)=(t^A,\phi^i(t),\phi^\alpha_A(t))$.

In this case, the condition of Lie morphism (\ref{morp cond}) is
written \begin{equation}\label{morpcond}
  \rho^i_
 \alpha\phi^\alpha_A = \ds\frac{\partial \phi^i}{\partial t^A}
 \quad ,\quad o=\ds\frac{\partial \phi^\alpha_A}{\partial t^B} -
\ds\frac{\partial \phi^\alpha_B}{\partial t^A}
+\mathcal{C}^\alpha_{\beta\gamma}\phi^\beta_B\phi^\alpha_A
\,.\end{equation}

\begin{remark}{\rm
 In the standard case where $E=TQ$,
the above morphism conditions reduce to
\[\phi^i_A=\ds\frac{\partial \phi^i}{\partial t^A} \quad
\makebox{and}\quad  \ds\frac{\partial \phi^i_A}{\partial
t^B}=\ds\frac{\partial \phi^i_B}{\partial t^A}\,.\] Then,
 by considering morphisms we are just considering
the first prolongation of fields $\phi:\rk\to Q$.}
\end{remark}

\paragraph{The Euler-Lagrange equations}\

For an arbitrary section $\xi:\rk\times\stackrel{k}{\oplus} E \to
\mathcal{T}^E_k(\stackrel{k}{\oplus} E) $ of
$\widetilde{\tau}_{\rk\times\stackrel{k}{\oplus}
E}:\mathcal{T}^E_k(\stackrel{k}{\oplus} E) \to
\rk\times\stackrel{k}{\oplus} E$,
consider the equations
\begin{equation}\label{EC E-L}
    \overline{\mathcal{Y}}_A^C(\xi)=\delta_A^C\quad ,\quad
    \imath_{\xi}\Omega_L^A =\ds\frac{1}{k}\Big(d^A E_L+
\ds\sum_{C=1}^k\frac{\partial L}{\partial
t^C}(\overline{\mathcal{Y}}_A^C\Big)\ ,
 \end{equation}
Writing $\xi= \xi_B^C\mathcal{Y}^B_C+\xi_B^\alpha
\mathcal{X}^B_\alpha + (\xi_B)^\alpha_C (\mathcal{V}^B)^C_\alpha $,
from (\ref{dif func}), (\ref{local omega}), (\ref{local ener}) we
have that  (\ref{EC E-L}) is equivalent to
{\small\begin{eqnarray}\label{eq ec el}
\xi_B^C = \delta_B^C\quad ,\quad
  \derpars{L}{t^C}{y^\alpha_A}\,\xi^\alpha_A =
\ds\frac{1}{k}\derpars{L}{t^C}{y^\beta_ B}\,y^\beta_B \quad & ,&\quad
  \derpars{L}{y^\gamma_C}{y^\alpha_A}\,\xi^\alpha_A =
\ds\frac{1}{k}\derpars{L}{y^\gamma_C}{y^\beta_ B}\,y^\beta_B
\\\noalign{\medskip}\nonumber
  \derpars{L}{t^C}{y^\alpha_A}\,\xi_A^C +\Big( \rho^i_\beta\,\derpars{L}{q^i}{y^\alpha_A}
-\rho^i_\alpha\,\derpars{L}{q^i}{y^\beta_A} + {\mathcal
C}^\gamma_{\alpha\beta}\derpar{L}{y^\gamma_A}\Big)\,\xi^\beta_A & +&
\derpars{L}{y^\beta_ B}{y^\alpha_A}\,(\xi_A)_B^\beta
=\ds\frac{1}{k}\Big(\derpar{L}{q^i}-\derpars{L}{q^i}{y^\beta_
B}\,y^\beta_ B\Big)\,\rho^i_\alpha
\end{eqnarray}}

\, From (\ref{eq ec el}) we obtain
 \begin{equation}\label{imp ec
el}\begin{array}{lcl} \xi_B^C = \delta_B^C\quad, \quad
  \ds\sum_{A=1}^k\derpars{L}{t^C}{y^\alpha_A}\,\xi^\alpha_A =
\derpars{L}{t^C}{y^\beta_ B}\,y^\beta_B\quad , \quad
  \ds\sum_{A=1}^k\derpars{L}{y^\gamma_C}{y^\alpha_A}\,\xi^\alpha_A =
\derpars{L}{y^\gamma_C}{y^\beta_ B}\,y^\beta_B\,,&&
\\\noalign{\medskip}
 \ds\sum_{A=1}^k \derpars{L}{t^A}{y^\alpha_A} +
 \ds\sum_{A=1}^k\Big( \rho^i_\beta\,\derpars{L}{q^i}{y^\alpha_A}
-\rho^i_\alpha\,\derpars{L}{q^i}{y^\beta_A} + {\mathcal
C}^\gamma_{\alpha\beta}\derpar{L}{y^\gamma_A}\Big)\,\xi^\beta_A +
\ds\sum_{A=1}^k\derpars{L}{y^\beta_ B}{y^\alpha_A}\,(\xi_A)_B^\beta
& &\\\noalign{\medskip}
=\Big(\derpar{L}{q^i}-\derpars{L}{q^i}{y^\beta_ B}\,y^\beta_
B\Big)\,\rho^i_\alpha \end{array}\ .
\end{equation}

\begin{definition}
$L$ is said to be a {\rm regular Lagrangian} if the matrix
$\left(\ds\frac{\partial ^2 L}{\partial y^\alpha_A\partial
y^\beta_B}\right)$ is regular.
\end{definition}

When $L$ is regular, from the three identity of (\ref{imp ec el}) we
obtain
\begin{equation}\label{cond sopde}
\xi^\alpha_A=y^\alpha_A\;.
\end{equation} In this case, the solution  $\xi$  to the equations (\ref{EC E-L}) is a {\sc sopde} and
 the functions $(\xi_A)_B^\beta$ are the solutions to the equation
\begin{equation}\label{eqla3}
\ds\sum_{A=1}^k\left(\derpars{L}{t^A}{y^\alpha_A} +
\rho^i_\beta\;\derpars{L}{q^i}{y^\alpha_A}\; y^\beta_A + {\mathcal
C}^\gamma_{\alpha\beta}\derpar{L}{y^\gamma_A}\;y^\beta_A +
\derpars{L}{y^\beta_B}{y^\alpha_A}\,(\xi_A)_B^\beta\right) =
\derpar{L}{q^i}\,\rho^i_\alpha\;.
\end{equation}

Let $\Phi=(\overline{\Phi},\underline{\Phi})$ be a morphism on
Lie algebroids between $\tau_{\rk}:T\rk\to\rk$ and $\tau:E\to Q$, and
$\widetilde{\Phi}:\rk\to\rk\times\stackrel{k}{\oplus}E$ the
associated map. If $\widetilde{\Phi}:\rk\to
\rk\times\stackrel{k}{\oplus} E\,$ is an integral section of the
{\sc sopde} $\xi$ locally given by
$\widetilde{\Phi}(t)=(\phi^B(t),\phi^i(t), \phi^\alpha_A(t))$. From
(\ref{integral sect}), (\ref{morpcond}), (\ref{imp ec el}),
(\ref{cond sopde}) and (\ref{eqla3}) we obtain
\begin{eqnarray*}
\ds\frac{\partial \phi^B}{\partial t^A}\Big\vert_{t} =
\delta^B_A\qquad , \qquad \ds\frac{\partial \phi^i}{\partial
t^A}\Big\vert_{t} =
\phi^\alpha_A(t)\rho^i_\alpha(\phi^j(t))\qquad,\qquad0&=&\ds\frac{\partial
\phi^\alpha_A}{\partial t^B} - \ds\frac{\partial
\phi^\alpha_B}{\partial t^A}
+\mathcal{C}^\alpha_{\beta\gamma}\phi^\beta_B\phi^\alpha_A \\
\ds\sum_{A=1}^k\left(\derpars{L}{t^A}{y^\alpha_A}\Big\vert_{\phi(t)}+\ds\frac{\partial
\phi^i}{\partial t^A}\Big\vert_{t} \ds\frac{\partial ^2 L}{\partial
q^i\partial y^\alpha_A}\Big\vert_{\phi(t)} + \ds\frac{\partial
\phi^\beta_B}{\partial t^A}\Big\vert_{t}\ds\frac{\partial^2
L}{\partial y^\alpha_A\partial y^\beta_B}\Big\vert_{\phi(t)}\right)
&=& \rho^i_\alpha \ds\frac{\partial L}{\partial
q^i}\Big\vert_{\phi(t)} -
y^\beta_A(\phi(t))\mathcal{C}^\gamma_{\alpha\beta}\ds\frac{\partial
L}{\partial y^\gamma_A}\Big\vert_{\phi(t)} \;,
\end{eqnarray*}
where the last equation is a consequence of $\Phi$ being
a Lie morphism. The above equations can be written as follows
\begin{equation} \label{eq E-L}
\begin{array}{rcl} \ds\frac{\partial \phi^B}{\partial t^A}\Big\vert_{t} =
\delta^B_A\qquad , \qquad \ds\frac{\partial \phi^i}{\partial
t^A}\Big\vert_{t} &=&
\phi^\alpha_A(t)\rho^i_\alpha(\phi^j(t))\qquad,\qquad 0
=\ds\frac{\partial \phi^\alpha_A}{\partial t^B} - \ds\frac{\partial
\phi^\alpha_B}{\partial t^A}
+\mathcal{C}^\alpha_{\beta\gamma}\phi^\beta_B\phi^\alpha_A \\
\ds\sum_{A=1}^k\ds\frac{\partial}{\partial t^A}\left(
\ds\frac{\partial  L}{\partial y^\alpha_A}\Big\vert_{\phi(t)}\right)
&=& \rho^i_\alpha \ds\frac{\partial L}{\partial
q^i}\Big\vert_{\phi(t)} -
y^\beta_A(\phi(t))\mathcal{C}^\gamma_{\alpha\beta}\ds\frac{\partial
L}{\partial y^\gamma_A}\Big\vert_{\phi(t)}\;,
\end{array}
\end{equation}
which are called {\sl the Euler-Lagrange field
equations written in terms of a Lie algebroid $E$} in the
$k$-cosymplectic setting.

The results of this section can be summarized in the following

\begin{theorem}\label{algeform}
Let $L:\rk\to\rk\times\stackrel{k}{\oplus}E$ be a regular Lagrangian
and let $\xi:\rk\times \stackrel{k}{\oplus}E \to
\mathcal{T}^E_k(\stackrel{k}{\oplus}E)$ be a section  of
$\widetilde{\tau}_{\rk\times\stackrel{k}{\oplus}E}$ such that
\begin{equation}\label{ec ge EL}
\overline{\mathcal{Y}}_B^C(\xi)=\delta_B^C\quad , \quad
\imath_{\xi}\Omega_L^A =\ds\frac{1}{k}\Big(d^A E_L+
\ds\sum_{C=1}^k\frac{\partial L}{\partial
t^C}(\overline{\mathcal{Y}}_A^C\Big)\ .
\end{equation}
\begin{enumerate}
\item Then $\xi$ is a {\sc sopde}.
\item Let $\Phi=(\overline{\Phi},\underline{\Phi})$ be a morphism on
Lie algebroids between $\tau_{\rk}:T\rk\to\rk$, and $\tau:E\to Q$, and
let  $\widetilde{\Phi}:\rk\to\rk\times\stackrel{k}{\oplus}E$ be the
associated map. If $\widetilde{\Phi}:\rk\to
\rk\times\stackrel{k}{\oplus} E\,$ is an integral section of the
{\sc sopde} $\xi$, then it is a solution of the {\it the
Euler-Lagrange field equations (\ref{eq E-L}) written in terms of a
Lie algebroid $E$}.
\end{enumerate}
\end{theorem}

\subsubsection{Relation with the standard Lagrangian $k$-cosymplectic formalism}\label{615}

As a final remark in this Subsection, it is interesting to point out
that the standard Lagrangian $k$-symplectic formalism   is a
particular case of the Lagrangian formalism on Lie algebroids,
 when $E=TQ$,  the anchor map $\rho$ is the identity
on $TQ$, and the structure constants $\mathcal{C}_{\alpha
\beta}^\gamma=0$.

 In this case we have:
\begin{itemize}
\item The manifold $\rk\times\stackrel{k}{\oplus}E$ is identified with $\rk\times T^1_kQ$
and $\mathcal{T}^{TQ}_k(T^1_kQ)$ with $T^1_k(\rk\times T^1_kQ)$.

\item The energy function $E_L:\rk\times T^1_kQ\to \r$ is given by $E_L=\ds\sum_{A=1}\triangle_A(L)-L$,
 where the vector fields $\triangle_A$ on $\rk\times T^1_kQ$ have been explained in Remark \ref{liovilleA} .

\item A section $\xi:\rk\times \stackrel{k}{\oplus}E \to
\mathcal{T}^E_k(\stackrel{k}{\oplus}E)$ corresponds to a $k$-vector
field $\xi=(\xi_1,\ldots,\xi_k)$ on $\rk\times T^1_kQ$, that is,
$\xi$ is a section of $\tau^k_{\rk\times T^1_kQ}:T^1_k(\rk\times
T^1_kQ) \to \rk\times T^1_kQ$.

\item Let $f$ be a function on $\rk\times T^1_kQ$, then
$$d^Af(Y_1,\ldots, Y_k)=df(Y_A)\ ,$$
where $df$ denotes the standard differential and $(Y_1,\ldots, Y_k)$
is a $k$-vector field on $\rk\times T^1_kQ$.

\item We have that
 \begin{eqnarray*}
\overline{\mathcal{Y}}_B^C((X_1,\ldots, X_k))&=& dt^B(X_C)\\
\Omega_L^A((X_1,\ldots, X_k),((Y_1,\ldots,
Y_k))&=&\omega_L^A(X_A,Y_A) \ ,
\end{eqnarray*}
 where
$\omega_L^A$ are the Poincar\'{e}-Cartan $2$-form of the standard
$k$-cosymplectic formalism defined in section \ref{212}.

\item Thus, in the standard $k$-cosymplectic formalism,
the equation (\ref{ec ge EL}) can be written as follows:
$$
dt^A(\xi_B)=\delta^A_B\quad ,\quad i_{\xi_A}\omega_L^A=
\ds\frac{1}{k}dE_L+ \,\ds\sum_{C=1}^k\ds\frac{\partial L}{\partial
t^C}dt^A\delta^\Delta_A\,,
$$
 which implies
$$dt^A(\xi_B)=\delta^A_B\quad ,\quad\ds\sum_{A=1}^ki_{\xi_A}\omega_L^A=dE_L+
 \,\ds\sum_{A=1}^k\ds\frac{\partial L}{\partial t^A}dt^A\,.$$

\item In the standard case, a map $\phi:\rk\to Q$ induces a morphism
on Lie algebroids $(T\phi,\phi)$ between $T\rk$ and $TQ$. In this
case, the associated map $\widetilde{\Phi}$ of this morphism is
defined as the first prolongation $\phi^{[1]}$ of $\phi$ given by
$$\widetilde{\Phi}(t)=(t,T\phi(\ds\frac{\partial}{\partial
t^1}\Big\vert_{t}),\ldots, T\phi(\ds\frac{\partial}{\partial
t^k}\Big\vert_{t}))\,.$$ Let us observe that
$\widetilde{\Phi}=\phi^{[1]}$ (see \ref{de652}).
\end{itemize}

Thus, from the Theorem \ref{algeform} and the above remarks, we
deduce the following corollary, which is a summary of the Lagrangian
$k$-cosymplectic formalism.

\begin{corollary}Let $L:\rk\to \rk\times T^1_kQ$ be a regular Lagrangian and
$\xi=(\xi_1,\ldots, \xi_k)$ a $k$-vector field on $\rk\times T^1_kQ$
such that
$$dt^A(\xi_B)=\delta^A_B\quad ,\quad\ds\sum_{A=1}^ki_{\xi_A}\omega_L^A=
dE_L+ \,\ds\sum_{A=1}^k\ds\frac{\partial L}{\partial t^A}dt^A\ .$$
\begin{enumerate}
\item Then $\xi$ is a {\sc sopde}.
\item  If $\widetilde{\Phi}$ is an integral section of
the $k$-vector field $\xi$, then it is a solution to the
Euler-Lagrange field equations (\ref{e-l-2}) in the standard
Lagrangian k-cosymplectic field theories. Let us observe that
$\widetilde{\Phi}=\phi^{[1]}$.\end{enumerate}
\end{corollary}

Finally, we introduce a comparative table between the standard
$k$-cosymplectic Lagrangian formalism and the case on Lie
algebroids:
 {\small
 $$\begin{array}{lll}
    & \makebox{\bf\underline{\qquad$k$-cosymplectic\qquad}} &
 \makebox{\bf \underline{\qquad Lie Algebroids\qquad}}
    \\
    \noalign{\bigskip}
  \mbox{Phase space} & \rk\times  T^1_kQ&  \mathbb{R}^k\times \stackrel{k}{\oplus}
E\\ \noalign{\medskip}
   \begin{array}{l} \mbox{Canonical} \\ \mbox{forms}\end{array}&
\omega^A\in\Lambda^2(\rk\times T^1_kQ )&
 \Omega^A\in
Sec((\mathcal{T}^E_k(\stackrel{k}{\oplus}E))^*)\wedge
Sec((\mathcal{T}^E_k(\stackrel{k}{\oplus}E))^*) \\
\noalign{\medskip}
 \mbox{Lagrangians}   &  L:\rk\times T^1_kQ\to \r &
L:\rk\times \stackrel{k}{\oplus} E\to\r\\
\noalign{\medskip}
  \begin{array}{l} \mbox{Geometric}\\ \mbox{equations}\end{array}   &
  {\small \left\{\begin{array}{l}
dt^A ((X_L)_B)  =   \delta^A_B\\
\noalign{\medskip} \ds\sum_{A=1}^k \, i_{(X_L)_A} \omega^A   = dE_L
+ \,\ds\sum_{A=1}^k\ds\frac{\partial L}{\partial t^A}dt^A
\end{array}\right.} & {\small \left\{
  \begin{array}{ccl}
    \overline{\mathcal{Y}}_B^C(\xi)&=&\delta_B^C\\\noalign{\medskip}
    \imath_{\xi}\Omega^A &=&\ds\frac{1}{k}\Big(d^A L+
\ds\sum_{C=1}^k\frac{\partial L}{\partial t^C}
\overline{\mathcal{Y}}_A^C\Big)
  \end{array}
\right.}\\\noalign{\medskip}&\begin{array}{c}
((X_L)_1,\ldots,(X_L)_K)\\\noalign{\bigskip} \makebox{$k$-vector
field on }\rk\times T^1_kQ\end{array} &\begin{array}{c} \xi\in
Sec(\mathcal{T}^E_k(\stackrel{k}{\oplus}E))\\\noalign{\medskip}
{\small\mathcal{T}^E_k(\stackrel{k}{\oplus}E)=(\rk\times
\stackrel{k}{\oplus}E)\times_{\rk\times T^1_kQ}
T^1_k(\rk\times\stackrel{k}{\oplus}E)}\end{array}
  \end{array}$$}

\subsection{Hamiltonian formalism}\label{ham form al}

In this subsection we develop on Lie algebroids the equivalent
to section \ref{ham form} in the standard $k$-cosymplectic
formalism.

We begin this section by introducing the manifold $\stackrel{k}{\oplus}
E^{\;*}$, which plays the role of $(T^1_k)^{\;*}Q$ in the classical
$k$-cosymplectic Hamiltonian setting.

 Let $(E,[\cdot,\cdot],\rho)$ be a Lie algebroid over a manifold
$Q$. For the Hamiltonian approach we consider the dual bundle,
$\tau^{\;*}:E^{\;*}\to Q$  of $E$.

\subsubsection{Geometric elements}\label{621}

\paragraph{The manifold $\stackrel{k}{\oplus} E^{\;*}$}\

The standard $k$-cosymplectic Hamiltonian formalism is developed on the
manifold $\rk\times (T^1_k)^*Q$. Considering a Lie algebroid $E$ as a substitute for the
tangent bundle, it is natural to consider that the analog of $(T^1_k)^*Q$
is the Whitney sum over $Q$ of $k$ copies of the dual space $E^*$.

We denote by $\stackrel{k}{\oplus} E^{\;*}=E^{\;*} \oplus
\stackrel{k}{\ldots} \oplus E^{\;*}$ the Whitney sum of $k$ copies
of the vector bundle $E^{\;*}$,  and the projection map
$\widetilde{\tau}^{\;*}:\oplus^k E^{\;*}\to Q$, which is
$\widetilde{\tau}^{\;*}({a_1}_\mathbf{q}^{\;*},\ldots,{a_k}_{\bf
q}^{\;*})= \mathbf{q}$.

\paragraph{Local basis of sections of $\widetilde{\tau}^{\;*}:\stackrel{k}{\oplus} E^{\;*}\to Q$}\

 Let
$a_\mathbf{q}^*=(a_{1_\mathbf{q}}^{\;*},\ldots,a_{k_\mathbf{q}}^{\;*})$
be an arbitrary point of $\stackrel{k}{\oplus} E^{\;*}$, since
$a_{A_\mathbf{q}}^{\;*}\in E^{\;*}$, and $\{e^\alpha\}$ is a local
basis of sections of $E^{\;*}$ ($\{e^\alpha\}$ is the dual basis of the
basis of sections of $E$, $\,\{e_\alpha\}$), we have
$a_{A_\mathbf{q}}^{\;*}=y_\alpha(a_{A_\mathbf{q}}^{\;*})
e^\alpha(\mathbf{q})$, then
$$
a_\mathbf{q}^*=y_\alpha(a_{1_\mathbf{q}}^*)
(e^\alpha(\mathbf{q}),0,\ldots, 0) + \ldots+
y_\alpha(a_{k_\mathbf{q}}^*) (0,\ldots, 0,
e^\alpha(\mathbf{q}))=\ds\sum_{A,\alpha}y_\alpha(a^*_{A_\mathbf{q}})\,
\widetilde{e}^{\;\alpha}_A(\mathbf{q})\;,
$$
where
$\widetilde{e}^{\;\alpha}_A(\mathbf{q})=(0,\ldots,\stackrel{A}{\widetilde{e^\alpha(\mathbf{q})}},\ldots,
0)$, and where $\stackrel{A}{\widetilde{\qquad}}$ indicates the
$A^{th}$ position of $\widetilde{e}^{\;\alpha}_ A(\mathbf{q})$.

Thus, $\{\widetilde{e}^{\;\alpha}_ A\}$ is a local basis of sections
of $\stackrel{k}{\oplus} E^{\;*}$, and if $(q^i,y_\alpha)$ are local
coordinates on $(\tau^{\;*})^{\;-1}(U)\subseteq E^{\;*}$, the
induced local coordinates $(q^i,y_\alpha^A)$ on
$(\widetilde{\tau}^{\;*})^{\;-1}(U)\subseteq \stackrel{k}{\oplus}
E^{\;*}$ are given by
$$
q^i(a_{1_\mathbf{q}}^{\;*},\ldots,a_{k_\mathbf{q}}^{\;*})=q^i(\mathbf{q})\,,\quad
y_\alpha^A(a_{1_\mathbf{q}}^{\;*},\ldots,a_{k_\mathbf{q}}^{\;*})=y_\alpha(a_{A_\mathbf{q}}^{\;*})\;.
$$

\paragraph{The vector bundle $\mathcal{T}^E_k(\stackrel{k}{\oplus}E^*)$}\

 We now consider the
$k$-prolongation $\mathcal{T}^E_k(\stackrel{k}{\oplus}E^*)\subset
\stackrel{k}{\oplus}E \times
T^1_k(\rk\times\stackrel{k}{\oplus}E^*)$
 of a Lie algebroid $E$ over the fibration
$\widetilde{\tau}^*:\stackrel{k}{\oplus}E^*\to Q$ ( let us observe
that in this case the fiber $\pi:P\to Q$ is $\widetilde{\tau}^*:
\stackrel{k}{\oplus} E^*\to Q$).
 The vector bundle $\mathcal{T}^E_k(\stackrel{k}{\oplus}E^*)$ plays the
role of $T^1_k( \rk\times (T^1_k)^*Q)\to  \rk\times (T^1_k)^*Q$, and
its sections corresponds to the $k$-vector fields on $ \rk\times
(T^1_k)^*Q$.

 Recalling section \ref{k-prol},
for this particular case we obtain that if $(q^i,y^A_\alpha)$ are
local coordinates on $\stackrel{k}{\oplus}E^*$, we have the local
coordinates $(t^A,q^i,y_\alpha^A,z^\alpha_A,v_A^B,(v_A)_\beta^B)$ on
$\mathcal{T}^E_k(\stackrel{k}{\oplus}E^*)$ given by (see (\ref{local
coord k-prol}))
$$
\begin{array}{lll}
t^A(a_\mathbf{q},W_{(\mathbf{t},b_\mathbf{q}^{\;*})})=t^A(\mathbf{t})\,,
& q^i(a_{\bf q},W_{(\mathbf{t},b_\mathbf{q}^{\;*})})=
q^i(\mathbf{q}) \,, &
y_\alpha^A(a_\mathbf{q},W_{(\mathbf{t},b_\mathbf{q}^{\;*})})=y^A_\alpha(b_{\bf
q}^{\;*})\,,
\\\noalign{\medskip}
z^\alpha_A(a_\mathbf{q},W_{(\mathbf{t},b_\mathbf{q}^{\;*})})=y_A^\alpha(a_{\bf
q})\,, &
v_A^B(a_\mathbf{q},W_{(\mathbf{t},b_\mathbf{q}^{\;*})})=
v_{A_{(\mathbf{t},b_\mathbf{q}^{\;*})}}(t^B)\,,
&
(v_A)_\beta^B(a_\mathbf{q},W_{(\mathbf{t},b_\mathbf{q}^{\;*})})=
v_A{_{(\mathbf{t},b_\mathbf{q}^{\;*})}}(y_\beta^B)\,, \\
\end{array}
$$
 and the local basis
$(\{\mathcal{X}_\alpha^A,\;\mathcal{Y}^A_B,\;(\mathcal{V}^A)_B^\beta\}$
of sections of $\tau_{\rk\times\stackrel{k}{\oplus}E^*}:
\mathcal{T}^E_k(\stackrel{k}{\oplus}E^*)\to \rk\times
\stackrel{k}{\oplus}E^*$,
defined in (\ref{base k-prol}), is written here as follows
\begin{equation}\label{base*}
\begin{array}{lcl}
\mathcal{X}_\alpha^A(\mathbf{t},b_{\mathbf{q}}^*)\; &=&
(\widetilde{e}^A_\alpha(\mathbf{q}),
(0,\ldots,\stackrel{A}{\widetilde{\rho^i_\alpha({\bf
q})\ds\frac{\partial }{\partial
q^i}\Big\vert_{(\mathbf{t},b_{\mathbf{q}}^*)}}},\ldots, 0))
\\\noalign{\medskip}
\mathcal{Y}^A_B(\mathbf{t},b_{\mathbf{q}}^*)& = &(
\mathbf{0}_q,(0,\ldots,
\stackrel{A}{\widetilde{\ds\frac{\partial}{\partial
t^B}\Big\vert_{(\mathbf{t},b_{\mathbf{q}}^*)}}},\ldots,0))
\\\noalign{\medskip}
(\mathcal{V}^A)_B^\beta(\mathbf{t},b_{\mathbf{q}}^*) &=&
(\mathbf{0}_q,(0,\ldots,
\stackrel{A}{\widetilde{\ds\frac{\partial}{\partial
y_\beta^B}\Big\vert_{(\mathbf{t},b_{\mathbf{q}}^*)}}},\ldots, 0))
\end{array}\ . \end{equation}

\subsubsection{Hamiltonian formalism}\label{623}

Let $(E,[\cdot,\cdot]_E,\rho)$ be a Lie algebroid on a manifold $Q$
and $H:\rk\times\stackrel{k}{\oplus} E^{\;*}\to \r$ a Hamiltonian
function.

\paragraph{The Liouville sections}\

We may introduce $k$ sections of the vector bundle
$(\mathcal{T}^E_k(\stackrel{k}{\oplus}E^*))^{\;*}\to
\rk\times\stackrel{k}{\oplus} E^{\;*}$ as follows.
$$\begin{array}{rcllcl}
\Theta^A:\rk\times\stackrel{k}{\oplus} E & \longrightarrow &
(\mathcal{T}^E_k(\stackrel{k}{\oplus}E^*))^{\;*} & & &
\\\noalign{\medskip}
(\mathbf{t},b_\mathbf{q}^{\;*}) & \longmapsto &
\Theta^A_{(\mathbf{t},b_{\bf q}^{\;*})}: &
(\mathcal{T}^E_k(\stackrel{k}{\oplus}E^*)
)_{(\mathbf{t},b_\mathbf{q}^{\;*})}& \longrightarrow & \r
\\\noalign{\medskip}
  &  &  & (a_\mathbf{q},W_{(\mathbf{t},b_\mathbf{q}^{\;*})}) & \longmapsto &
  \Theta^A_{(\mathbf{t},b_\mathbf{q}^{\;*})}(a_\mathbf{q},W_{(\mathbf{t},b_\mathbf{q}^{\;*})})=
b_{A_\mathbf{q}}^{\;*}(a_{A_\mathbf{q}})\;.
\end{array}$$

In local coordinates we have
\begin{equation}\label{theta*}
\Theta^A=\ds\sum_{\beta}y^A_\beta\mathcal{X}^\beta_A\;,
\end{equation} where $\{\mathcal{X}^\alpha_A,\,\overline{\mathcal{Y}}_A^B,\,
(\mathcal{V}_A)^B_\beta\}$ is the local  basis of sections of
$(\mathcal{T}^E_k(\stackrel{k}{\oplus} E)^* )^{\;*}$, which is the
dual basis of the local basis
$\{\mathcal{X}_\alpha^A,\,\mathcal{Y}^A_B,\,
(\mathcal{V}^A)_B^\beta\}$ of sections of
$\mathcal{T}^E_k(\stackrel{k}{\oplus} E) $.

Now for each $A$ we define the $2$-section
$\Omega^A:\rk\times\stackrel{k}{\oplus}
E^{\;*}\to(\mathcal{T}^E_k(\stackrel{k}{\oplus}E^*))^*\wedge
(\mathcal{T}^E_k(\stackrel{k}{\oplus}E^*))^*$ as
 \[\Omega^A =-d^A\Theta^A\;,\]
 where $d^A$ denotes the derivation introduced in
 (\ref{dif k-prol}) with $P=\stackrel{k}{\oplus}E^*$.

By a straightforward computation from  (\ref{k-vector asoc k-prol}),
(\ref{lie brack k-prol}), (\ref{dif k-prol}), (\ref{base*}) and
(\ref{theta*}), we obtain
\begin{equation}\label{omega A*}
\Omega^A=\sum_{\beta}\mathcal{X}^\beta_A\wedge(\mathcal{V}_A)^A_\beta
+ \ds\frac{1}{2} \sum_{\beta,\gamma,\delta}
\mathcal{C}^\delta_{\beta\gamma}y^A_\delta\mathcal{X}_A^\beta\wedge
\mathcal{X}_A^\gamma\ .
\end{equation}

\paragraph{Hamilton's equations}\

 Let $H:\rk\times\stackrel{k}{\oplus} E^{\;*}\to \r$ be a Hamiltonian function.
For an arbitrary section $\xi:\rk\times\stackrel{k}{\oplus} E^{\;*}
\to \mathcal{T}^E_k(\stackrel{k}{\oplus}E^*)$ of
$\widetilde{\tau}_{\rk\times\stackrel{k}{\oplus}E^*}:\mathcal{T}^E_k(\stackrel{k}{\oplus}E^*)\to
\rk\times\stackrel{k}{\oplus} E^{\;*}$,
we consider the system of equations
\begin{equation}\label{EC H}
    \overline{\mathcal{Y}}_A^C(\xi)=\delta_A^C\quad ,\quad
    \imath_{\xi}\Omega^A =\ds\frac{1}{k}\Big(d^A H-
\ds\sum_{B=1}^k\frac{\partial H}{\partial
t^B}\overline{\mathcal{Y}}_A^B\Big)\; .
  \end{equation}

Writing $\xi= \xi_B^C\mathcal{Y}^B_C+\xi_B^\alpha
\mathcal{X}^B_\alpha + (\xi_B)_\alpha^C (\mathcal{V}^B)_C^\alpha $,
from (\ref{dif func}), (\ref{base*}) and (\ref{omega A*}) we obtain
that (\ref{EC H}) is equivalent to the equations
\begin{equation}\label{localxi-1}
\xi_B^C = \delta^B_C\quad,\quad\delta^A_B\xi^\alpha_A
=\ds\frac{1}{k} \ds\frac{\partial H}{\partial y^B_\alpha}\quad
,\quad (\xi_A)^A_\beta - \mathcal{C}^\gamma_{\alpha\beta}y^A_\gamma
\xi_A^\alpha = -\ds\frac{1}{k} \rho^i_\beta\ds\frac{\partial
H}{\partial q^i}\ .
\end{equation}

From (\ref{localxi-1}) we obtain
\begin{equation}\label{localxi-2}
\xi_B^C = \delta^B_C\quad,\quad
 \ds\sum_{A=1}^k\delta^A_B\xi^\alpha_A= \xi^\alpha_B= \ds\frac{\partial H}{\partial y^B_\alpha}\quad
,\quad \ds\sum_{A=1}^k(\xi_A)^A_\beta - \ds\sum_{A=1}^k
\mathcal{C}^\gamma_{\alpha\beta}y^A_\gamma \xi_A^\alpha =-
\rho^i_\beta\ds\frac{\partial H}{\partial q^i}\,.
\end{equation}

Substituting the second identities of (\ref{localxi-2}) in the three equations, we have
\begin{equation}\label{localxi-3}
\xi_B^C = \delta^B_C\quad,\quad
 \xi^\alpha_B= \ds\frac{\partial H}{\partial y^B_\alpha}\quad
,\quad \ds\sum_{A=1}^k(\xi_A)^A_\beta= - \Big(
\rho^i_\beta\ds\frac{\partial H}{\partial q^i}+\ds\sum_{A=1}^k
\mathcal{C}^\gamma_{\alpha\beta}y^A_\gamma \ds\frac{\partial
H}{\partial y^A_\alpha}\Big) \,.
\end{equation}

Let $\underline{\psi}:\rk\to
\stackrel{k}{\oplus}E^*,\;\underline{\psi}(t) =
(\psi^A(t),\psi^i(t),\psi^A_\alpha(t)) $ be  an integral section of
 $\xi$ ,that is, $\underline{\psi}$ is an integral section
of the associated $k$-vector field
$\widetilde{\tau}_2(\xi)=(\widetilde{\tau}_2^1 (\xi),\ldots,
\widetilde{\tau}_2^k (\xi))$ on $\rk\times\stackrel{k}{\oplus} E^*$.
 Thus
 \begin{equation}\label{sint1} \xi_A^B=\ds\frac{\partial
\psi^B}{\partial
 t^A}\;,\;
 \xi^\beta_A \rho^i_\beta = \ds\frac{\partial \psi^i}{\partial
 t^A}\;,\;
  (\xi_A)_\beta^B = \ds\frac{\partial \psi^B_\beta}{\partial
  t^A}\;.
\end{equation}

From (\ref{localxi-3}) and (\ref{sint1}) we obtain
{\small\begin{equation}\label{h eq al}
\derpar{\psi^{\;C}}{t^B}\Big\vert_{t} = \delta^B_C\;,\;
\derpar{\psi^{\;i}}{t^A}\Big\vert_{t}
=\rho^i_\alpha\ds\frac{\partial H}{\partial
y^B_\alpha}\Big\vert_{\psi(t)}\;,\;
\ds\sum_{A=1}^k\derpar{\psi^{\;A}_\beta}{t^A}= - \Big(
 \rho^i_\beta\ds\frac{\partial H}{\partial
q^i}\Big\vert_{\psi(t)}+\ds\sum_{A=1}^k
\mathcal{C}^\gamma_{\alpha\beta}\psi^A_\gamma(t) \ds\frac{\partial
H}{\partial y^A_\alpha}\Big\vert_{\psi(t)}\Big)\,.
\end{equation}}

In the standard case, a solution of the Hamilton equations is a
section $\psi:\r^k\to \rk\times(T^1_k)^*Q$ locally given by
$\psi(t)=(t,\psi^i(t),\psi^A_i(t))$ which satisfies the equations
(\ref{he}).
Let us observe that giving a map $\psi:\r^k\to \rk\times(T^1_k)^*Q$
is equivalent to giving the following morphism of Lie  algebroids:
\[\xymatrix{T\rk \ar[r]^-{T\psi}\ar[d]_-{\tau_{\rk}}& T(\rk\times(T^1_k)^*Q)\equiv
\mathcal{T}^{TQ}(\rk\times(T^1_k)^*Q)\ar[d]^-{\tau_{\rk\times(T^1_k)^*Q}}\\
\rk \ar[r]_-{\psi}& \rk\times(T^1_k)^*Q }\]

In our case, a solution to the Hamilton equations must
be a morphism $\psi=(\overline{\psi},\underline{\psi})$ of Lie
algebroids between $\tau_{\rk}:T\rk\to\rk$ and
$\tau^E_{\rk\times\stackrel{k}{\oplus}
E^*}:\mathcal{T}^E(\rk\times\stackrel{k}{\oplus} E^*)\subset E\times
T(\rk\times\stackrel{k}{\oplus} E^*) \to
\rk\times\stackrel{k}{\oplus} E^*$, where
$\mathcal{T}^E(\rk\times\stackrel{k}{\oplus} E^*)$ is the
prolongation of the Lie algebroid $E$ over the fibration $\rk\times
\stackrel{k}{\oplus}E^*\to Q$.
\[\xymatrix{T\rk \ar[r]^-{\overline{\psi}}\ar[d]_-{\tau_{\rk}}&
\mathcal{T}^E(\rk\times\stackrel{k}{\oplus} E^*)
\ar[d]^-{\tau^E_{\rk\times\stackrel{k}{\oplus} E^*}}\\
\rk \ar[r]_-{\underline{\psi}}& \rk\times\stackrel{k}{\oplus} E^*}\]

Let $\{e_\alpha\}$ and
$\{\mathcal{X}_\alpha,\mathcal{V}_A,\mathcal{V}^\alpha_A\}$ be a
local basis of $Sec(\tau_{\rk})$ and
$Sec(\tau^E_{\rk\times\stackrel{k}{\oplus} E^*})$, respectively, and
$\{e^\alpha\}$ and $\{\mathcal{X}^\alpha,\mathcal{V}^A,\mathcal{V}_
\alpha^A\}$ their dual basis. (Here
$\mathcal{X}_\alpha(t,b_\mathbf{q}^*)=(e_\alpha(\mathbf{q}),\rho^i_\alpha(\mathbf{q})\ds\frac{\partial}{\partial
q^i}\Big\vert_{(t,b_\mathbf{q}^*)}),\;
\mathcal{V}_A(t,b_\mathbf{q}^*)=(0_{\mathbf{q}},\ds\frac{\partial}{\partial
t^A}\Big\vert_{(t,b_\mathbf{q}^*)})$ and
$\mathcal{V}^\alpha_A(t,b_\mathbf{q}^*)=(0,\ds\frac{\partial}{\partial
y^A_\alpha}\Big\vert_{(t,b_\mathbf{q}^*)}))$.
If $\psi=(\underline{\psi},\overline{\psi})$ is locally given by the
relations
$$
\begin{array}{lcl}
  \underline{\psi}(t) = (t^A,\psi^i(t),\psi^A_\alpha(t)) &\quad ,\quad &
  \psi = (\psi^\alpha_ A\mathcal{X}_\alpha + \mathcal{V}_A + \psi^B_{\beta
  A}\mathcal{V}^\beta_B)\otimes e^A
\end{array}$$ then the morphism condition (\ref{morp cond}) can be written
\begin{equation}\label{morpcondham}
 \rho^i_
 \alpha\psi^\alpha_A = \ds\frac{\partial \psi^i}{\partial t^A} \; ,\;
 \psi^B_{\beta A}=\ds\frac{\partial \psi^B_\beta}{\partial t^A}\; ,\;
 0=\ds\frac{\partial \psi^\alpha_A}{\partial t^B} -
\ds\frac{\partial \psi^\alpha_B}{\partial t^A}
+\mathcal{C}^\alpha_{\beta\gamma}\psi^\beta_B\psi^\gamma_A \ .
\end{equation}
\,From (\ref{h eq al}) and (\ref{morpcondham}) we obtain the
following:

\begin{theorem}\label{alhamform}
Let $H:\rk\times\stackrel{k}{\oplus}E^*\to \r$ be a  Hamiltonian and
$\xi:\rk\times\stackrel{k}{\oplus}E \to
\mathcal{T}^E_k(\stackrel{k}{\oplus}E^*)$ be a section  of
$\widetilde{\tau}_{\stackrel{k}{\oplus}E^*}$ such that
\[
 \overline{\mathcal{Y}}_A^C(\xi)=\delta_A^C\quad ,\quad
    \imath_{\xi}\Omega^A =\ds\frac{1}{k}\Big(d^A H-
\ds\sum_{B=1}^k\frac{\partial H}{\partial
t^B}\overline{\mathcal{Y}}_A^B\Big)\ .\]
 Let $\psi=(\overline{\psi},\underline{\psi})$ be  a morphism  of Lie
algebroids between $\tau_{\rk}:T\rk\to\rk$ and
$\tau^E_{\rk\times\stackrel{k}{\oplus}
E^*}:\mathcal{T}^E(\stackrel{k}{\oplus} E^*)\subset E\times
T(\rk\times\stackrel{k}{\oplus} E^*) \to
\rk\times\stackrel{k}{\oplus} E^*$. If
$\underline{\psi}:\rk\to\rk\times \stackrel{k}{\oplus}E^*\,$  is an
integral section of $\xi$, then $\psi$ is a solution of the system
of partial differential equations
\beann
 \label{Hamilton eq}\nonumber\derpar{\psi^{\;i}}{t^A}\Big\vert_{t}
=\rho^i_\alpha\ds\frac{\partial H}{\partial
y^B_\alpha}\Big\vert_{\psi(t)}\;,\;
\ds\sum_{A=1}^k\derpar{\psi^{\;A}_\beta}{t^A} &=& - \Big(
 \rho^i_\beta\ds\frac{\partial H}{\partial
q^i}\Big\vert_{\psi(t)}+\ds\sum_{A=1}^k
\mathcal{C}^\gamma_{\alpha\beta}\psi^A_\gamma(t) \ds\frac{\partial
H}{\partial
y^A_\alpha}\Big\vert_{\psi(t)}\Big)\,.\\\noalign{\medskip} \rho^i_
 \alpha\psi^\alpha_A = \ds\frac{\partial \psi^i}{\partial t^A} \; ,\;
 \psi^B_{\beta A}=\ds\frac{\partial \psi^B_\beta}{\partial t^A}\; ,\;
 0 &=&\ds\frac{\partial \psi^\alpha_A}{\partial t^B} -
\ds\frac{\partial \psi^\alpha_B}{\partial t^A}
+\mathcal{C}^\alpha_{\beta\gamma}\psi^\beta_B\psi^\gamma_A \ ,
\eeann
which are called the {\rm Hamilton field equations on Lie algebroids}.
\end{theorem}

\subsubsection{Relation with the standard Hamiltonian $k$-cosymplectic formalism}\label{624}

As a final remark, it is interesting to point out that the standard
Hamiltonian $k$-symplectic formalism   is a particular case of the
Hamiltonian formalism on Lie algebroids.

 In this case we have:
 \begin{itemize}
\item The manifold $\rk\times \stackrel{k}{\oplus}E^*$ is identified with $\rk\times (T^1_k)^*Q$
and $\mathcal{T}^{TQ}_k((T^1_k)^*Q)$ with $T^1_k(\rk\times
(T^1_k)^*Q)$.
\item A section $\xi:\rk\times \stackrel{k}{\oplus}E^* \to
\mathcal{T}^E_k(\stackrel{k}{\oplus}E^*)$ corresponds to a $k$-vector
field $\xi=(\xi_1,\ldots,\xi_k)$ on $\rk\times (T^1_k)^*Q$, that is,
$\xi$ is a section of $\tau^k_{\rk\times (T^1_k)^*Q}:T^1_k(\rk\times
(T^1_k)^*Q) \to \rk\times (T^1_k)^*Q$.
\item Let $f$ be a function on $\rk\times (T^1_k)^*Q$, then
$$d^Af(Y_1,\ldots, Y_k)=df(Y_A)\  ,$$
where $df$ denotes the standard differential and $(Y_1,\ldots, Y_k)$
is a $k$-vector field on $\rk\times (T^1_k)^*Q$.
\item
 We have that \begin{eqnarray*}
\overline{\mathcal{Y}}_B^C((X_1,\ldots, X_k))&=& dt^B(X_C)\\
\Omega^A((X_1,\ldots, X_k),((Y_1,\ldots, Y_k))&=&\omega^A(X_A,Y_A)\
.\end{eqnarray*}
\item
 Thus, in the standard $k$-cosymplectic formalism
the equation (\ref{EC H}) can be written as follows
$$dt^A(\xi_B)=\delta^A_B\quad ,\quad i_{\xi_A}\omega^A=
\ds\frac{1}{k}dH - \,\ds\sum_{C=1}^k\ds\frac{\partial H}{\partial
t^C}dt^A\delta^\Delta_A\ ,$$
 which implies
$$dt^A(\xi_B)=\delta^A_B\quad ,\quad\ds\sum_{A=1}^ki_{\xi_A}\omega^A=
dH - \,\ds\sum_{A=1}^k\ds\frac{\partial H}{\partial t^A}dt^A\,.$$
\end{itemize}

Thus, from the Theorem \ref{alhamform} and the above remarks, we
deduce the following corollary, which summarizes the Hamiltonian
$k$-cosymplectic formalism.

\begin{corollary}Let $H:\rk\times\stackrel{k}{\oplus}E^*\to \r$ be a Hamiltonian function and
$\xi=(\xi_1,\ldots, \xi_k)$ a $k$-vector field on $\rk\times
(T^1_k)^*Q$ such that
$$dt^A(\xi_B)=\delta^A_B\quad ,\quad
\ds\sum_{A=1}^ki_{\xi_A}\omega^A=dH -\,\ds\sum_{A=1}^k\ds\frac{\partial H}{\partial t^A}dt^A\ .
$$
Thus, if $\psi:\rk\to\rk\times (T^1_k)^*Q$ is an integral section of
the $k$-vector field $\xi$, then it is a solution of the
 Hamilton equations (\ref{he}).
\end{corollary}

In the following table we compare the $k$-cosymplectic Hamiltonian
formalism in the standard case and on Lie algebroids:
 {\small
$$\begin{array}{lll}
    & \makebox{\bf\underline{\qquad$k$-cosymplectic\qquad}} &
 \makebox{\bf \underline{\qquad Lie Algebroids\qquad}}
    \\
    \noalign{\bigskip}
  \mbox{Phase space} & \rk\times  (T^1_k)^*Q&  \mathbb{R}^k\times \stackrel{k}{\oplus}
E^*\\ \noalign{\medskip}
   \begin{array}{l} \mbox{Canonical} \\ \mbox{forms}\end{array}&
\omega^A\in\Lambda^2(\rk\times(T^1_k)^*Q )&
 \Omega^A\in
Sec((\mathcal{T}^E_k(\stackrel{k}{\oplus}E^*))^*)\wedge
Sec((\mathcal{T}^E_k(\stackrel{k}{\oplus}E^*))^*) \\
\noalign{\medskip}
 \mbox{Hamiltonians}   &  H:\rk\times (T^1_k)^*Q\to \r &
H:\rk\times \stackrel{k}{\oplus} E^*\to\r\\
\noalign{\medskip}
  \begin{array}{l} \mbox{Geometric}\\ \mbox{equations}\end{array}   &
  {\small \left\{\begin{array}{l}
dt^A ((X_H)_B)  =   \delta^A_B \\
\noalign{\medskip} \ds\sum_{A=1}^k \, i_{(X_H)_A} \omega^A   = dH -
\,\ds\sum_{A=1}^k\ds\frac{\partial H}{\partial t^A}dt^A
\end{array}\right.} & {\small \left\{
  \begin{array}{ccl}
    \overline{\mathcal{Y}}_B^C(\xi)&=&\delta_B^C \\\noalign{\medskip}
    \imath_{\xi}\Omega^A &=&\ds\frac{1}{k}\Big(d^A H-
\ds\sum_{C=1}^k\frac{\partial H}{\partial t^C}
\overline{\mathcal{Y}}_A^C\Big)
  \end{array}
\right.}\\\noalign{\medskip}&\begin{array}{c}
((X_H)_1,\ldots,(X_H)_K)\\\noalign{\bigskip} \makebox{$k$-vector
field on }\rk\times (T^1_k)^*Q\end{array} &\begin{array}{c} \xi\in
Sec(\mathcal{T}^E_k(\stackrel{k}{\oplus}E^*))\\\noalign{\medskip}
{\small\mathcal{T}^E_k(\stackrel{k}{\oplus}E^*)=(\rk\times
\stackrel{k}{\oplus}E)\times_{\rk\times T^1_kQ}
T^1_k(\rk\times\stackrel{k}{\oplus}E^*)}\end{array}
  \end{array}$$}

\subsection*{Acknowledgments}

 We acknowledge the partial financial support of {\sl
Ministerio de Educaci\'on y Ciencia}, Project MTM2006-27467-E/. The
author NRR also acknowledges the financial support of {\sl
Ministerio de Educaci\'on y Ciencia}, Project MTM2005-04947. We wish
to thank to Mr. Jeff Palmer for his assistance in preparing the
English version of the manuscript.

\end{document}